\documentclass[useAMS,usenatbib]{mnras}
\usepackage{aecompl}
\usepackage{graphicx}
\usepackage{amssymb}
\usepackage{amsmath}
\usepackage{multicol}
\usepackage{longtable}
\usepackage{longtable}
\usepackage{float}
\usepackage{enumitem}
\usepackage{footmisc}
\usepackage{rotating}
\usepackage{changepage}
\usepackage{subcaption}

\usepackage{BibDef}

\usepackage{etoolbox}
\makeatletter
\patchcmd\@combinedblfloats{\box\@outputbox}{\unvbox\@outputbox}{}{%
   \errmessage{\noexpand\@combinedblfloats could not be patched}%
}%
 \makeatother

\usepackage{xcolor,hyperref}

\newcommand{\krome}{\textsc{Krome}}

\title[Disk thermochemistry I: thermochemical model]{Modelling thermochemical processes in protoplanetary disks I: numerical methods.}

\author[T. Grassi et al.]{\parbox{\textwidth}{T.~Grassi$^{1,2}$\thanks{Corresponding author: tgrassi@usm.lmu.de}\href{https://orcid.org/0000-0002-3019-1077}{\includegraphics[scale=0.04]{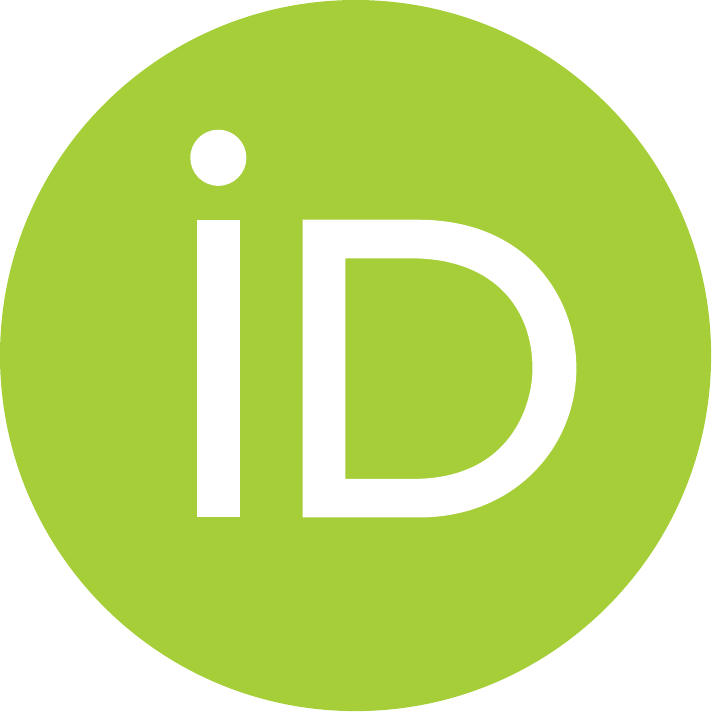}}, B.~Ercolano$^{1,2}$, L.~Sz\H{u}cs$^{3}$, J.~Jennings$^{4}$\href{https://orcid.org/0000-0002-7032-2350}{\includegraphics[scale=0.04]{figs/orcid}}, and G.~Picogna$^{1,2}$\href{https://orcid.org/0000-0003-3754-1639}{\includegraphics[scale=0.04]{figs/orcid}}}\vspace{0.4cm}\\
\parbox{\textwidth}{
$^{1}$Universit\"ats-Sternwarte M\"unchen, Scheinerstr. 1, D-81679 M\"unchen, Germany\\
$^{2}$Excellence Cluster Origin and Structure of the Universe, Boltzmannstr.2, D-85748 Garching bei M\"unchen, Germany\\
$^{3}$Max-Planck-Institut f\"{u}r extraterrestrische Physik, Giessenbachstrasse 1, 85748 Garching, Germany\\
$^{4}$Institute of Astronomy, University of Cambridge, Madingley Road, Cambridge, CB3 0HA, UK\\
}}

\begin{document}

\newcommand{\enzo}{\textsc{Enzo}}
\newcommand{\flash}{\textsc{Flash}}
\newcommand{\ramses}{\textsc{Ramses}}
\newcommand{\fortran}{\textsc{Fortran}}
\newcommand{\python}{\textsc{Python}}
\newcommand{\kida}{\textsc{KIDA}}
\newcommand{\dlsodes}{\textsc{DLSODES}}
\newcommand{\linktocode}{\url{https://bitbucket.org/tgrassi/lemongrab/}}
\newcommand{\ith}{$i$th }
\newcommand{\jth}{$j$th }
\newcommand{\nth}{$n$th }
\newcommand{\kth}{$k$th }
\newcommand{\ellth}{$\ell$th }

\newcommand{\dd}{\mathrm d}
\newcommand{\mA}{\mathrm A}
\newcommand{\mB}{\mathrm B}
\newcommand{\mC}{\mathrm C}
\newcommand{\mD}{\mathrm D}
\newcommand{\mE}{\mathrm E}
\newcommand{\mH}{\mathrm H}
\newcommand{\mHe}{\mathrm{He}}
\newcommand{\me}{\mathrm e}
\newcommand{\mSi}{\mathrm Si}
\newcommand{\mO}{\mathrm O}
\newcommand{\mX}{\mathrm X}
\newcommand{\cmc}{\mathrm{cm}^{-3}}
\newcommand{\real}{\mathbb R}
\newcommand{\superscript}[1]{\ensuremath{^{\scriptscriptstyle\textrm{#1}\,}}}
\newcommand{\trader}{\superscript{\textregistered}}

\newcommand{\eqn}[1]{Eq.~(\ref{#1})}
\newcommand{\theireqn}[1]{Eq.~(#1)}
\newcommand{\eqnrange}[2]{Eqs.~(\ref{#1}) to (\ref{#2})}
\newcommand{\sect}[1]{Sect.~\ref{#1}}
\newcommand{\fig}[1]{Fig.~\ref{#1}}
\newcommand{\tab}[1]{Tab.~\ref{#1}}
\newcommand{\appx}[1]{Appendix~\ref{#1}}
\newcommand{\file}[1]{\texttt{#1}}

\newcommand{\beq}{\begin{equation}}
\newcommand{\eeq}{\end{equation}}
\newcommand{\beqa}{\begin{eqnarray}}
\newcommand{\eeqa}{\end{eqnarray}}

\newcommand{\codename}{\textsc{prizmo}}

\newcommand{\tgas}{T}
\newcommand{\tdust}{T_{\rm d}}
\newcommand{\ngas}{n_{\rm gas}}
\newcommand{\dtog}{\mathcal{D}}
\newcommand{\dex}{{\rm dex}}

\newcommand{\arl}[1]{\url{#1}}

\renewcommand\arraystretch{1.7}

\newcommand{\tgcomment}[1]{{\bf #1}}
\newcommand{\jprcomment}[1]{[\textcolor{magenta}{#1 /JPR}]}

\newcommand\osu{osu\_01\_2007}
\def\tm{\leavevmode\hbox{$\rm {}^{TM}$}}

\date{Accepted *****. Received *****; in original form ******}

\pagerange{\pageref{firstpage}--\pageref{lastpage}} \pubyear{2099}

\maketitle

\label{firstpage}

\begin{abstract}
The dispersal phase of planet-forming disks via winds driven by irradiation from the central star and/or magnetic fields in the disk itself is likely to play an important role in the formation and evolution of planetary systems. Current theoretical models lack predictive power to adequately constrain observations. We present \codename{}, a code for evolving thermochemistry in protoplanetary disks capable of being coupled with hydrodynamical and multi-frequency radiative transfer codes. We describe the main features of the code, including gas and surface chemistry, photochemistry, microphysics, and the main cooling and heating processes. The results of a suite of benchmarks, which include photon-dominated regions, slabs illuminated by radiation spectra that include X-ray, and well-established cooling functions evaluated at different temperatures show good agreement both in terms of chemical and thermal structures.
The development of this code is an important step to perform quantitative spectroscopy of disk winds, and ultimately the calculation of line profiles, which is urgently needed to shed light on the nature of observed disk winds.
\end{abstract}


\section{Introduction}

Planets form from the dust and gas in the circumstellar discs surrounding young, low mass stars. The surface density evolution of these planet-forming disks as well as the mechanisms to finally disperse the gas are likely to play an important role in the formation of planetary systems (e.g.~\citealt{Throop2005, Drazkowska2016, Carrera2018, Ercolano2018}) and the evolution and migration of young planets \citep{Alexander2012, Ercolano2015, Jennings2018}.

Photoevaporation by high energy radiation from the central star as well as magnetic fields are thought to drive vigorous disk winds capable of shaping the evolution and the final dispersal of planet-forming material (see \citealt{Armitage2011, Alexander2014, Ercolano2017} for recent reviews and \citealt{Kunitomo2020, Rodenkirch2020} for a discussion of the interaction between these processes). However, the efficiency of these winds as predicted by various models spans several order of magnitudes with theoretical calculations thus far yielding fairly weak observational constraints.

Current studies have been mainly limited to a comparison of the X-ray and EUV-only photoevaporation model \citep{Alexander2006b, Alexander2006a, Owen2010, Owen2011, Owen2012, Picogna2019} of forbidden lines from singly ionised neon, neutral oxygen and singly ionised sulphur and nitrogen (e.g.~\citealt{Glassgold2007, Alexander2008, Ercolano2010, Schisano2010, Ercolano2016}) to observational surveys (e.g.~\citealt{Pascucci2009, Rigliaco2013, Natta2014, Simon2016, Banzatti2019}). Unfortunately, the strong temperature dependence of collisionally excited lines makes them unsuitable to probe the bulk of the wind in the launching regions. Indeed, the theoretical calculations, while being partially successful in matching some of the properties of the observed emission lines, also highlight the need to consider molecular diagnostics which may be able to better sample the wind launching regions. Unfortunately, the predictive power of molecular line intensities and profiles from current theoretical models is not yet sufficient for this task (see discussion in \citealt{Ercolano2017}).

This work is the first in a series of papers that will allow modellers to perform synthetic observations of disk winds (magnetic and/or photoevaporative) to identify and analyse diagnostics and determine origins in the disc. Paper~II of this series (Sz\H{u}cs et al.) describes the relevant chemical processes in disk winds and atmospheres and provides a more detailed description of the currently available chemical codes and their limitations for disk winds.

Here we present \codename{}\footnote{The code will be publicly available at \url{https://bitbucket.org/tgrassi/prizmo/} together with Paper~III, where the code will be employed to compute the thermal structure of a disk.}, a code designed to advance chemical abundances and temperature by a time-step (and the impinging radiation flux accordingly)  in a single cell that can be part either of a hydrodynamical or of a multifrequency radiative transfer code. \codename{} is a flexible yet relatively fast code that can be adapted to a set of astrophysical problems requiring gas- and dust-phase chemistry, photochemistry, and the evaluation of a wide range of thermochemical processes, i.e.~heating and cooling.

The study of chemistry in protoplanetary disks, when coupled with radiative transfer and/or hydrodynamics, has been undertaken by several codes with different levels of complexity. Notable examples are \citet{Ilgner2006}, where chemistry has been employed to determine the ionization fraction of disks, or more recently \citet{Wang2017}, where winds driven by ultraviolet and X-ray radiation are studied via 2D hydrodynamic simulations coupled with simplified radiative transfer and thermochemistry. \citet{Ilee2017} studied the fragmentation of a disk coupling smoothed particle hydrodynamics, including radiative transfer, and time-dependent chemical evolution, while \citet{Booth2019} studied the interplay between chemical evolution and pebble drift in planet-forming disks.

Other well-established numerical frameworks for combined radiative transfer and chemistry in disks include \textsc{ProDiMo} \citep{Woitke2009}, a code that includes chemistry, X-rays and FUV radiative transfer, heating and cooling, and the capability of determining the equilibrium disk structure; \textsc{DALI} \citep{Bruderer2009}, a code with dust radiative transfer, chemistry, heating and cooling balance, and disk structure calculation; \textsc{ANDES} \citep{Akimkin2013}, with 1+1D frequency-dependent radiative transfer, gas-grain chemical evolution, thermal energy balance, and dust grain evolution; \citet{Cleeves2013} developed a 2D disk model of a T-Tauri star system, including FUV and X-ray photons, grain settling, isotope chemistry, and a detailed discussion on the effect of the cosmic-rays; \textsc{TORUS-3DPDR} \citep{Bisbas2015} couples hydrodynamics, radiation transport, and PDR chemical and physical calculation, to compare observational and theoretical results.

In \sect{sect:chemistry} we present the algorithm employed to model chemical reactions in the gas phase and on the dust grains, as well as photochemistry. In \sect{sect:thermo} we review the thermochemical processes included in \codename{}, i.e.~cooling and heating, and in \sect{sect:benchmarks} we compare the output from our code with the results from benchmarks of photo-dominated regions. The chemical and thermochemical databases employed are reported in \sect{sect:database}, the limitations are discussed in \sect{sect:limitations}, while the summary is in \sect{sect:conclusions}.

\section{Chemistry}\label{sect:chemistry}

\subsection{Methodology}
The core of the code is the \textsc{DLSODES} solver \citep{Hindmarsh2005}, that evolves the following system of ordinary differential equations (ODE)
\beqa\label{eqn:ode}
  \frac{\dd n_i}{\dd t} & = & -n_i\sum_j k_{ij}(T, \bar J) n_j + \sum_{jl} k_{jl}(T, \bar J) n_j n_l\\
  \frac{\dd \tgas}{\dd t} & = & (\gamma - 1) \frac{\Gamma(\bar n, \tgas, \bar J) - \Lambda(\bar n, \tgas, \tdust)}{k_{\rm B} \ngas}\label{eqn:ode_thermo}
\eeqa
where $n_i$ is the number density of the \ith chemical species (either gas- or dust-phase), $k_{ij}$ is the reaction rate coefficient of the reaction occurring between the \ith and the \jth chemical species, $\tgas$ is the kinetic temperature of the gas, $\bar n$ is the set of the chemical abundances, $\bar J$ is the array of the impinging radiation intensities in the different energy bins, $\tdust$ is the dust temperature that is solved using radiative equilibrium and described in \sect{sect:dust_cooling}, $\gamma$ the adiabatic index, $\Gamma$ and $\Lambda$ the heating and cooling processes respectively, and $k_{\rm B}$ is the Boltzmann's constant.

Since \codename{} is employed as a library called by each cell of a framework code (e.g.~hydrodynamical or radiative transfer), it is necessary to provide some information about the global geometry of the problem, depending if multifrequency or standard Draine's field \citep{Draine1978} is used. In particular, the framework code needs to provide at runtime the column density of H$_2$, CO, and N$_2$ integrated from the radiation source to the evolving cell, as well as the column density of  H$_2$, CO, and H$_2$O from the current cell toward the radiation escape surface, e.g.~the vertical column density when dealing with a protoplanetary disk. These quantities are required to compute self-shielding and cooling efficiency (see~\sect{sect:CO_H2O_cooling}) when these molecules are present in the network.

The code also needs to know the cosmic-rays ionization rate, and some information about the grain size distribution, i.e.~the limits in size, the slope of the power law, and the dust-to-gas mass ratio.

Following the approach of \krome{} \citep{Grassi2014}, \codename{} uses a \python{} pre-processor to write optimized \fortran{} code, taking advantage of several numerical methods to reduce the global computational footprint. We remark that \codename{} is designed to evolve chemistry \emph{coupled with thermochemical processes and multifrequency radiation}. When employed to solve standard chemistry (i.e.~without coupled thermal processes) it only has similar computational performances to codes that use \dlsodes{} or analogous ODE solvers (e.g.~\citealt{Semenov2010, Wakelam2012, Ruaud2016}).

\subsection{Standard gas chemistry}
Chemical networks are provided to the code in an easily human-readable format. The reactions and the species are parsed to be converted into corresponding \python{} objects, that share a set of attributes (mass, charge, etc\dots) and methods (checking mass conservation, parsing textual format, etc\dots). Most of the information are obtained from databases rather than from the user. For example, rate coefficients can be defined directly by their analytical expression, but also obtained from well-established databases, such as \kida{} for chemical reaction rates\footnote{\url{http://kida.astrophy.u-bordeaux.fr/}} \citep{Wakelam2012} or Verner's astrophysical data collection\footnote{\url{http://www.pa.uky.edu/~verner/atom.html}} \citep{Verner1996}. Analogously, chemical species' thermochemical properties are from Burcat's database\footnote{\url{http://garfield.chem.elte.hu/Burcat/burcat.html}} \citep{Burcat1984}, and the binding energies are taken from recent works \citep{Penteado2017}.

Chemical reaction rates can be taken from different databases at the same time, as well as overridden  with user-defined expressions. Parsing is also very flexible, allowing to include strings in the chemical network written in \kida{} and \textsc{UMIST} \citep{McElroy2013} format at the same time, as well as in its own format.

More details and examples on how the user can customize the chemical network are reported in \appx{appx:chemical_network}.

\subsection{Surface chemistry}\label{sect:surface_chemistry}
\codename{} is also capable of generating surface-chemistry reactions from the information present in the internal database, as sublimation, freeze-out, and surface-only rate reactions by using the corresponding expressions. In particular, for sublimation we use the Polanyi-Wigner model of thermal desorption (e.g.~\citealt{Stahler1981})
\beq\label{eqn:evaporation}
 k_{{\rm e},i} = \nu_0 \exp\left(- \frac{E_{{\rm b},i}}{k_{\rm B} \tdust}\right)\,,
\eeq
where $\nu_0=10^{12}$~s$^{-1}$ is the Debye frequency \citep{Draine2009}, $E_{{\rm b},i}$ the binding energy of the \ith species, $k_{\rm B}$ the Boltzmann's constant, and $\tdust$ the dust surface temperature.

Freeze-out is computed using \citep{Hollenbach1979}
\beq
 k_{{\rm f},i} = S \, v_i \, f_{\rm d}\,,
\eeq
where the sticking probability is
\beq\label{eqn:stick}
 S = \left(1 + 4\times10^{-2} \sqrt{\tgas + \tdust} + 2\times10^{-3} \tgas + 8\times10^{-6} \tgas^2\right)^{-1}\,,
\eeq
the scaling factor that takes into account the grain size distribution is
\beq\label{eqn:fdust}
 f_{\rm d} = \frac{\rho_{\rm d}}{4/3\pi\rho_0} \cdot \frac{a_{\rm max}^{p+3} - a_{\rm min}^{p+3}}{a_{\rm max}^{p+4} - a_{\rm min}^{p+4}} \cdot \frac{p+4}{p+3}\,,
\eeq
$v_{\rm i}$ is the thermal velocity of the \ith species in the gas pahse, $\rho_{\rm d}$ the dust mass density, $\rho_0$ its bulk density, $a_{\rm min}$ and $a_{\rm max}$ are the limits of the dust size distribution radii, that is represented by a power-law with exponent $p$ \citep{Mathis1977}, i.e.~$\varphi(a)\propto a^p$, normalized in order to have $\rho_{\rm d}=4/3\pi\rho_0\int_a a^3 \varphi(a) \dd a$, where the integral is defined in the range $a_{\rm min}$ to $a_{\rm max}$.

At the present stage, we employed the generic sticking coefficient in \eqn{eqn:stick}, aware that more accurate expressions are available for specific grain substrates and molecules (e.g.~\citealt{LeitchDevlin1985}), and we refer the reader to Sect.~3 of \citet{Cuppen2017}. We plan to upgrade this aspect of the code in the future.

For the reactions that occur on the surface of grains we use \citep{Hocuk2015}
\beq
 k_{{\rm s}, ij} =  P_a\nu_0 f_d \frac{a_p^2}{4\pi} \left[\exp\left(-q_h \frac{E_{{\rm b},i}}{k_{\rm B} \tdust}\right) + \exp\left(- q_h \frac{E_{{\rm b},j}}{k_{\rm B} \tdust}\right) \right]\,,
\eeq
where $a_p=3\times10^{-8}$~cm is the distance between binding sites, $q_h=2/3$ is the hopping factor, and the tunnelling probability $P_a=\exp\left(-4\pi a_r\sqrt{2\mu_{ij} E_a} / h\right)$ is determined by the size of the barrier $a_r=10^{-8}$~cm, $\mu_{ij}$ is the reduced mass of the two reactants, $E_a$ the energy barrier that depends on the specific reaction, and $h$ the Planck's constant. Here we assume that the binding energies of the species are not affected by the presence of ices on the substrate (i.e.~grains are considered always bare), and that the products obtained from the chemical reactions on the surface of the grains remains on the surface (see e.g.~\citealt{Cazaux2016,Minissale2016}), and that can only return to the gas phase via evaporation, with the rate in \eqn{eqn:evaporation}. We are aware of these limitations, and we are planning to upgrade this particular aspect of the code in a future release.

We do not include automatic reaction rate creation for photodesorption, but the processes can be added manually in the chemical network, following for example \citet{Hollenbach2009}, that consider a species-dependent yield (see \citealt{Andersson2008, Cuppen2017} for a discussion on the uncertainties). Analogously, cosmic-rays desorption is not included, but can be added using e.g.~\citet{Hasegawa1993} based on \citet{Leger1985}.

\subsection{H$_2$ formation on dust}\label{sect:H2_dust}
To model the formation of molecular hydrogen on dust grains we follow the results of the model described in \citet{Cazaux2009}, that include physisorption and chemisorption, tunnelling, and realistic grain surface barriers. For a single silicate dust grain of radius $a$, the rate coefficient of the reaction \mbox{H+H$\to$H$_2$} is
\beq\label{eqn:H2_dust_rate}
 k_{\rm d} = \frac{1}{2} \pi a^2 n_{\rm d} S\, \epsilon_{\rm d}\, v_{\rm g}\,,
\eeq
where the gas thermal velocity of the hydrogen atoms is
\beq
 v_{\rm g} = \sqrt{\frac{8 k_{\rm B} \tgas}{\pi m_{\rm H}}}\,,
\eeq
and
\beq
 \epsilon_{\rm d} = \frac{1}{1+\epsilon_1} + \frac{\epsilon_2}{\epsilon_3}\,,
\eeq
with $k_{\rm B}$ the Boltzmann constant, $m_{\rm H}$ the mass of the hydrogen, and
\beqa
 \epsilon_1 &=& \frac{16\, \tdust}{E_{\rm ch} - E_{\rm s}} \exp\left(-\frac{E_{\rm ph}}{\tdust} + 4\times 10^9 a_{\rm pc} \sqrt{E_{\rm ph} - E_{\rm s}}\right)\,,\nonumber\\
 \epsilon_2 &=& 2\, \exp\left(\frac{E_{\rm s} - E_{\rm ph}}{E_{\rm ph} + \tgas}\right)\,,\nonumber\\
 \epsilon_3 &=& \left(1 + \sqrt{\frac{E_{\rm ch} - E_{\rm s}}{E_{\rm ph} - E_{\rm s}}}\right)^2\,,
\eeqa
with $E_{\rm ph}=700$~K, $E_{\rm ch}=1500$~K, $E_{\rm s}=-1300$~K, $a_{\rm pc}=1.7\times10^{-10}$~m (Cazaux, private comm.).
For a dust model as the one discussed in \sect{sect:surface_chemistry}, \eqn{eqn:H2_dust_rate} can be written as
\beq\label{eqn:H2_dust_k}
 k_{\rm d} = f_{\rm d} S\, \epsilon\, v_{\rm g} = f_{\rm d} \Phi_{\rm d}(\tgas, \tdust)\,,
\eeq
where $f_{\rm d}$ is defined in \eqn{eqn:fdust}, and $\Phi_{\rm d}$ is precomputed during the preprocessor stage and linearly interpolated at runtime as a function of $\tgas$ and $\tdust$. A more extensive discussion on the formation of molecular hydrogen on dust grains can be found e.g.~in \citet{Wakelam2017}.

\subsection{Photochemistry}
The code is designed to compute photochemistry rates either with a radiation spectrum with multiple bins discretized in energy, or with visual extinction $A_v$ and radiation intensity (Habing factor) $G_0$ normalized over the Draine's FUV field \citep{Draine1978,TielensBook}.
\subsubsection{Multiple energy bins}\label{sect:photochemistry_bins}
To compute the reaction rates we use the classic approach that assumes
\beq\label{eqn:ph_sigma}
    k_{\rm ph} = \frac{1}{h} \int_{E_{\rm th}}^\infty \frac{\sigma(E) J(E)}{E}\dd E\,,
\eeq
where the $E_{\rm th}$ is the energy threshold of the given reaction, $\sigma$ its cross section, and $J$ the radiation, and where the chosen discretization of the integral takes advantage of arrays vectorization as explained in \appx{appx:integrals}.
Cross sections are taken from different sources, \citet{Verner1996} for atomic data, and \citet{Heays2017} for molecular data (see \sect{sect:database}).

In the case of energy binning, the user decides the energy range $[E_{\min}, E_{\max}]$ and the number of bins $N_{\rm ph}$ that will be employed. Analogously to what the Monte Carlo radiative transfer code \textsc{Mocassin} \citep{Ercolano2003,Ercolano2005} does, to ensure that the cross-section value at the threshold is correctly captured, the code automatically divides the selected energy range, first using three bins per each reaction energy threshold (e.g.~hydrogen has $E_{\rm th, H}=13.6$~eV), i.e.~$E_{\rm th}$ and $E_{\rm th}\pm\Delta E$, where $\Delta E=0.00408$~eV, and then distributing the remaining grid points on a log-spaced grid from $E_{\min}$ to $E_{\max}$. The code verifies that at least $N_{\rm ph}/2$ grid points are used for the log-spaced grid. Once the grid has been defined, it remains fixed, the cross-sections are integrated over this, and every process that deals with radiation (see e.g.~photoelectric heating) is optimized accordingly.

\subsubsection{Visual extinction approximation}\label{sect:visual_extinction}
If the shape of the radiation resembles a Draine field
\beqa\label{eqn:j_draine}
  J(E) &=& h \left(1.658\times10^{6} E^2 - 2.152\times10^{5} E^3\right.\nonumber\\
       &+& \left.6.919\times10^{3} E^4\right)\,,
\eeqa
with $E$ in eV and $J(E)$ in eV$\,$cm$^{-2}$Hz$^{-1}$s$^{-1}$, it is safe to assume that the reaction rates are simply represented by the following expression
\beq\label{eqn:ph_av}
  k_{\rm ph} = G_0\, a_i \exp\left(- c_i A_v \right)\,,
\eeq
where $G_0$ uniformly scales the radiation without changing the energy distribution, $a_i$ is the result of integrating \eqn{eqn:ph_sigma} with \eqn{eqn:j_draine}, and $c_i$ a coefficient that takes into account the attenuation due to dust (including scattering) as a function of the visual extinction $A_v$ \citep{Heays2017}.

This approach is not the default, but it is useful when the code needs to be coupled with frameworks that employ frequency-independent radiation, e.g.~most hydrodynamical codes. Note that using \eqn{eqn:ph_av} instead of \eqn{eqn:ph_sigma} is slightly faster, since there is no significant overhead from solving the integral (however computational time is reduced by use of vectorization, see \appx{appx:integrals}).

\subsubsection{Self-shielding}\label{sect:self_shielding}
Both in the cases decribed in \sect{sect:photochemistry_bins} and \ref{sect:visual_extinction}, it is necessary to take into account the self-shielding for H$_2$ and CO photodissociation reactions. To be fully consistent the code should compute the absorption from the rotovibrational lines of these two molecules, but for the present set-up this operation is too computationally expensive, since it involves a large number of molecular lines \citep{Visser2009}. We therefore make use of approximations that are designed to work with the Draine field. In particular for H$_2$ we use \citep{Draine1996,Richings2014}
\beqa\label{eqn:selfshielding_H2}
f_{\rm H_2} & = & \frac{1-\omega}{\left(1+x/b_5\right)^a} \exp\left[-5\times10^{-7} (1+x)\right]\nonumber\\
 & + & \frac{\omega}{\sqrt{1+x}} \exp\left(-8.5\times10^{-4}\sqrt{1+x}\right)
\eeqa
where
\beq
 \omega = 0.013 \left[1 + \left(\frac{\tgas}{2700~{\rm K}}\right)^{1.3}\right]^{1/1.3} \exp\left[-\left(\frac{\tgas}{3900{\rm K}}\right)^{14.6}\right]
\eeq
and $x=N_{\rm H_2} / N_{\rm crit}$, where
$N_{\rm crit}=1.3\times10^{14} \left[1+\left(\frac{\tgas}{600{\rm K}}\right)^{0.8}\right]$ and $a=1.4$ when $\tgas<3000$~K, $N_{\rm crit}=2\times10^{14}$ and $a=1.1$ when $\tgas<4000$~K, and finally, $N_{\rm crit}=10^{14} \left(\frac{\tgas}{4760{\rm K}}\right)^{-3.8}$ and $a=\left(\frac{\tgas}{4500{\rm K}}\right)^{-0.8}$ otherwise. 

For CO self-shielding we employ the table from \citet{Visser2009}, where $f_{\rm CO}(N_{\rm H_2}, N_{\rm CO})$ is interpolated on the fly from a table. Analogously, N$_2$ self-shielding is computed by interpolating the expression $f_{\rm N_2}(N_{\rm H_2}, N_{\rm N_2}, N_{\rm H})$ from the tables described in \citet{Heays2014}  (details employed for these and CO tables are discussed in \sect{sect:database}).

Accounting for self-shielding, the rates for the photodissociation of molecular hydrogen, CO, and N$_2$ are respectively
\beqa
 k_{\rm H_2} & = & k_{\rm H_2}^{\rm ref} f_{\rm H_2}\\
 k_{\rm CO} & = & k_{\rm CO}^{\rm ref} f_{\rm CO}\\
 k_{\rm N_2} & = & k_{\rm N_2}^{\rm ref} f_{\rm N_2}\,,
\eeqa
where $k_{i}^{\rm ref}$ can be computed with \eqn{eqn:ph_sigma} or \eqn{eqn:ph_av}.

\subsubsection{Attenuation of the radiation}
Multifrequency radiation $J(E)$ is attenuated to $J'(E)$ after crossing a cell of size $\Delta x$ accordingly to the chemical composition as
\beq
  J'(E) = J(E) \exp\left[- \kappa(E) \rho_{\rm d} \Delta x -\sum_i\sigma_i(E)n_i \Delta x \right]\,,
\eeq
where the sum runs over the chemical species which have photochemical cross sections $\sigma_i(E)$ and abundance $n_i$, while the dust with mass density $\rho_{\rm d}$ and frequency-dependent dust opacity $\kappa(E)$ further attenuates the radiation. Dust opacity is loaded at runtime after being pre-calculated during the preprocessor stage. For simplicity, we do not include additional processes in the attenuation of $J(E)$, such as absorption resulting in molecular excitations. In the attenuation of $J(E)$ we also do not include H$_2$ and CO self-shielding from photodissociation, that are taken into account following \sect{sect:self_shielding}.

\subsection{Optimizing the calculation of rate coefficients}\label{sect:rate_optimization}
When thermochemistry is computed in the same set of differential equations, rate coefficients need to be evaluated every time the solver calls the function that evaluates their right-hand side, i.e.~\eqn{eqn:ode}. This represents a considerable computational overhead, since many reaction rates contain complex (i.e.~expensive) operations (logarithms, exponentials, etc\dots). To overcome this problem we have divided the rate coefficients into three categories: (i) interpolatable in a log-log space as a function of the gas temperature, (ii) not-interpolatable because the rate depends on other factors than just the gas temperature, and (iii) single-evaluated, that do not depend on the temperature and that can be evaluated only during the first call to the solver, as for example photochemistry and cosmic ray ionization. The code automatically splits the reactions into these three groups in order to minimize the cost of rate evaluation at runtime.

\section{Thermochemistry}\label{sect:thermo}
\codename{} is capable of evolving the temperature of the gas and the dust consistently with the chemical evolution and with the impinging radiation by using the set of processes that are described in this Section. Solving the thermal component together with the chemical evolution ensures the consistency of the results, but on the other hand reduces the computational efficiency, since the ODE system in \eqn{eqn:ode} and \eqn{eqn:ode_thermo} might become numerically stiffer, and the Jacobian becomes less sparse.
For these reasons it is fundamental that all the expressions of the thermochemical processes included (as well as their derivatives) do not present discontinuities with respect to the variables of the ODE system (e.g.~temperature and density). This holds, for example, for CO and H$_2$O cooling tables, or for the rate coefficients that are involved in the collisional excitation cooling.

\subsection{Cosmic-ray heating}
Cosmic-ray heating is modelled following the approach of \citet{Galli2015}, where the main sources of energy are H and H$_2$ reactions with comsic rays, namely
\beq
  \Gamma_{\rm CR} = \zeta \left(5.5\times10^{-12} n_{\rm H} + 2.5\times10^{-11} n_{\rm H_2}\right)\,,
\eeq
where $\zeta$ is the cosmic-rays ionization rate per H$_2$ molecule in s$^{-1}$, $n_{\rm H}$ and $n_{\rm H_2}$ are atomic and molecular hydrogen number densities in cm$^{-3}$, and the total heating $\Gamma_{\rm CR}$ is in units of erg~s$^{-1}$~cm$^{-3}$.

\subsection{PAH heating}
Heating from polycyclic aromatic hydrocarbons (PAH) is related to the dust photoelectric heating, however, due to the uncertainties in the PAH chemistry we use \citep{Bakes1994,Woitke2009}
\beq
  \Gamma_{\rm PAH} = 10^{-22}\, f_{\rm PAH}\, y_{\rm H}\, \varepsilon_{\rm PAH}\, G_0\, \dtog\,,
\eeq
where $f_{\rm PAH}=0.02$ (default value), $y_{\rm H}$ is the H nuclei number density, $\dtog$ is the dust-to-gas mass ratio, and
\beq
  \varepsilon_{\rm PAH} = \frac{0.0487}{1 + 4\times10^{-3} x_{\rm PAH}^{0.73}}\,,
\eeq
with $x_{\rm PAH}=G_0 \sqrt{\tgas} n_{\rm e}^{-1}$, where $n_{\rm e}$ is the electron number density. The resulting heating is in units of erg~s$^{-1}$~cm$^{-3}$.

\subsection{Photochemical heating}
The machinery for the photochemical heating depends on the impinging multifrequency radiation $J(E)$ and it is conceptually similar to the one employed to calculate the reaction rates in \sect{sect:photochemistry_bins}, therefore it takes advantage of the same vectorization optimization. In particular for each photochemical rate we have
\beq
  \Gamma_{{\rm ph}, i} = \eta_{\rm ph}\, n_i \int_{E_{{\rm ph}, i}}^\infty \frac{J(E)\sigma_i(E)}{E}(E-E_{{\rm ph}, i})\dd E\,,
\eeq
that can be written as
\beq
  \Gamma_{{\rm ph}, i} = \eta_{\rm ph}\, n_i\left[\int_{E_{\rm th}}^\infty J(E)\sigma_i(E)\dd E - E_{{\rm ph}, i}k_{\rm ph}\right]\,,
\eeq
where the first term of the right hand side will be vectorized according to \appx{appx:integrals} and $k_{\rm ph}$ is the corresponding photochemical rate, that has been already computed following \sect{sect:photochemistry_bins}. $E_{{\rm ph}, i}$ is the energy threshold of the photochemical reaction rate (e.g.~13.6~eV for H photoionization).

The fraction of energy deposited in the gas depends on the ionization fraction $x_{\rm e} = n_{\rm e} / n_{\rm tot}$ as
\beq
 \eta_{\rm ph} = 10^{0.25452\,\log\left(x_{\rm e}\right)}\,,
\eeq
a fit with 10\% error with respect to the expression from \citet{Xu1991}, employed here to have a function with no discontinuities in $x_{\rm e}$.

\subsection{Dust photoelectric heating}
Photoelectric heating is computed taking into account the multifrequency radiation spectrum as in \citet{Weingartner2001} and \citet{Weingartner2006}. In principle, this requires the code to track the charge distribution on the dust grains. For the sake of the code efficiency, and to reduce the complexity of the algorithms, we compute the charge distributions assuming equilibrium between the charged grains and the ions and the electrons in the gas phase \citep{Okuzumi2009,Fujii2011,Grassi2019}.

The main processes involved are described by (i) $k_{\rm pe}$ the rate at which the valence band electrons are removed from the grains,  (ii) $k_{\rm pd}$ the rate of the photodetachment of the attached electrons, and (iii) the electrons-grains interactions $k_{\rm e}$, as well as (iv) the cations-grains $k_{\rm i}$. These components are employed to solve the linear system for $n(Z)$
\beq
  n(Z) \left[k_{\rm pd}(Z) + k_{\rm pe}(Z) + \sum_i n_i k_i\right] = n(Z+1) n_{\rm e} k_{\rm e}
\eeq
and $\sum_Z n(Z)=n_{\rm d}$ to find the fraction of grains $n(Z)$ in each charged state $Z$ (see \appx{appx:superdiagonal}), where the sum on $i$ runs over the cations\footnote{The maximum charge $|Z|$ is defined by the user during the pre-processor stage. In the benchmarks presented in \sect{sect:benchmarks} we use $Z=[-4, 4]$.}.

Once the fraction of grains in each charged level $Z$ is known, it is possible to compute the total heating by using
\beq\label{eqn:pe_heat}
 \Gamma_{\rm pe}(Z) = n(Z)\frac{\int_a \varepsilon(a, Z) \int_E J(E) E^{-1}   \Phi(a, E, Z)\dd E\dd a }{h \int_a a^p \dd a}\,,
\eeq
with the integrals ranging from $a_{\rm min}$ to $a_{\rm max}$ and $E_{\rm th}$ to infinity, and
\beq
 \Phi(a, E, Z) = \pi a^{2+p} Q(a, E) Y(a, E, Z)\,,
\eeq
where $Q$ is the dust absorption coefficient\footnote{In principle, the optical properties of a grain depend on the charge $Z$ \citep{BohrenBook1983,Li2000}, but large uncertainties in the process make it hard to constrain its dependence on $Z$.}, that is computed by \codename{} with Mie theory \citep{BohrenBook1983}, as discussed in \sect{sect:database}. The yield of the process is
\beq
 Y(a,E,Z) = y_2(a,E,Z) \min\left[y_0(a,E,Z)  y_1(a,E), 1\right]\,.
\eeq
For $y_1$ and $y_2$ we follow the equations from \citet{Weingartner2006}, while for $y_0$ we employ the data from Fig.~2 of their paper. We assume silicate grains with work function $W=8$~eV, band gap $E_{\rm bg}=5$~eV, photo attenuation length $l_{\rm a}(\lambda)=\lambda [4\pi {\rm Im}(m_\lambda)]^{-1}$, where $m_\lambda$ is the refractive index at $\lambda=c(Eh)^{-1}$, electron escape length $l_{\rm e}=10^{-7}$~cm if $E<211$~eV or $3.27\times10^{-11} (E / 10^5\,{\rm eV})^{1.5}$~cm otherwise \citep{Weingartner2006}.

In \eqn{eqn:pe_heat} the efficiency is
\beq
  \varepsilon(a, Z) = 6 \int_{E_{\rm min}}^{E_{\rm max}}\frac{E(E-E_{\rm low}) (E_{\rm high}-E)}{\left(E_{\rm high}-E_{\rm low}\right)^3 y_2(a,E,Z)} \dd E
\eeq
with $E_{\rm min}=0$ when $Z\geq0$ and from Eq.~(3) of \citet{Weingartner2006} otherwise, while $E_{\rm max}=E-E_{\rm pet}+E_{\rm min}$ with $E_{\rm pet}$ that depends on the valence band ionization potential; see Eq.~(2) and Eq.~(6) in \citet{Weingartner2006}. When $Z<0$ we have $E_{\rm low} = E_{\rm min}$ and $E_{\rm high} = E_{\rm max}$, and $Z\geq0$ instead $E_{\rm low}=-(Z+1)e^2/a$ and $E_{\rm high}=E-E_{\rm pet}$, with $e$ the elementary charge.

The rate at which electrons are removed from the dust grains $k_{\rm pe}(Z)$ is given by \eqn{eqn:pe_heat} without the efficiency term $\varepsilon$ and $n(Z)$, and the photodetachment rate is
\beq
  k_{\rm pd}(Z) = \frac{\int_E J(E) E^{-1}  \int_a a^{p+2} \sigma_{\rm pd}(E)\dd a \dd E}{h \int_a a^p \dd a}\,
\eeq
where the photo detachment cross-section $\sigma_{\rm pd}(E)$ is as in Eq.~(20) of \citet{Weingartner2001} with $\Delta E = 3$~eV.
Finally, the interplay with the gas-phase is controlled by the electron-grain interactions rates
\beq\label{eqn:phe_ke}
  k_{\rm e} = S_{\rm e} v_{\rm g}(T) \frac{\int_a a^{p+2} \widetilde{J}_{\rm e}(a, T, Z)\dd a}{\int_a a^p \dd a}\,,
\eeq
where $S_{\rm e}=0.1$ is the electron sticking efficiency, that here we assume to be constant, but that in principle is a function of the grain size, the dust temperature, and the depth of the potential well between electrons and grains due to polarization interaction (see \citealt{Draine1986,Nishi1991,Bai2011,Grassi2019}). The thermal speed of the electrons is
\beq\label{eqn:vgas}
  v_{\rm g} = \sqrt{\frac{8k_{\rm B}T}{\pi m_{\rm e^-}}}\,,
\eeq
where $m_{\rm e^-}$ is the mass of the electron, and $\widetilde{J}_{\rm e}(a, T, Z)$ is a function that depends on the size and on the charge of the dust grain's reaction-partner, and represents the integral of the electron-grain capture cross section with the Maxwellian velocity distribution (see Sect.~II.b and III.a of \citealt{Draine1986}). To find $k_i$ for cations we use \eqn{eqn:phe_ke}, assuming unitary sticking efficiency and $v_{\rm g}$ with the mass of the specific cation instead of the electron, and using $\widetilde{J}_i(a, T, Z)$. i.e.~the analogue of $\widetilde{J}_{\rm e}$ for cations.

As discussed in the previous sections, we pre-compute the coefficients of the integrals to benefit from vectorization at runtime.

\subsection{Atomic radiative cooling}\label{sect:radiative_atomic_cooling}
We compute the radiative atomic cooling by assuming equilibrium between the electronic levels of the atoms included in the chemical network. This is solved by considering the collisional excitation with hydrogen atoms and electrons present in the gas, or other atoms and molecules (e.g.~He and H$_2$) when collisional rate coefficients are available. For a generic atom with $N$ levels we have to satisfy the following linear system \citep{Maio2007,Woitke2009}
\beq\label{eqn:emission_system}
  n_i \sum_{j\neq i}R_{ij} = \sum_{j\neq i} n_j R_{ji}\,,
\eeq
where $R_{ij} = A_{ij} + \sum_\ell k_{ij}^\ell n_\ell$ if $i>j$ and $R_{ij} = \sum_\ell k_{ij}^\ell n_\ell$ if $i<j$. The Einstein coefficients $A_{ij}$ represent the spontaneous transition probability from the \ith to the \jth level, while $k_{ij}^\ell$ is the excitation rate coefficient to excite the atom from the \ith to the \jth level, with the \ellth collider that has number density $n_\ell$. The temperature-dependent rate coefficients for collisions with protons and electrons are from the \textsc{Chianti} database (see \sect{sect:database}). The rates are linearly interpolated at runtime to reduce the computational cost. Additional rate coefficients for the first three excited levels of C$^{(+)}$, Si$^{(+)}$, and O$^{(+)}$ and H$_2$ and He colliders are included as in \krome{} \citep{Grassi2014}. To solve the linear system in \eqn{eqn:emission_system} we employ \texttt{dgesv} from \textsc{LAPACK}\footnote{\url{http://www.netlib.org/lapack/}} for systems with more than three levels, otherwise we solve the system analytically to save computational time.

When the level population $n_i$ of each excited level is computed, the resulting cooling of a transition $i\to j$ of a given atom is $\Lambda_{{\rm line},ij}=n_i \Delta E_{ij} A_{ij} \beta_{ij}$, where $\Delta E_{ij}$ is the difference in energy between the levels, and the escape probability is \citep{TielensBook}
\beq
  \beta_{ij} = \left[4\tau_{ij} \sqrt{\ln\left(\frac{\tau_{ij}}{\sqrt{\pi}}\right)}\right]^{-1}\,,
\eeq
if $\tau_{ij}>7$ and
\beq
  \beta_{ij} = \frac{1-e^{-2.34\tau_{ij}}}{4.68\tau_{ij}}\,,
\eeq
otherwise. Here
\beq\label{eqn:tau_ij}
  \tau_{ij} = \frac{c^3 h^3}{8\pi} \frac{A_{ij} n_i}{\Delta E_{ij}^3 \dd_z v} \left(\frac{n_j g_i}{n_i g_j} -1\right)\,,
\eeq
where $\dd_z v$ is the velocity gradient along the $z$-component, for which we have assumed that the velocity gradient is large when compared to the thermal motion, as discussed in \citet{TielensBook}.

This approach allows not only to compute the total collisional emission cooling by summing $\Lambda_{{\rm line}, ij}$ for all the available electronic transitions, but also to have access to the individual emission lines from the gas and to track their evolution in time.
At runtime we do not compute the shape of the emitted line (Lorentzian, Gaussian, etc.), since it is not relevant to compute the total cooling.

This formalism is valid also for molecules, for which we employ the data from the \textsc{LAMDA} database\footnote{\url{https://home.strw.leidenuniv.nl/~moldata/}} \citep{Schoier2005} to evaluate the emission of the different lines, while for cooling we use precomputed tables, as discussed in the next sections.

\subsection{Bremsstrahlung cooling}
Bremsstrahlung produces cooling from the radiation emitted by charged particles that decelerate when deflected by the presence of other charged particles, and following \citet{Cen1992} we have
\beq
 \Lambda_{\rm BS} = g_{\rm ff}\, n_{\rm e^-} \sum_i Z_i^2 n_i\,,
\eeq
where g$_{\rm ff}=1.5$ is the Gaunt factor, $n_{\rm e^-}$ the number density of the electrons, $Z_i$ and $n_i$ the charge and the number density of the \ith{} species, and where $i$ runs on all the ions. The final cooling $\Lambda_{\rm BS}$  is in units of erg~s$^{-1}$~cm$^{-3}$.

\subsection{Chemical cooling/heating}
Chemical cooling or heating is determined by the endothermicity or exothermicity of the given reaction, i.e.~if a given reaction requires or releases energy. The amount of energy for the \ellth reaction is determined by the reaction rate coefficient ($k_{\ell}$) and the difference in enthalpy of formation ($\Delta H_{\rm f}^{\circ}$) between the reactants and the products that have abundances $n_i$, $\Lambda_{{\rm chem}, \ell}=k_{\ell}\left(\sum_{\rm react} \Delta H_{{\rm f},i}^{\circ} - \sum_{\rm prod} \Delta H_{{\rm f},i}^{\circ}\right)\prod_{\rm prod} n_j$. In principle, analogously to the rate coefficient, the enthalpy of formation of a given compound is a function of the temperature, that can be computed by using the thermodynamic data in polynomial from \citep{Burcat1984}. However, since the variation with the temperature of $\Delta H_{\rm f}^{\circ}$ is small compared to the variation of $k_{\ell}$, and since several coefficients have a limited temperature range,  we compute only the standard enthalpy, i.e.~when $T=298.15$~K. The coefficients of the polynomials are taken from Burcat's table and employed to compute $\Delta H_{\rm f}^{\circ}$ during the preprocessor stage, and multiplied at runtime by the reaction rate and the abundances of the products.

Since not all the reactions play a key role in the total $\Lambda_{\rm chem}$, it is possible to limit this calculation to a subset of chemical species, for example by employing only hydrogen- and helium-based species, that are the most abundant. Collisional ionization are always considered endothermic, while cation-electron recombinations cooling is $\Lambda_{{\rm rec}, i}=k_{\rm B} \tgas\, k_{i,{\rm e^-}} n_{i} n_{\rm e^-}$, where $k_{i,{\rm e^-}}$ is the corresponding rate coefficient \citep{Cen1992}.

\subsection{CO and H$_2$O cooling}\label{sect:CO_H2O_cooling}
In principle, molecular cooling can be computed by using the same machinery presented in \sect{sect:radiative_atomic_cooling}, however given the availability of pre-computed tables, we prefer to use the latter to reduce the computational time.
CO cooling (and analogously water cooling) is obtained from the tables in the Appendix~B of \citet{Omukai2010}, that are functions of the gas temperature ($T$), the amount of H nuclei ($y_{\rm H}$), and of the column density from the cell to the surface from which the radiation is assumed to escape ($N_{\rm CO}$), namely $\Lambda_{\rm CO}=n_{\rm CO}f_{\rm CO}(T, y_{\rm H}, N_{\rm CO})$. These tables are computed using the method of \citet{Neufeld1993}, that assume level populations in statistical equilibrium, and employ data from the \textsc{LAMDA} database. These are interpolated at runtime by \codename{} using a three-dimensional linear interpolation routine. The limits of the CO tables are $T=[3, 10^4]$~K, $y_{\rm H}=[10^{-2}, 10^{14}]$~cm$^{-3}$, and $N_{\rm CO}=[10^{-18}, 10^{25}]$~cm$^{-2}$. Anlogously, water cooling limits are $T=[10, 10^3]$~K, $y_{\rm H}=[10^{2}, 10^{12}]$~cm$^{-3}$, and $N_{\rm H_2O}=[10^{9}, 10^{19}]$~cm$^{-2}$. Outside these limits we assume that when the gas temperature is small the cooling is inefficient, since collisions are less effective in populating the ro-vibrational molecular levels, as well as when the local density is high. Conversely, when the column density toward the escape surface is small the cooling is more efficient, since the radiation is capable of escaping from the simulation domain.

\subsection{Dust cooling and dust temperature}\label{sect:dust_cooling}
When the temperature of the dust is less (greater) than the temperature of the gas, the dust cools (heats) the surrounding medium, since they are in radiation balance (Kirchhoff's law) and exchange kinetic energy with molecules and atoms. In particular, we assume \citep{Hollenbach1979,DraineBook2011,Grassi2017}
\beq\label{eqn:radiation_balance}
  \Gamma_{\rm em} = \Gamma_{\rm abs} + \Lambda_{\rm d-g}\,,
\eeq
where the first term is the thermal radiative emission of the dust, the second the absorption that depends on the impinging radiation, and the last term is the dust-gas thermal exchange. In our case these terms are
\beq
  \Gamma_{\rm em} = \frac{4\pi}{h}\int_E B(E,\tdust) \int_a Q(a, E) a^{2+p}  \dd a\, \dd E\,,
\eeq
where $B(E,\tdust)$ is the spectral radiance of a blackbody with temperature $\tdust$. The integral on the energy is computed in the range of validity of $Q$, while the integral on the dust size from $a_{\rm min}$ to $a_{\rm max}$ that we assume constant during the evolution. $\Gamma_{\rm em}$ is pre-computed and interpolated at runtime as a function of $\tdust$ only. The absorption
\beq
  \Gamma_{\rm abs} = \frac{\pi}{h}\int_E J(E) \int_a Q(a, E) a^{2+p} \dd a\, \dd E\,,
\eeq
depends on the impinging radiation $J(E)$, and since this term changes at runtime, we take advantage of vectorization as discussed in \appx{appx:integrals}, pre-computing the rest of the integral. In this case the limits on the integral on energy are the limits of $J(E)$ as defined by the problem boundary conditions (the energy grid is not changed during runtime).
Finally, the interaction between gas and dust is
\beq\label{eqn:lambda_dg}
  \Lambda_{\rm d-g} = 2\pi\, v_{\rm g} k_{\rm B}\alpha_{\rm g}  \ngas (T-\tdust) \int_a a^{p+3} \dd a \,,
\eeq
where the definition of the gas thermal velocity $v_{\rm g}$ in \eqn{eqn:vgas} has $\mu m_{\rm p}$, i.e.~the mean molecular weight of the gas and the mass of the proton, instead of just the mass of the electron, and $\alpha_{\rm g}=0.5$ takes into account the composition of the gas to evaluate the momentum exchange \citep{Hollenbach1979}, but that in our case we keep constant.

Since $\Gamma_{\rm em}$ and $\Lambda_{\rm d-g}$ both depend on $\tdust$, we can use \eqn{eqn:radiation_balance} and a bisection method to find $\tdust$.

With $\tdust$ known, and by considering that \eqn{eqn:lambda_dg} holds for a single grain, the total cooling/heating from the dust-gas interaction is
\beq\label{eqn:lambda_dg_gamma}
  \Lambda_{\rm d-g} = \frac{\mu m_{\rm p} \ngas \mathcal{D}}{4/3\pi\rho_0 \int_a a^{3+p} \dd a} \left(\Gamma_{\rm em} - \Gamma_{\rm abs}\right)\,,
\eeq
where $m_{\rm p}$, and where we use the difference of emission and absorption instead of \eqn{eqn:lambda_dg} directly to avoid employing the difference $(T-\tdust)$ that may cause instability in the solver due to the high precision required \citep{Grassi2017}.

\subsection{H$_2$ cooling}
For molecular hydrogen cooling we employ temperature-dependent look-up tables for several colliders, including H, H$^+$, H$_2$ (assuming ortho-to-para ratio 3:1), e$^-$, and He. These tables \citep{Glover2008,Glover2015} are represented by piece-wise functions (see~\appx{appendix:cool_H2}), each one defined on a $T_{\rm min}$ to $T_{\rm max}$ range, that are multiplied to have continuous derivative by a window function
\beq
 w(x) = 10^{200 \left[\sigma_f(x, -0.2, 50) \sigma_f(-x, -1.2, 50) - 1\right]}\,,
\eeq
where
\beq
  x = \frac{\log(\tgas)-\log(T_{\rm min})}{\log(T_{\rm max})-\log(T_{\rm min})}\,,
\eeq
and
\beq
  \sigma_f(x, x_0, s) = \frac{10}{10 + \exp[-s (x - x_0)]}\,.
\eeq

From the look-up functions (see \appx{appendix:cool_LDL}) we obtain the low-density cooling $\Lambda_{\rm H_2}^{\rm low}$ that determines the total cooling as
\beq
  \Lambda_{\rm H_2} = n_{\rm H_2}\frac{\Lambda_{\rm H_2}^{\rm low}\Lambda_{\rm H_2}^{\rm high}}{\Lambda_{\rm H_2}^{\rm low}+\Lambda_{\rm H_2}^{\rm high}}
\eeq
where the high-density cooling $\Lambda_{\rm H_2}^{\rm high}$ is reported in \appx{appendix:cool_HDL}.

\section{Benchmark models}\label{sect:benchmarks}
To verify the results produced by the code we selected a set of benchmarks that cover a reasonable range of physical and chemical configurations. We limit the present tests to photon-dominated regions (PDR), and we do not provide any disk benchmark, since these will be discussed more in detail in Paper~III (despite PDRs representing a good model to test disk physics, see e.g.~\citealt{Bruderer2009}). PDRs are studied in detail, especially by the well-established benchmark\footnote{\url{https://zeus.ph1.uni-koeln.de/site/pdr-comparison/}} of \citet{Rollig2007} (R07), where several codes are compared, and where detailed instructions and final results are provided in order to reproduce the results with relative ease. The PDR models discussed in R07 have two different densities ($\ngas=10^3$~cm$^{-3}$ and $\ngas=10^{5.5}$~cm$^{-3}$) and two radiation intensities ($\chi=10$ and $\chi=10^5$), and the resulting four models are evolved at constant or variable temperature (i.e.~with or without thermochemistry). In our case we evolve the models with variable temperature, for $\ngas=10^3$~cm$^{-3}$ and  radiation intensity $\chi=10$ (\texttt{V1}) and for $\ngas=10^{5.5}$~cm$^{-3}$ and $\chi=10^5$ (\texttt{V4}). Instead of using $A_{\rm v}$-based photochemical reaction rates as in R07, we employ the multi-frequency binning with the same Draine-like radiation source valid in the range 6~to 13.6~eV.

We then extend these tests to a wider radiation spectrum that includes X-rays (\texttt{XDR}), and verify the validity of our results  by comparing the temperature profiles obtained by \citet{Picogna2019} using a pure atomic chemical network. Finally, we validate our calculation of the equilibrium \texttt{cooling} function for different temperatures as in \citet{Gnat2012}.

\subsection{PDR: \citet{Rollig2007} \texttt{V1}}\label{sect:bench_v1}
We follow the set-up of R07, that consists of a 1D semi-infinite slab with constant gas density $\ngas =10^3$~cm$^{-3}$, and with a plane-parallel radiation source emanating from the left side of the simulation box, that has a Draine spectrum with $\chi=10$. We scaled the the CO and H$_2$ photoionization rates to match the one required by the benchmark at $A_{\rm v}=0$. The cosmic-rays ionization rate is $\zeta=5\times10^{-17}$~s$^{-1}$, and the initial conditions for the chemistry are as in \tab{table:initial_conditions}. The grain size distribution has $p=-3.5$ in the range $a_{\rm min}=5\times10^{-7}$~cm, $a_{\rm max}=2.5\times10^{-5}$~cm,  silicate grains with $\rho_0=3$~g~cm$^{-3}$, and dust-to-gas mass ratio $\mathcal{D}=10^{-2}$ (see also \sect{sect:surface_chemistry}). The factor for the self-shielding in \eqn{eqn:selfshielding_H2} is $b_5=1$, while the velocity gradient for the escape probability in \eqn{eqn:tau_ij} is $\dd_z v=10^{-5}$~s$^{-1}$. We use 200~logarithmically-spaced grid points from $A_{\rm v}=10^{-6}$ to $A_{\rm v}=30$, assuming $A_{\rm v} =6.289\times10^{-22} N$ (as in R07), where $N$ is the column density in units of cm$^{-2}$. We do find equilibrium chemistry and temperature by running our code for $t=10^8$~yr, and we manually verify that equilibrium is reached. The ending time is explicitly set longer than the expected equilibrium time.

For this benchmark the formation of molecular hydrogen on dust grains is not modelled as in \sect{sect:H2_dust}, but we employ the expression in R07, i.e.~$k_{\rm d} = 3\times10^{-18}\sqrt{\tdust}$~cm$^3$~s$^{-1}$. This is because our aim is to test thermal processes and photochemistry, and the different expressions for H$_2$ formation on grains will considerably affect the final results, complicating their analysis (especially for \texttt{V4}, described later) . Additional details about this issue in \appx{appendix:H2_dust_cfr}.

\begin{table}
    \centering
    \setlength{\tabcolsep}{6pt} 
    \renewcommand{\arraystretch}{1} 
    \begin{tabular}{llllll}
        \hline
         & \texttt{V1} & \texttt{V4} & \texttt{cooling} & \texttt{XDR}  & Units\\
        \hline
        $\ngas$          & $1(3)$                & $1(5.5)$              & 1                 & $1(2)$      & cm$^{-3}$\\
        $\zeta$          & $5(-17)$              & $5(-17)$              & -                 & $5(-17)$    & s$^{-1}$\\
        $\chi$           & 10                    & $1(5)$                & 0                 & see text    & -\\
        $n_{\rm H_2}$    & $0.5$                 & $0.5$                 & -                 & -           & $\ngas$\\
        $n_{\rm H}$      & 0                     & 0                     & $0.8$             & 1           & $\ngas$\\
        $n_{\rm H^+}$    & 0                     & 0                     & $0.2$             & 0           & $\ngas$\\
        $n_{\rm O}$      & $3(-4)$               & $3(-4)$               & $4.41(-4)$        & $4.90(-4)$  & $\ngas$\\
        $n_{\rm O^+}$    & 0                     & 0                     & $4.90(-5)$        & 0           & $\ngas$\\
        $n_{\rm C}$      & $1(-4)$               & 0                     & 0                 & $2.69(-4)$  & $\ngas$\\
        $n_{\rm C^+}$    & 0                     & $1(-4)$               & $2.69(-4)$        & 0           & $\ngas$\\
        $n_{\rm He}$     & $1(-1)$               & $1(-1)$               & $8.51(-2)$        & $8.51(-2)$  & $\ngas$\\
        $n_{\rm Ne^+}$   & 0                     & 0                     & $8.51(-5)$        & 0           & $\ngas$\\
        $n_{\rm Ne}$     & 0                     & 0                     & 0                 & $8.51(-5)$  & $\ngas$\\
        $n_{\rm N}$      & 0                     & 0                     & 0                 & $6.76(-5)$  & $\ngas$\\
        $n_{\rm S^+}$    & 0                     & 0                     & 0                 & $1.32(-5)$  & $\ngas$\\
        $n_{\rm Mg^+}$   & 0                     & 0                     & 0                 & $3.98(-5)$  & $\ngas$\\
        $\tgas$          & 50                    & 50                    & see text          & 50          & K\\
        \hline
    \end{tabular}\caption{Initial conditions values for the different code benchmarks. Electron abundance $n_{\rm e^-}$ is initialized to ensure global charge neutrality. Note $a(b)=a\times10^b$.}
    \label{table:initial_conditions}
\end{table}

We employ a different chemical network, but using the same species as in the original benchmark, namely, H, H$^+$, H$_2$, H$_2^+$, H$_3^+$, O, O$^+$, OH$^+$, OH, O$_2$, O$_2^+$, H$_2$O, H$_2$O$^+$, H$_3$O$^+$, C, C$^+$, CH, CH$^+$, CH$_2$, CH$_2^+$, CH$_3$, CH$_3^+$, CH$_4$, CH$_4^+$, CH$_5^+$, CO, CO$^+$, HCO$^+$, He, He$^+$, e$^-$. The chemical network is reported in \appx{appendix:network} and it will be discussed in detail in Paper~II. Our aim is to fairly reproduce the gas and dust temperature profile, within the spread found by the different codes that participated in the original benchmark.

\begin{figure*}
 \begin{minipage}[t]{0.49\textwidth}
       \includegraphics[width=\textwidth]{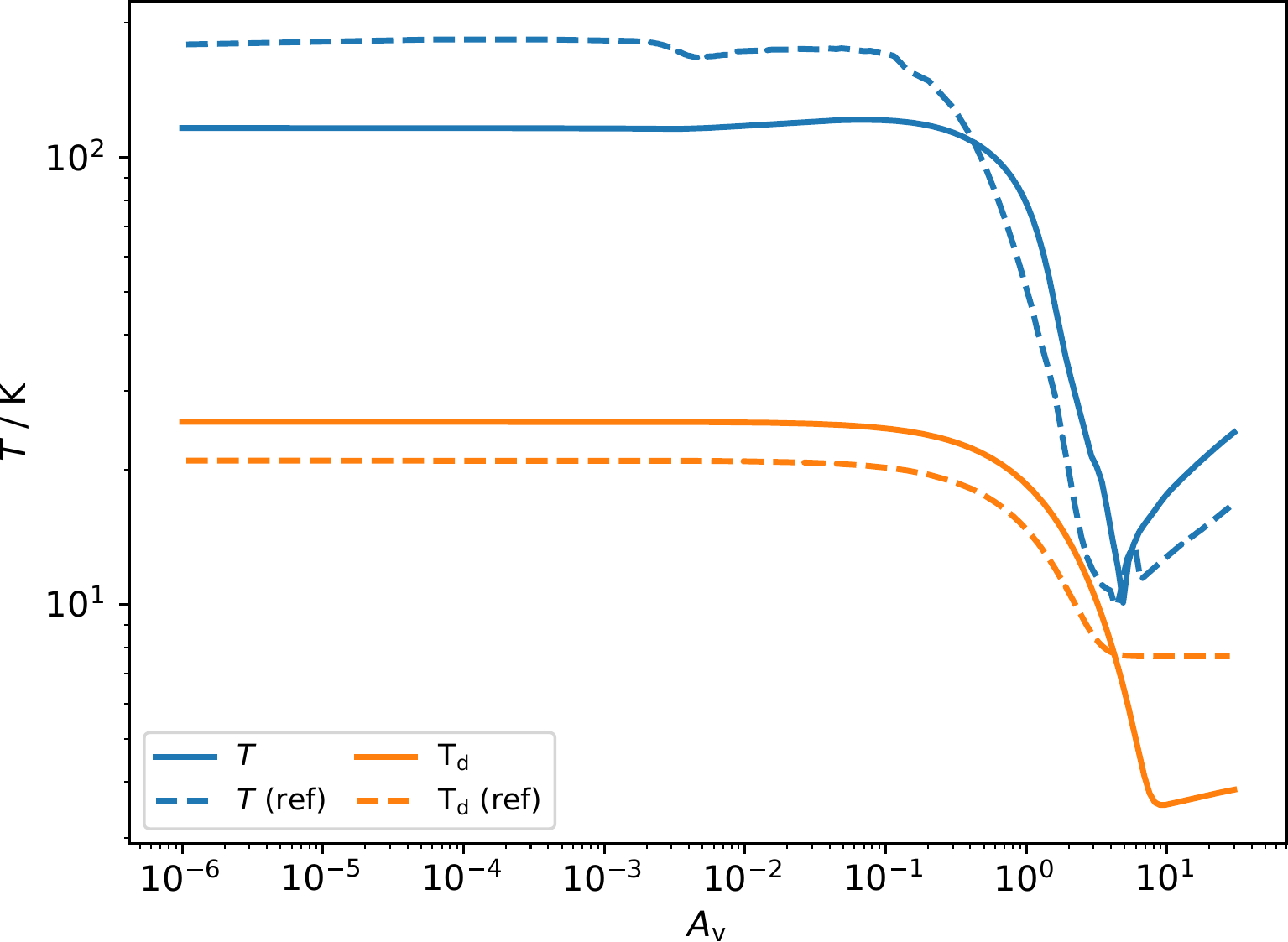}
 \end{minipage}
 \begin{minipage}[t]{0.49\textwidth}
       \includegraphics[width=\textwidth]{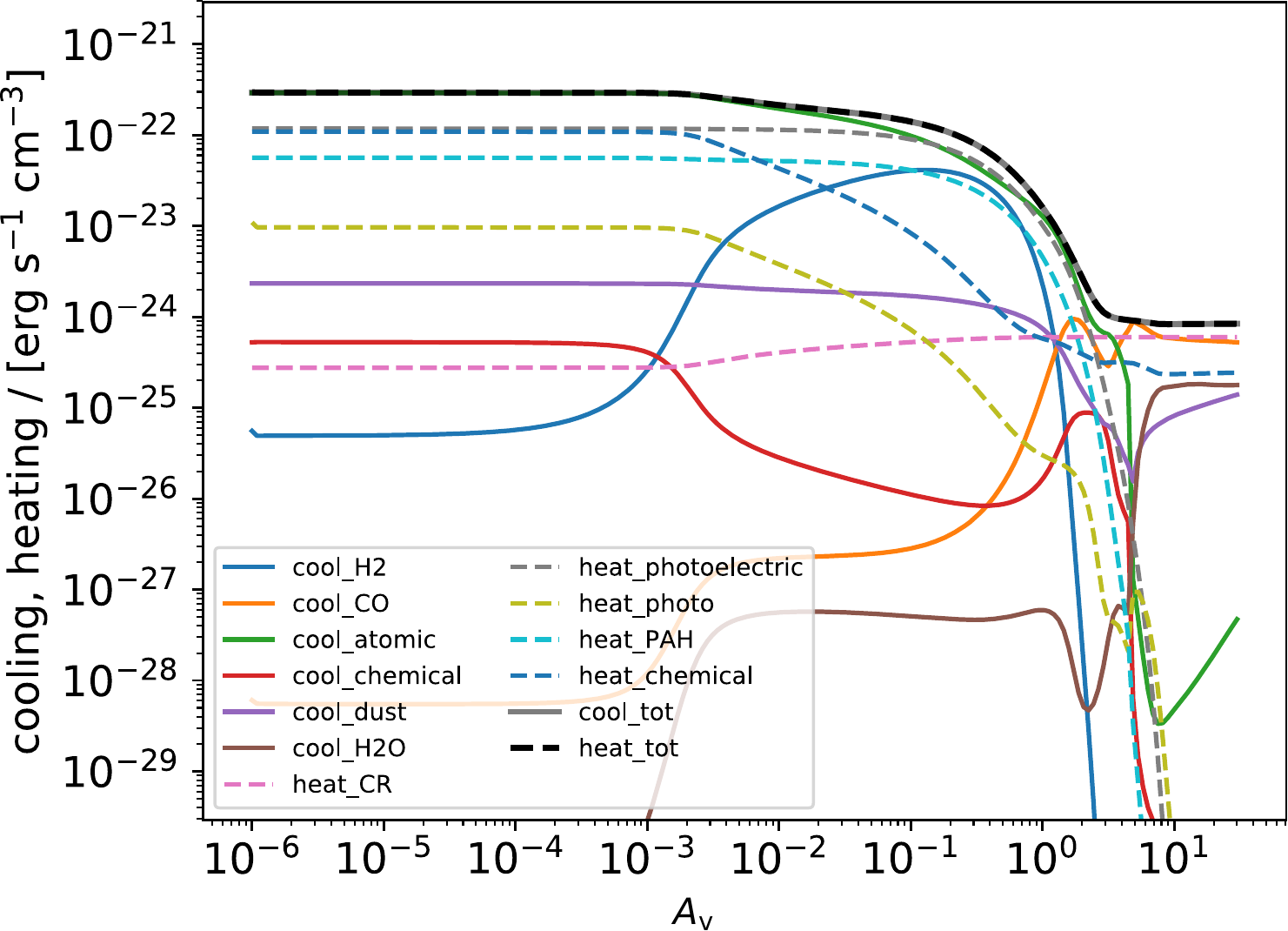}
 \end{minipage}

\vspace{-0pt}
 \begin{minipage}[t]{0.49\textwidth}
       \includegraphics[width=\textwidth]{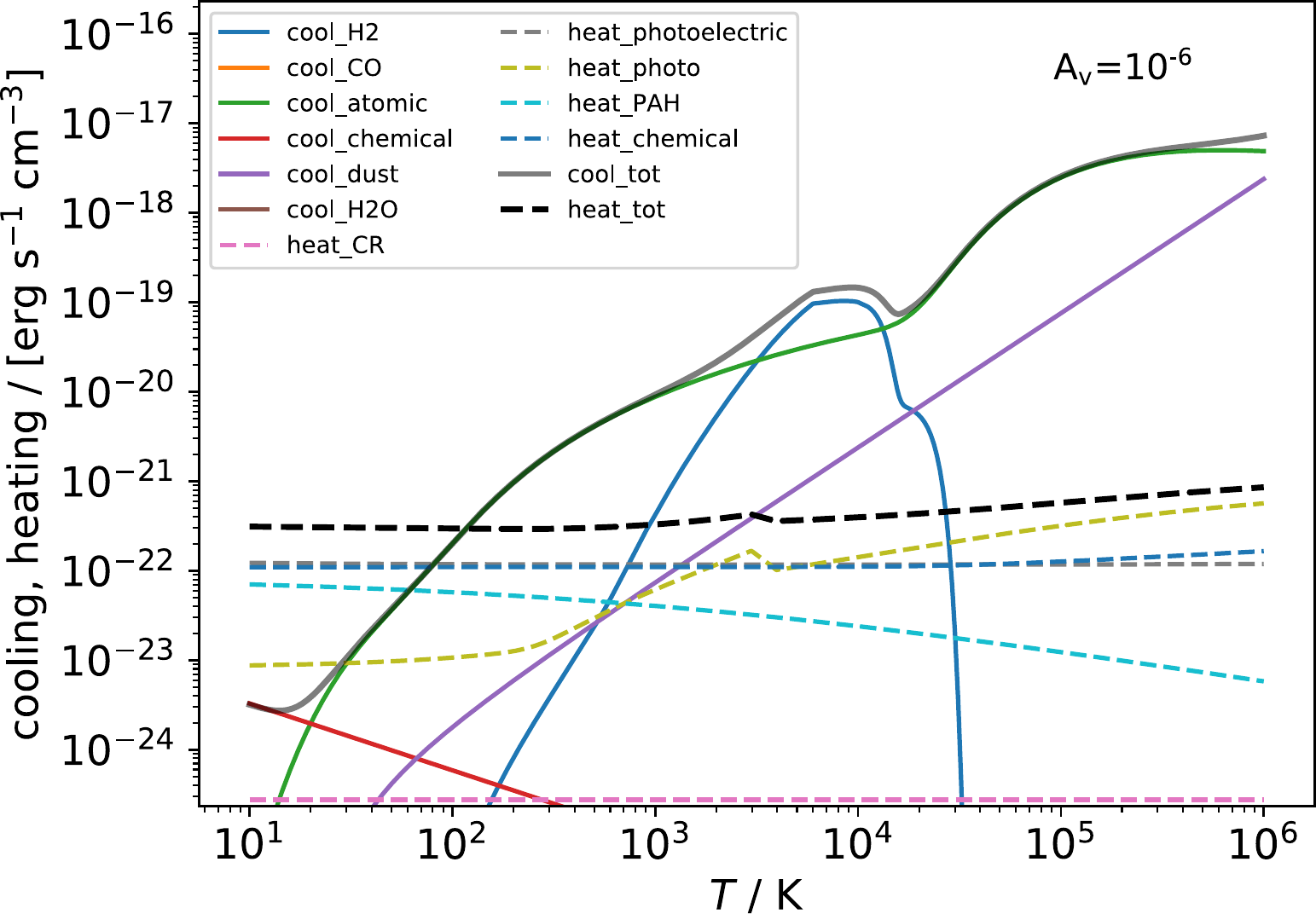}
 \end{minipage}
 \begin{minipage}[t]{0.49\textwidth}
       \includegraphics[width=\textwidth]{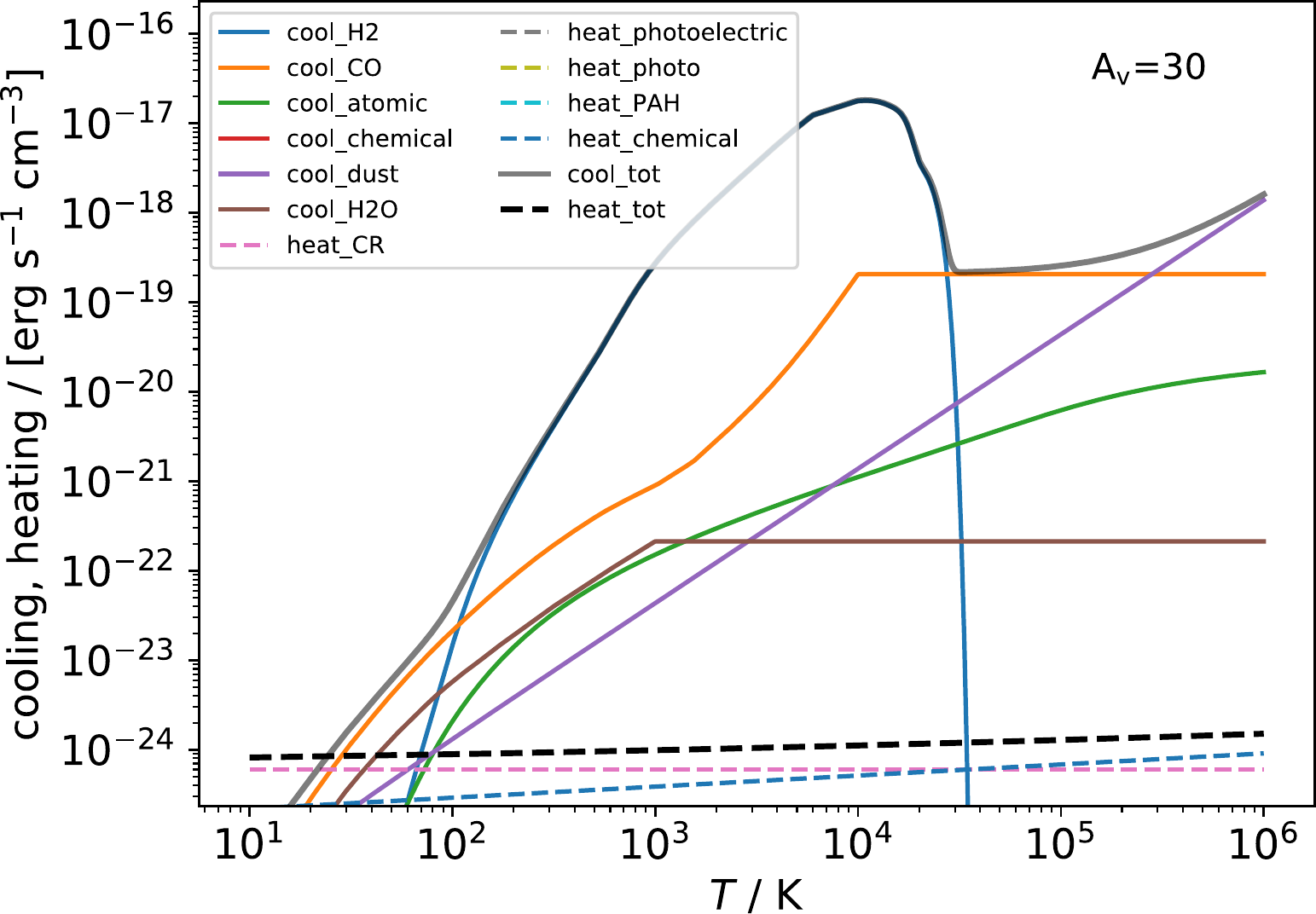}
 \end{minipage}
\caption{Model \texttt{V1}. \textbf{Top left panel}: gas (blue) and dust (orange) temperature profiles as a function of the visual extinction $A_{\rm v}$. Solid lines are the results from our code, while dashed indicate model ``UCL-chem'' from R07. \textbf{Top right panel}: cooling (solid) and heating (dashed) functions at different visual extinction values. Grey solid and black dashed lines respectively indicate the sum of the cooling (\texttt{cool\_tot}) and heating terms (\texttt{heat\_tot}). The functions are respectively molecular hydrogen cooling (\texttt{cool\_H2}), CO cooling (\texttt{cool\_CO}), collisional radiative atomic cooling (\texttt{cool\_atomic}), dust-gas interaction cooling (\texttt{cool\_dust}), water cooling (\texttt{cool\_H2O}), cosmic-ray heating (\texttt{heat\_CR}), photoelectric heating from dust grains (\texttt{heat\_photoelectric}), photoionization heating (\texttt{heat\_photo}), heating from PAH (\texttt{heat\_PAH}), and chemical heating (\texttt{heat\_chemical}). \textbf{Bottom left panel}: cooling (solid) and heating (dashed) functions as at different temperatures at $A_{\rm v}=10^{-6}$, keeping the chemical abundances fixed to the values found at equilibrium by the code at the given $A_{\rm v}$. Grey solid and black dashed lines respectively indicate the sum of the cooling (\texttt{cool\_tot}) and heating terms (\texttt{heat\_tot}). \textbf{Bottom right panel}: cooling (solid) and heating (dashed) functions as at different temperatures at $A_{\rm v}=30$, keeping the chemical abundances fixed to the values found at equilibrium by the code at the given $A_{\rm v}$. Grey solid and black dashed lines respectively indicate the sum of the cooling (\texttt{cool\_tot}) and heating terms (\texttt{heat\_tot}).}\label{fig:bench_v1_temperature}\label{fig:bench_v1_cooling_at}\label{fig:bench_v1_cooling_av_low}\label{fig:bench_v1_cooling_av_high}
\end{figure*}

We report the dust and grain temperature profiles as a function of the visual extinction $A_{\rm v}$ in the top left panel of \fig{fig:bench_v1_temperature}, where the dashed line is the model ``UCL-chem'' from R07, model \texttt{V1}. We note that the gas temperature profile is similar to the one from the benchmark, still keeping in mind that their profile is higher with respect to the other models, so that our temperature profile is within the uncertainty from the benchmark. Analogously, the dust temperature profile presents similar features and it is within the results obtained by the codes of the R07 benchmark. Compared to the findings in \citet{Hocuk2017}, we obtained a lower dust temperature at higher $A_{\rm v}$ (e.g.~their Fig.~3); we verified that this discrepancy derives from the different radiation field employed, i.e.~a Draine field limited to $6-13.6$~eV in our case, while they include a broader spectrum that accounts for the mid-infrared component, as discussed in their Sect.~B. The behavior of $\tdust$ at higher densities is discussed in \appx{appendix:vhigh}.

To understand the temperature profile, we report in the top right panel of \fig{fig:bench_v1_cooling_at} the cooling and the heating rates for each component. At low $A_{\rm v}$ the dominating coolant is \texttt{cool\_atomic}, i.e.~the collisional radiative cooling from atoms which is several orders of magnitude greater than the second most important \texttt{cool\_dust}, while the heating is controlled by photoelectric, chemical, and PAH heating. In the inner part of the cloud (higher $A_{\rm v}$), CO and water are the main coolants, while heating is dominated by cosmic-rays (\texttt{heat\_CR}) and chemical heating.

The bottom panels of \fig{fig:bench_v1_cooling_av_low} show the cooling and heating functions at different temperatures, keeping the chemical composition fixed, and in principle the intersection between the two total functions should indicate the equilibrium temperature. We note that for $A_{\rm v}=10^{-6}$, atomic line cooling is the dominating factor, apart from a small bump around $10^4$~K, where molecular hydrogen cooling is dominating (note again that in this plot for all the temperatures the chemical composition is the one found at the equilibrium for the given $A_{\rm v}$). At lower temperatures heating is dominated by photoelectric, chemical, and PAH heating, while at higher temperatures photoionization and photodissociation heating becomes dominant.
For $A_{\rm v}=30$, CO cooling dominates at lower temperatures, then is replaced by molecular hydrogen cooling from around 100~K to $3\times10^4$~K, where CO cooling becomes dominating again. The plateau in the CO cooling function there is given by the limits on our cooling functions. However, in this temperature range we do not expect to find a relevant amount of CO molecules. We could then expect that dust cooling becomes dominant, depending on whether the grains are thermally coupled with the gas.

\begin{figure*}
 \includegraphics[width=\textwidth]{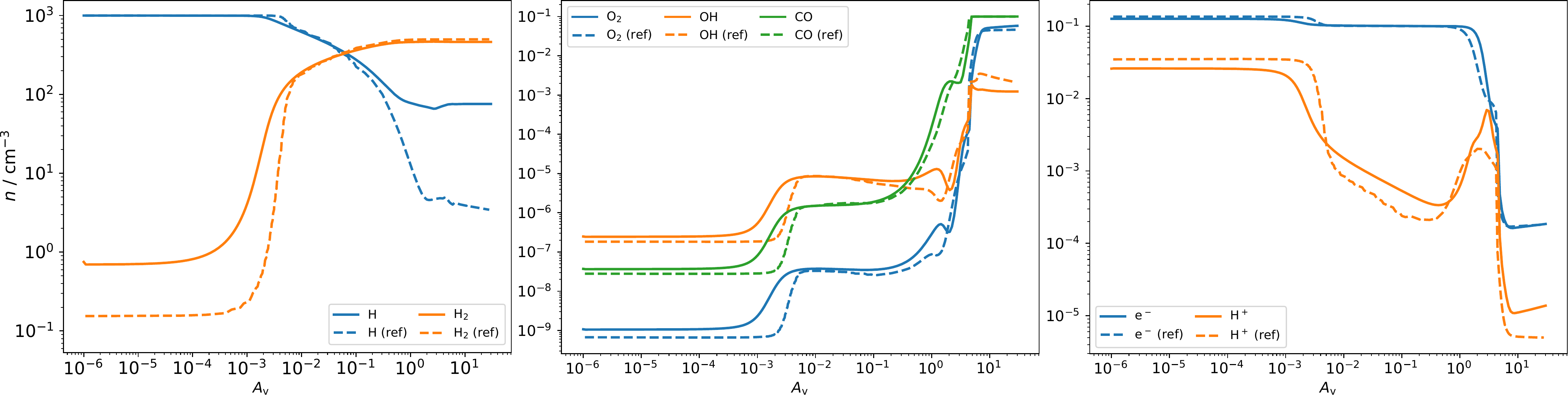}
 \caption{Model \texttt{V1}. \textbf{Left panel}: H (blue), and H$_2$ (orange) density profiles as a function of the visual extinction $A_{\rm v}$. Solid lines are the results from our code, while dashed indicate model ``UCL-chem'' from R07. \textbf{Middle panel}: O$_2$ (blue), OH (orange), and CO (green) density profiles as a function of the visual extinction $A_{\rm v}$. Solid lines are the results from our code, while dashed indicate model ``UCL-chem'' from R07. \textbf{Right panel}: e$^-$ (blue) and H$^+$ (orange) density profiles as a function of the visual extinction $A_{\rm v}$. Solid lines are the results from our code, while dashed indicate model ``UCL-chem'' from R07.}\label{fig:bench_v1_hydrogen}\label{fig:bench_v1_oxygen}\label{fig:bench_v1_electrons}
\end{figure*}

We also compare the results from some of the key chemical species, as for example the transition between molecular and atomic hydrogen (left panel of \fig{fig:bench_v1_hydrogen}), OH, O$_2$, and CO (middle panel), and electrons and H$^+$ (right panel). We note a general agreement with R07, apart from some discrepancies that come from the different chemical network and from the slightly different temperatures found. As expected, the molecular component is predominant at higher $A_{\rm v}$, and the ionization fraction is larger at lower $A_{\rm v}$, where radiation is dominating.

\subsection{PDR: \citet{Rollig2007} \texttt{V4}}\label{sect:bench_v4}

This test has the same initial conditions as the \texttt{V1} discussed in \sect{sect:bench_v1} (see \tab{table:initial_conditions}), apart from the total gas density and the radiation field strength, that now are $\ngas=10^{5.5}$~cm$^{-3}$ and $\chi=10^5$. We also assume that carbon is fully ionized, to avoid problems during the first call of the photoelectric heating routine (however, even a small amount of electrons is sufficient). This does not affect the final results, which reach equilibrium in any case.

\begin{figure*}
 \begin{minipage}[t]{0.48\textwidth}
       \includegraphics[width=\textwidth]{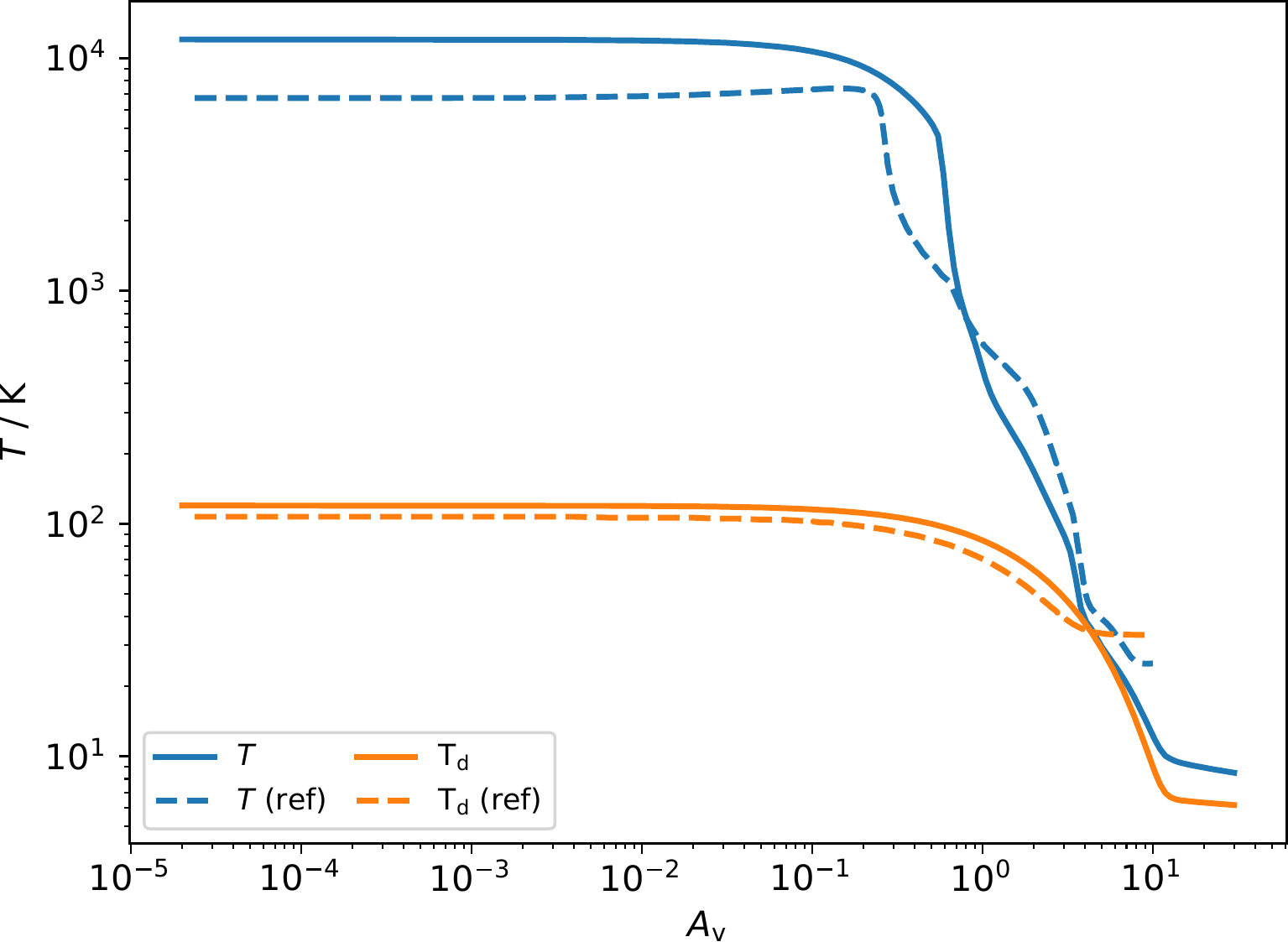}
 \end{minipage}
 \begin{minipage}[t]{0.48\textwidth}
       \includegraphics[width=\textwidth]{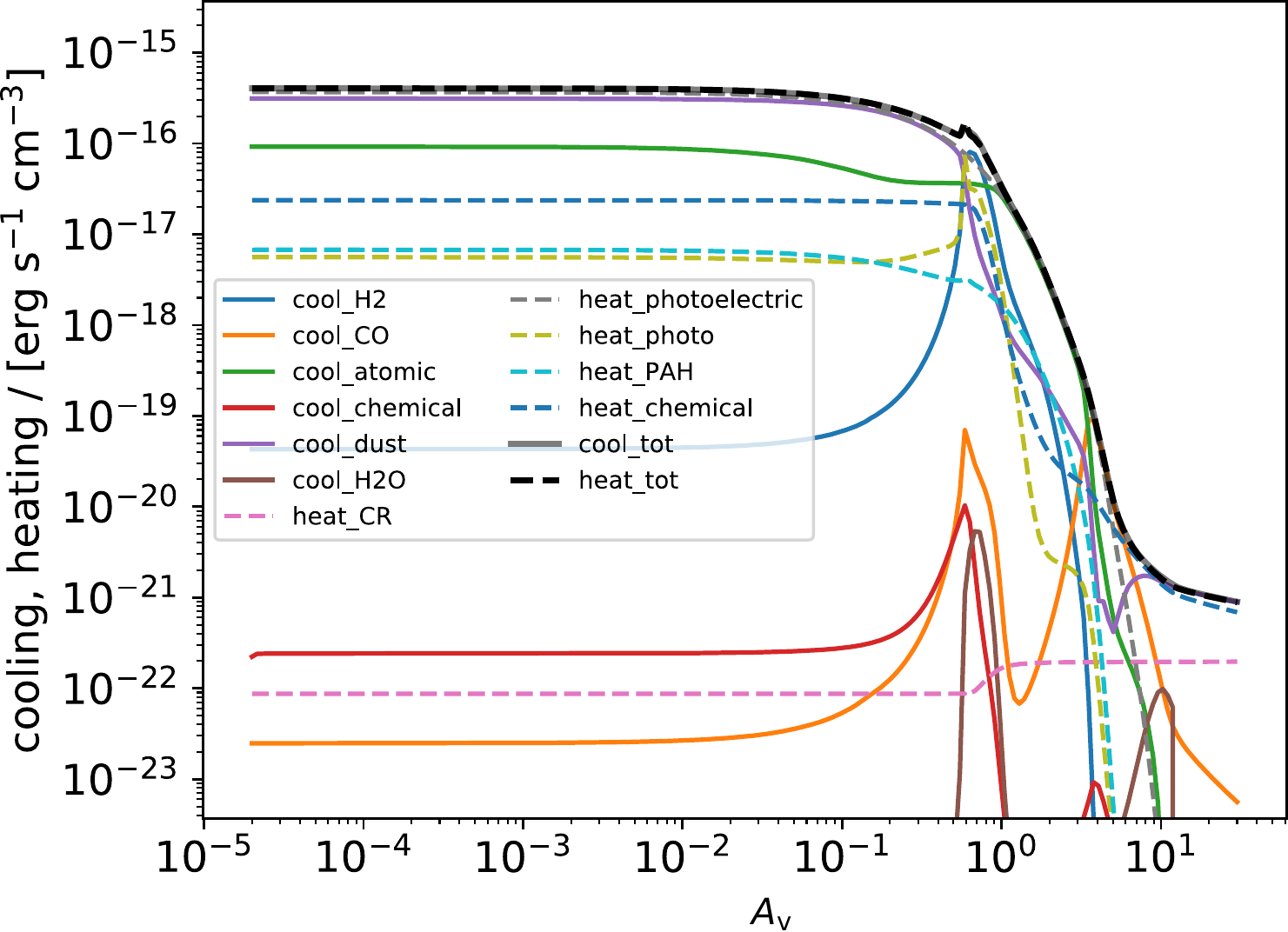}
 \end{minipage}
 \caption{Model \texttt{V4}. \textbf{Left panel}: gas (blue) and dust (orange) temperature profiles as a function of the visual extinction $A_{\rm v}$. Solid lines are the results from our code, while dashed indicate model ``Cloudy'' from R07. \textbf{Right panel}: cooling (solid) and heating (dashed) functions at different visual extinction values. Grey solid and black dashed lines indicate the sum of the cooling (\texttt{cool\_tot}) and heating terms (\texttt{heat\_tot}), respectively.}
 \label{fig:bench_v4_cooling_at}\label{fig:bench_v4_temperature}
\end{figure*}

Analogously to \sect{sect:bench_v1}, in the left panel of \fig{fig:bench_v4_temperature} we report the temperature profile as a function of the visual extinction for the gas and the dust components, for which we find general agreement between our results and the results from R07 (here the reference is ``Cloudy''). As discussed in the case \texttt{V1}, our results agree with the range of values found amongst the benchmark codes. The temperatures are higher at lower visual extinctions, i.e.~closer to the radiation source, and become lower when moving inside the cloud.

This trend is confirmed when in the right panel of \fig{fig:bench_v4_cooling_at} we plot the detailed cooling and heating functions as a function of the visual extinction; in the outer part of the slab, photoelectric heating and dust cooling are the two dominant components, while in the inner part chemical heating dominates. The transition between these two zones is controlled by the collisional atomic radiative cooling, and photoelectric and chemical heating.

\begin{figure*}
   \centering
       \includegraphics[width=\textwidth]{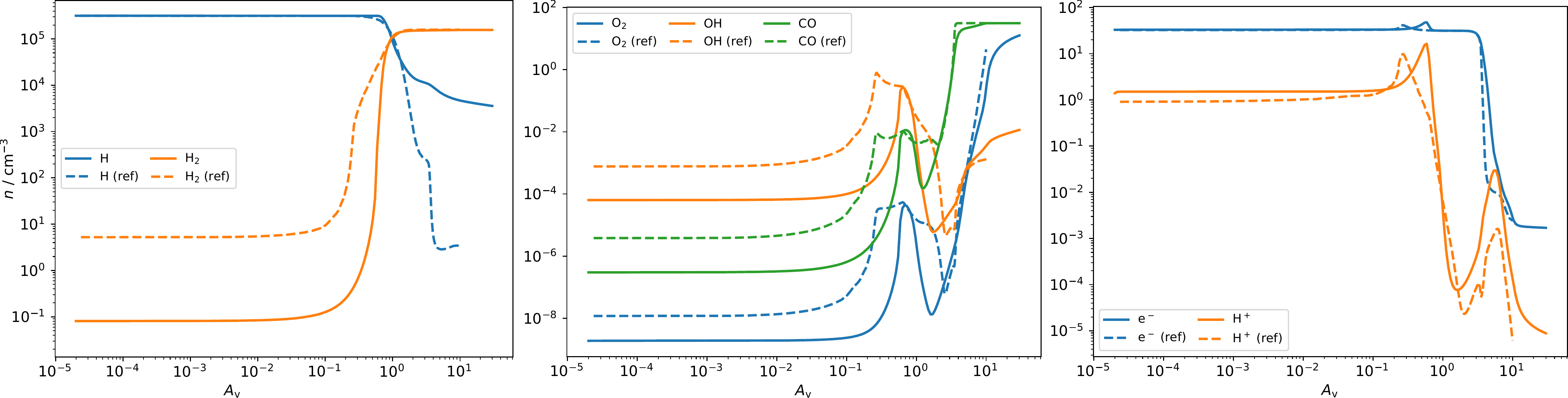}
 \caption{Model \texttt{V4}. \textbf{Left panel}: H (blue) and H$_2$ (orange) density profiles as a function of the visual extinction $A_{\rm v}$. Solid lines are the results from our code, while dashed indicate model ``Cloudy'' from R07.
  \textbf{Middle panel}: O$_2$ (blue), OH (orange), and CO (green) density profiles as a function of the visual extinction $A_{\rm v}$. Solid lines are the results from our code, while dashed indicate model ``Cloudy'' from R07. \textbf{Right panel}: e$^-$ (blue) and H$^+$ (orange) density profiles as a function of the visual extinction $A_{\rm v}$. Solid lines are the results from our code, while dashed indicate model ``Cloudy'' from R07.}
 \label{fig:bench_v4_electrons}\label{fig:bench_v4_hydrogen}\label{fig:bench_v4_oxygen}
\end{figure*}

Our results agree with the benchmark for where atomic hydrogen becomes molecular (\fig{fig:bench_v4_hydrogen}, left panel), but we find that differences in the gas temperature affect the chemistry, as shown by the middle panel of \fig{fig:bench_v4_oxygen}, where O$_2$, OH, and CO have lower molecular abundances because of the higher temperature, while in the inner part of the slab the abundances agree better, since temperatures are comparable. In the right panel of \fig{fig:bench_v4_electrons} the source of the difference we found at lower $A_{\rm v}$ can be explained by the fact that the ionization is controlled mainly by the photoionization rates, that are similar to the ones from the benchmark, while the discrepancies found in H$^+$ are due to the different gas-phase rate coefficients from our chemical network.

\subsection{\texttt{Cooling} benchmark}
\begin{figure}
   \centering
       \includegraphics[width=.48\textwidth]{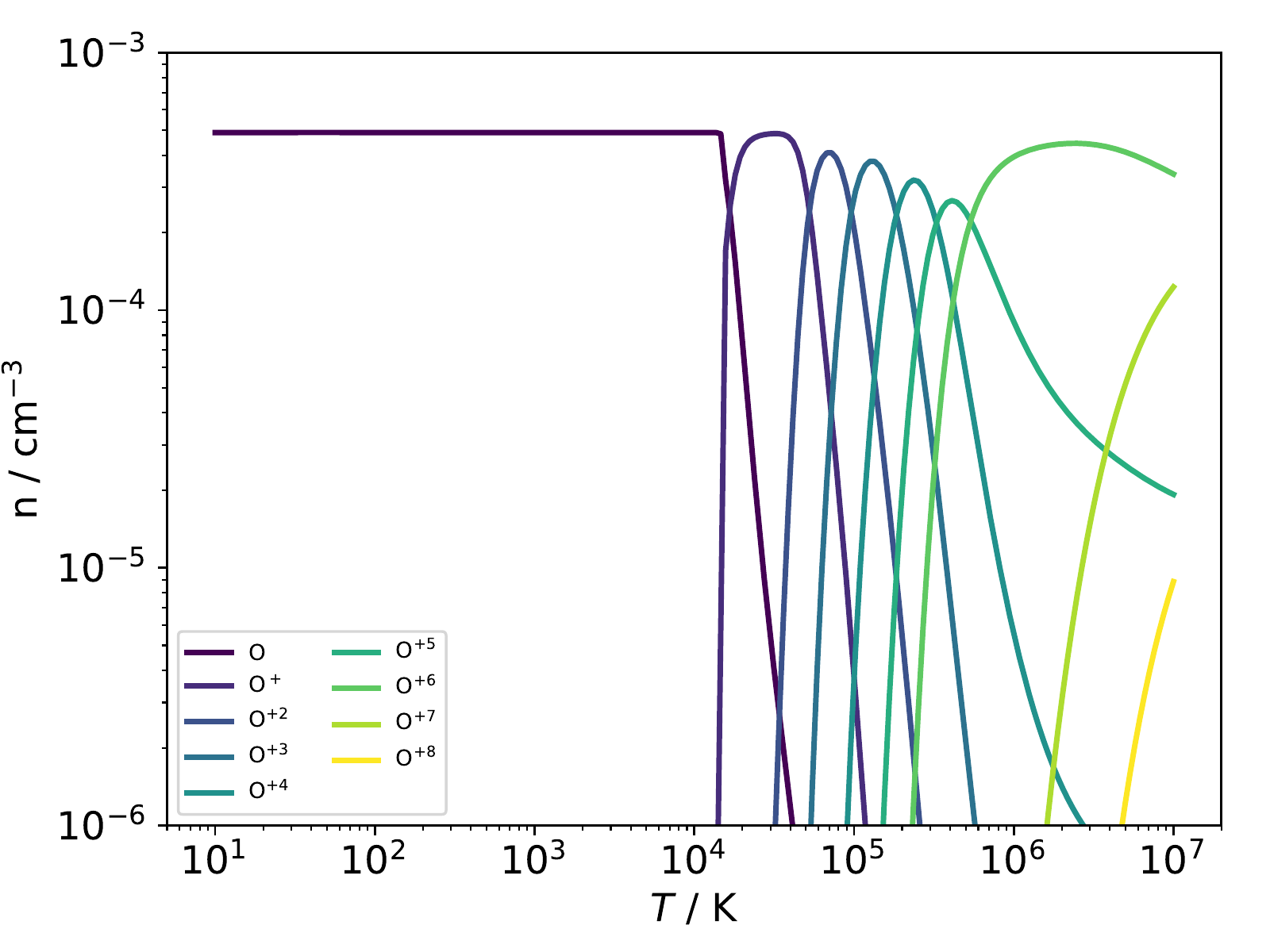}
 \caption{Model \texttt{Cooling}: abundances of oxygen ions at different temperatures assuming equilibrium chemistry.}
 \label{fig:bench_cool_oxygen}
\end{figure}

\begin{figure}
   \centering
       \includegraphics[width=.48\textwidth]{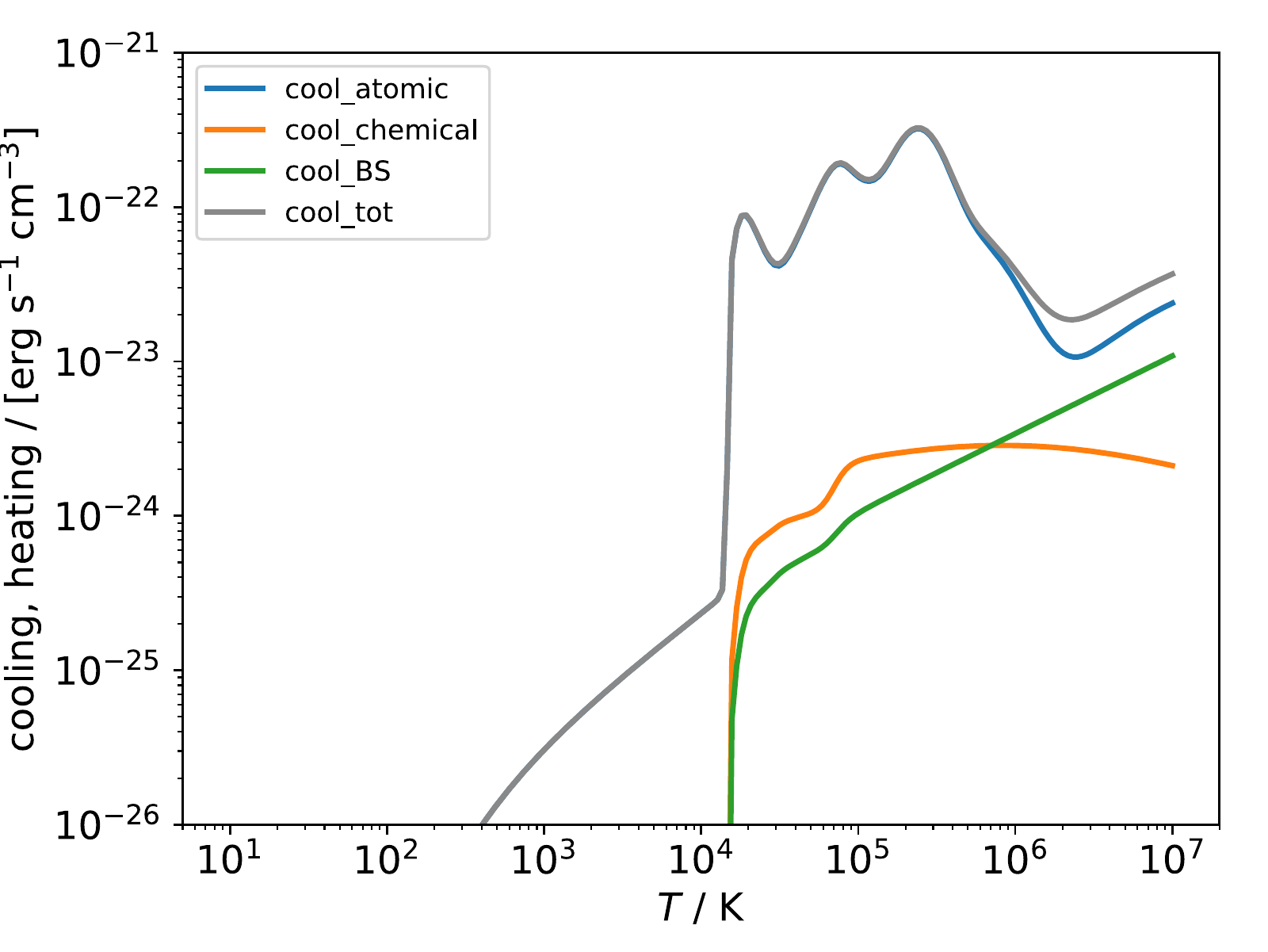}
 \caption{Model \texttt{Cooling}: cooling profile for different temperatures assuming chemical equilibrium. \texttt{cool\_atomic} (blue solid line) is the collisional radiative cooling, \texttt{cool\_chemical} (orange) is the cooling from recombination and collisional ionization, \texttt{cool\_BS} (green) is the bremsstrahlung from ions, and \texttt{cool\_tot} (grey) is the total cooling.}
 \label{fig:bench_cool_cooling}
\end{figure}

The aim of the \texttt{cooling} benchmark is to compute the equilibrium chemistry at different gas temperatures and evaluate the cooling, in order to obtain a function similar to~Fig.~3 in \citet{Gnat2012} (see also \citealt{Sutherland1993}). Their results show the contributions of the different chemicals species to the total cooling, and consist of a temperature-dependent cooling function from $T=10^4$ to $10^8$~K for atomic species in chemical equilibrium without any impinging external radiation. Since we extend our test to $T<10^4$~K, we compare our results also to Fig.~4 from \citet{Maio2007}, that produced an analogous cooling function that includes that specific temperature range. We employ H, He, C, O, Ne, and no dust, with the initial abundances as in \tab{table:initial_conditions}, and we evolve the system for $10^6$~yr. The chemical network includes ion-electron recombinations, charge exchange reactions, and collisional ionizations, for all the available atomic ionization levels by using the reactions available in the internal database of \codename{} (see~\sect{sect:database}). Once the chemical equilibrium is reached for all the chemical species (see e.g.~oxygen ions in \fig{fig:bench_cool_oxygen}), we evaluate the cooling function, as reported in \fig{fig:bench_cool_cooling}. We note that \texttt{cool\_atomic}  (i.e.~the cooling from collisional excitation lines) dominates at every temperature, except where bremsstrahlung contributes ($T>5\times10^5$~K). Below $10^4$~K the cooling is dominated by C and O and their respective ions, as pointed out by \citet{Maio2007} and \citet{Gnat2012}. The first peak in \fig{fig:bench_cool_cooling} is due to hydrogen, and the subsequent peaks are caused by the interplay of carbon and oxygen cooling emission lines, with Ne contributing around $5\times10^5$ to $10^6$~K (cfr.~Fig.~3 in \citealt{Gnat2012}).

To ensure that independently from the chemical network the machinery that produces the atomic cooling is working, we verified that the intensity from our emission lines matches the results obtained with \textsc{ChiantiPy}\footnote{\url{https://github.com/chianti-atomic/ChiantiPy/}} for specific temperatures, and protons and electrons densities. Despite some small difference caused by the different chemical network, the results reported in \fig{fig:bench_cool_cooling} are similar to what was obtained by \citet{Gnat2012}, both in term of cooling function and chemical equilibrium\footnote{\url{http://wise-obs.tau.ac.il/~orlyg/cooling/CIEion/tab2.txt}}.

\subsection{\texttt{XDR}, atomic}
To test the effects of a spectrum that includes X-rays we reproduce the set-up from \citet{Picogna2019}, where the temperature of a slab of gas is computed for different ionization rates and column densities. In particular, we refer to the dashed black lines in their Fig.~2, where the gas temperature is calculated as a function of the ionization parameter $\xi$ and for three different column densities $N$. Their results are computed by using the Monte Carlo radiative transfer code \textsc{Mocassin}.

The ionization parameter is defined as \citep{Tarter1969,Owen2010}
\beq
    \xi = \frac{L_{\rm x}}{n R^2}\,,
\eeq
where $L_{\rm x}$ is the X-ray luminosity of the central source, $n=10^2$~cm$^{-3}$ is the local gas density, and $R$ is the distance from the source.

\begin{figure}
   \centering
       \includegraphics[width=.48\textwidth]{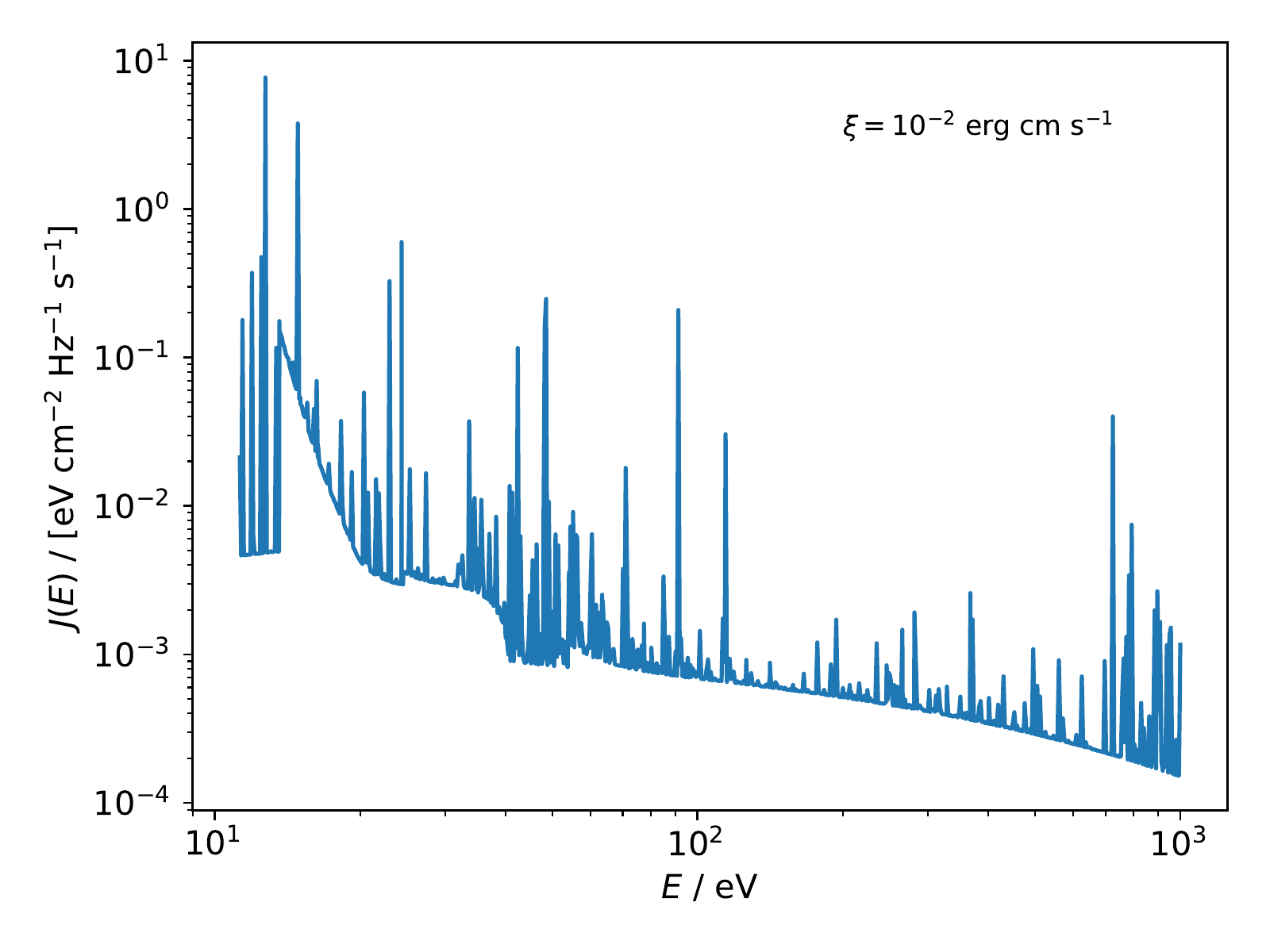}
 \caption{Model \texttt{XDR}, atomic: the radiation spectrum employed for the XDR test as a function of the energy, here reported for $\xi=2\times10^{-3}$~erg~cm~s$^{-1}$. The spectrum intensity scales linearly with $\xi$.}
 \label{fig:bench_xray_spectrum}
\end{figure}

The set-up is similar to \sect{sect:bench_v4}, i.e.~a semi-infinite slab, with constant density, an  emitting source on the left side. However, in this test we only employ atomic species (H, He, C, O, N, S, Si, Ne, and Mg), we have no dust, the line escape probability is disabled (i.e.~$\beta=1$ in \sect{sect:radiative_atomic_cooling}), and the spectrum (that scales linearly with $\xi$) is reported in \fig{fig:bench_xray_spectrum}. The reaction rates are chosen using the internal database by following the same criteria as in the \texttt{cooling} test, but extended to the current chemical species. The initial conditions are in \tab{table:initial_conditions}. For the present test we adopt a lower limit in temperature of 30~K as in \citet{Picogna2019}.

\begin{figure}
   \centering
       \includegraphics[width=.48\textwidth]{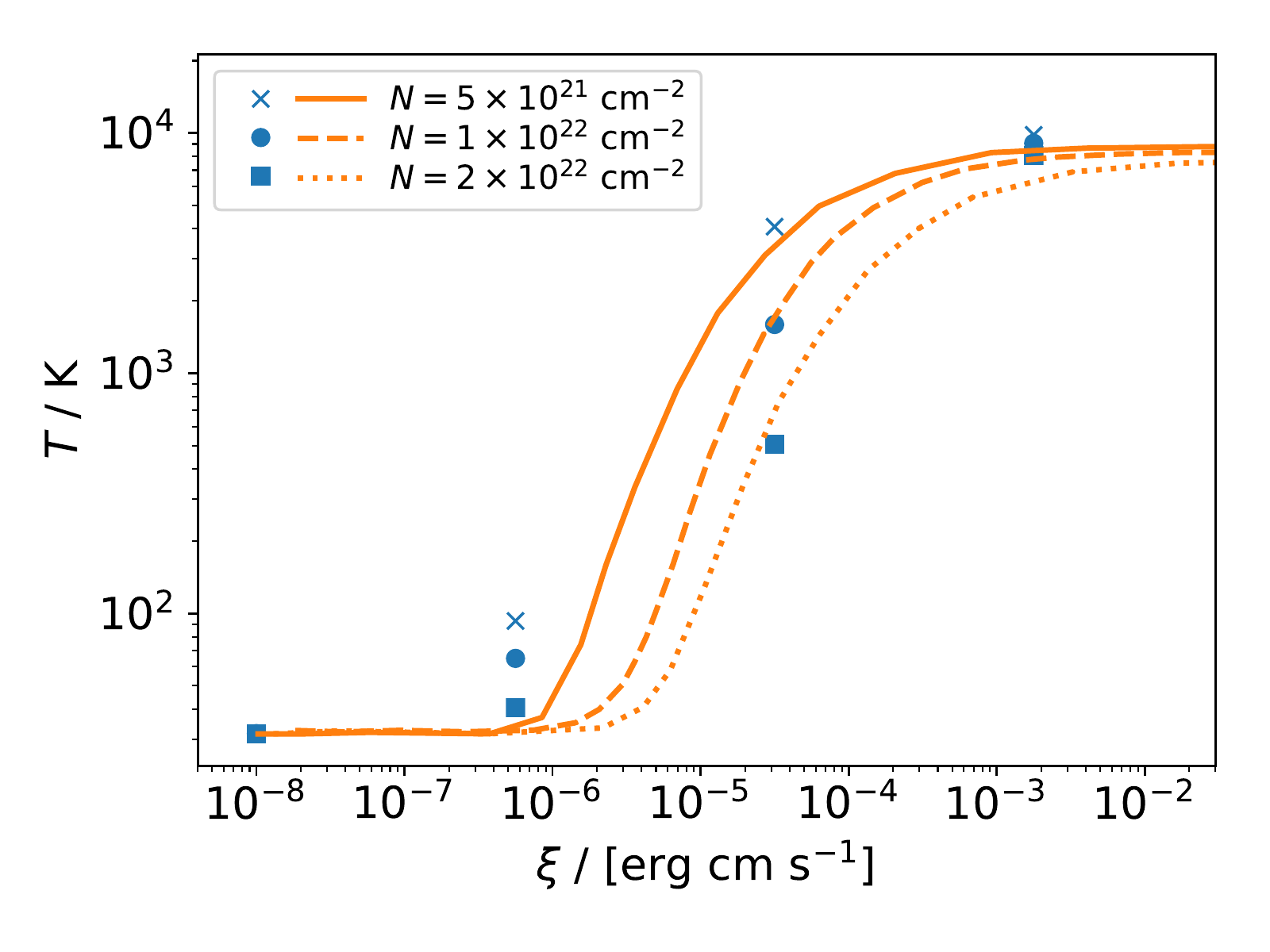}
 \caption{Model \texttt{XDR}, atomic: equilibrium temperature found by our code for different ionization parameter values $\xi$, at different column densities $N=5\times10^{21}$ (cross), $N=1\times10^{22}$ (circle), and $N=2\times10^{22}$~cm$^{-2}$ (square), compared with the fit from \citet{Picogna2019} for the corresponding column densities (respectively solid, dashed, and dotted lines).}
 \label{fig:bench_xray_xit}
\end{figure}

The results obtained in \fig{fig:bench_xray_xit} by \codename{} are compared to Fig.~2 of \citet{Picogna2019},
where three selected column densities are reported, i.e.~$N=5\times10^{21}$, $N=1\times10^{22}$, and $N=2\times10^{22}$~cm$^{-2}$, computed as $N=n\,z$, where $z$ is the distance from the left side of the slab, and $n$ is the constant gas density.
We note that \citet{Picogna2019} reports a fitting function of the original results from \textsc{Mocassin}; these models have a larger variance than the fitting functions, especially in the range $5\times10^{-7}\lesssim\xi\lesssim10^{-5}$~erg~cm~s$^{-1}$. In addition to this, in that interval, the variation of the temperature with $N$ is quite rapid where $5\times10^{21}\lesssim N\lesssim\times10^{22}$~cm$^{-2}$, that explains the discrepancy between the fitting functions from \citet{Picogna2019} and our results around $\xi\approx10^{-6}$~erg~cm~s$^{-1}$.

\begin{figure}
   \centering
       \includegraphics[width=.48\textwidth]{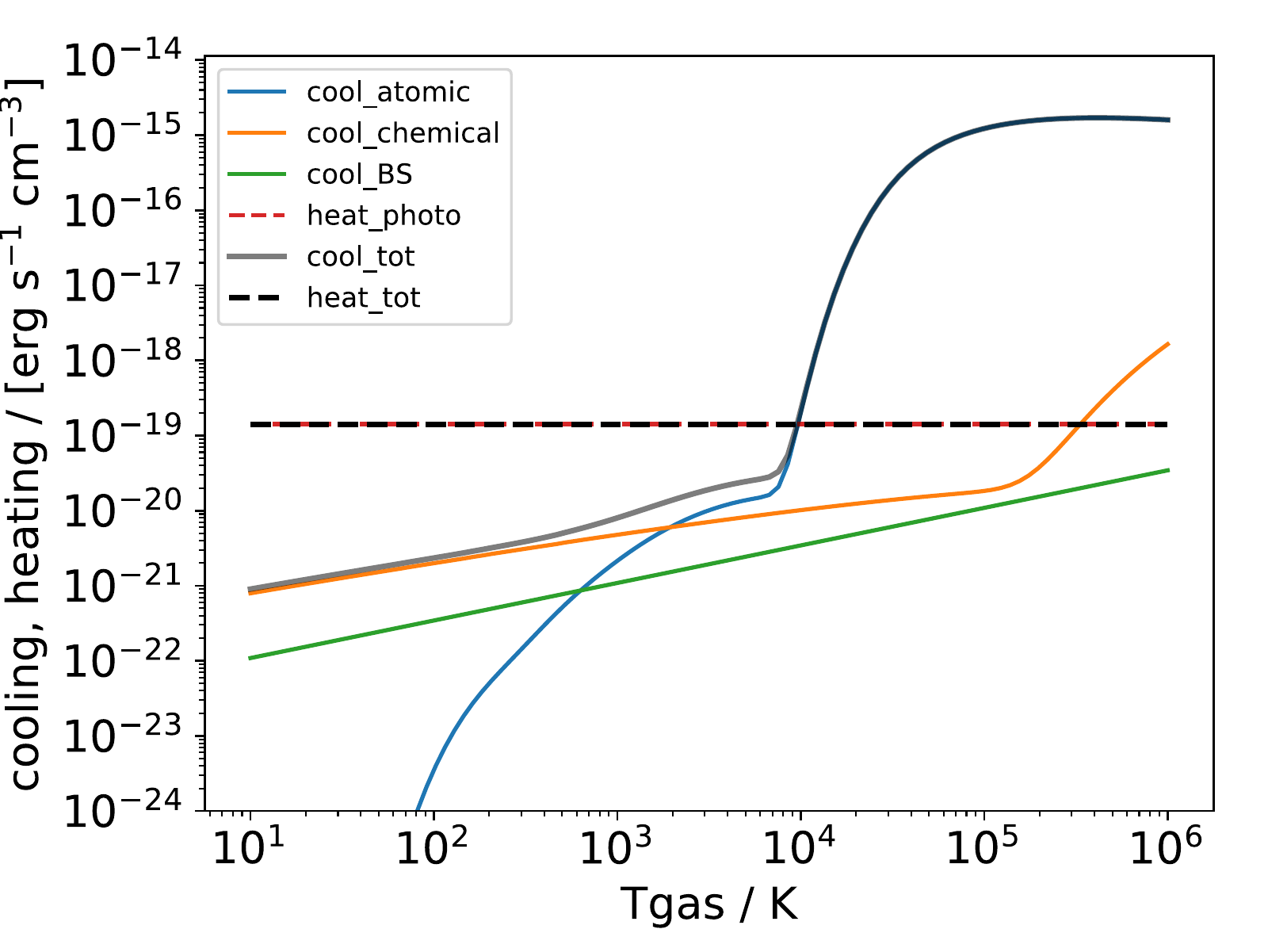}
 \caption{Model \texttt{XDR}, atomic: Cooling function for $\xi=2\times10^{-2}$~erg~cm~s$^{-1}$ and $N=5\times10^{21}$~cm$^{-2}$, where the individual functions are collisional radiative atomic cooling (solid blue), chemical cooling (solid orange), bremsstrahlung (solid green), and heating from photochemistry (dashed red). Solid grey line is the total cooling. Note that total heating (dashed black) overlaps photoheating.}
 \label{fig:bench_xray_cooling}
\end{figure}

To understand what are the dominant thermal processes, we report in \fig{fig:bench_xray_cooling} the detailed cooling functions for the model with $\xi=2\times10^{-3}$~erg~cm~s$^{-1}$, where we notice that the chemical cooling from recombination dominates at lower temperatures, collisional excitation radiative cooling starts to be relevant when $T>10^4$~K, bremsstrahlung is less important, while heating is dominated by photoionization.

This test shows that \codename{} produces correct results, similar to what has been obtained by \textsc{Mocassin} in \citet{Picogna2019}.

\section{Databases used by prizmo}\label{sect:database}
\codename{} requires a large amount of astrochemical data from several sources. For this reason \codename{} contains an internal database where all the required information is stored and used during the pre-processor stage and at runtime. These data are handled by the different \python{} objects and employed to write parts of the \fortran{} code needed at runtime. For the sake of clarity, we describe in this Section the different databases.

For the collisional atomic line cooling we employ data from \krome{} and from the \textsc{Chianti} database for H, He, Li, Mg, N, C, O, Ne and Si. \textsc{Chianti} is the default, unless we have rate collisions from the former, since some of the rates are updated and include additional colliders. We manually inspect the rate coefficients to guarantee that they do not present sharp transitions in temperature, and to extend the range of validity where possible.

Radiative recombination rates are taken from \citet{Verner1996}, for H-like, He-like, Li-like, and Na-like ions, and from \citet{Shull1982}, \citet{Landini1990}, \citet{Landini1991}, and \citet{Pequignot1991} for C, N, O, Ne, Na, Al, F, P, Cl, and Fe.

Atomic photoionization cross-sections are taken from \citet{Verner1996}, where the data is proivded as fitting functions of the photon energy.

Charge exchange data are from \citet{Arnaud1985} for collisions of H$^{(+)}$ and He$^{(+)}$ with Li, C, N, O, Na, Mg$^{(+)}$, Si$^{(+)}$, S, Mn, and Fe$^{(+)}$, while collisional ionizations employ the fit provided by \citet{Voronov1997} for a variety of atoms including H, He, Li, C, N, O, Ne, Na, Mg, Al, Si, S, Cl, Fe, and others less relevant for the current problems.

CO cooling is based on \citet{Omukai2010}, and consists of a three-dimensional look-up tables of the local density, the CO column density, and the gas temperature. A similar approach is employed for H$_2$O.
Even if molecular cooling is evaluated by using tables, the code employs data from the \textsc{LAMDA} database to evaluate the molecular lines emission with a machinery similar to the one employed to compute atomic cooling and emissions.

Dust refractive indexes are stored as in \citet{Laor1993}, i.e.~with the frequency-dependent real and imaginary parts of the dielectric function for different grain materials. These data are used to compute the photoelectric heating, and to compute the dust temperature.
Dust opacity can be evaluated by the code using the dielectric functions of the grain material and Mie theory\footnote{See e.g.~\url{http://scatterlib.wikidot.com/mie} and \url{https://bitbucket.org/tgrassi/compute_qabs/}.} \citep{BohrenBook1983,Giuliano2019}. However, we also include some pre-computed dust opacity tables to reduce the pre-processing time when dust grains have some default material composition (e.g.~silicates) and grains size distribution. In this case we employ some of the look-up tables\footnote{\url{https://www.astro.princeton.edu/~draine/}} from \citet{Draine2003b}, that have also a more complicated treatment including for example PAH.

Apart from specific rate coefficients, e.g.~charge exchange, we include the \verb+kida.uva.2014+ network from the \textsc{KIDA} database, to include selected rates that are not provided with a \fortran{} expression by the user. However, we note that including reaction rates blindly might cause unpredictable results.

CO self-shielding is taken from \citet{Visser2009} as a function of $N_{\rm CO}$ and $N_{\rm H_2}$. Although it is possible to retrieve cooling tables for a large set of parameters\footnote{\url{https://home.strw.leidenuniv.nl/~ewine/photo} \label{footnote:leiden}}, we limit our data to $b_{\rm CO}=0.3$~km~s$^{-1}$, $b_{\rm H_2}=3$~km~s$^{-1}$, $b_{\rm H}=5$~km~s$^{-1}$, $T_{\rm ex, CO}=5$~K, $T_{\rm ex, H_2}=11.18$~K, $[^{12}\rm C]/[^{13}\rm C]=69$, $[^{16}\rm O]/[^{18}\rm O]=557$, and $[^{18}\rm O]/[^{17}\rm O]=3.6$. For N$_2$ self-shielding we follow an analogous by using the tables\footref{footnote:leiden} from \citet{Heays2014}, with $b_{\rm N_2}=0.17$~km~s$^{-1}$, $b_{\rm H_2}=3$~km~s$^{-1}$, $b_{\rm H}=5$~km~s$^{-1}$, $T_{\rm ex, {}^{14}N_2}=T_{\rm ex, {}^{14}N{}^{15}N}=T_{\rm ex, H_2}=50$~K, and $\rm[{}^{14}N_2]/[{}^{14}N{}^{15}N]=225$.

Thermochemical data are taken from Burcat's polynomials, that are functions of the temperature. For each species these consist of two individual polynomials for two different contiguous temperature ranges. These $a_i$ coefficients can be employed to compute the enthalpy of formation using
\beq
    \frac{\Delta H_{\rm f}^{\circ}(T)}{RT}=\sum_{i=1}^5 \frac{a_i T^{i-1}}{i}+ \frac{a_6}{T}\,,
\eeq
where $R$ is the gas constant.

Energy-dependent molecular cross-sections for photoionization and photodissociation reactions are taken from the \textsc{Leiden} database\footref{footnote:leiden} and limited to the branching ratios indicated there. We do employ the version of the database that instead of using specific lines has a regular grid spacing of 0.1~nm, where the area of the cross-section in each bin is equivalent to the area of the sum of the cross-sections of the lines within that bin. \codename{} can use the cross-sections from the database but with different branching ratios, using a specific decorator in the chemical network file (see \appx{appx:chemical_network}).

For X-ray reactions with molecules we need the partial cross-sections for the different atomic shells, and the list of cross-sections needed to compute the total cross-sections of a given reaction, so that for example ${\rm CO \to C^+ + O^+ + 2e^-}$ has $\sigma_{\rm CO} = 0.5(\sigma_{\rm C,K}+\sigma_{\rm C,L}+\sigma_{\rm O,K}+\sigma_{\rm O,L})$, where the first subscript is the atomic species, while the second indicates the shell. These data are taken from \citet{Adamkovics2011} and collected with the advice of C.~Rab (private comm.); see \appx{appendix:xray_mols}.


\section{Limitations}\label{sect:limitations}
\codename{} presents a set of limitations due to the fact that many microphysical processes are required to model gas and dust in a protoplanetary disk, and which can span a large range of temperatures, densities, and radiation spectra (see e.g.~\citealt{Haworth2016}). There are three types of approximations: (i) the limited knowledge of the physics and the chemistry in disks, for example some of the reaction rate coefficients are available only within small temperature ranges, and their extrapolation outside the interval is arbitrary. (ii) The simplification required to model complicated physical processes that are tightly interconnected, or that present a large number of parameters and variables. Finally, (iii) solving time-dependent chemistry and microphysics coupled with multi-frequency radiative transfer, or with hydrodynamics, presents a demanding computational challenge, and for this reason we are forced to simplify the description even of some processes with well-understood mechanisms, but with a large numerical footprint, in order to make these simulations feasible.

We will not discuss the problems related to the physical/chemical uncertainties, since they have been already discussed in the previous sections, but we will explain the caveats and the limitations related to our specific implementation.
However, processes that are currently simplified might be improved in the future, thanks to the development of numerical techniques, dedicated hardware, and a better understanding of chemistry, microphyiscs, and protoplanetary disks.

A consistent limitation of our model is the problem of dealing with multi-frequency radiation, and includes not only FUV, but also X-rays. Unfortunately, it is very demanding to model radiation transport with an energy binning that matches exactly the features a large number of molecular and atomic cross-sections, and so we must instead limit our methodology to a reduced spectrum. This limitation might over or underestimate the photochemical rate coefficients (see Paper~I), affecting the chemical and (consequently) the thermal evolution.

Analogously, some of the key reactions that require the knowledge of the intensity of the radiation in several energy bins are treated with a reduced method, that is based on pre-computed radiation with some specific spectral shape, e.g.~the Draine radiation field. Computing the equivalent integrated intensity in a specific band is clearly a limitation, since the effects due to ``non-Draine'' shapes are completely lost. This limitation applies to H$_2$ and CO photodissociation. Similarly, using pre-computed methods for self-shielding of the same molecules, reduces the capability of modelling the thermochemistry of a disk consistently.

Although \codename{}'s capability to compute the emission from the chemical species included in the network, we assume that the radiation produced is lost, and therefore does not feedback on the resulting radiation spectra that come out of the simulated cell, i.e.~after being attenuated by dust and photochemical reactions. This is because we do not compute the shape of the lines that are emitted, but in principle this limitation might be overcome in the future.

Another limit comes from the (currently) fixed grain size distribution. Although, this constraint allows the code to pre-compute a large set of quantities that save considerable computational resources at runtime, the size distribution of dust grains can vary drastically over the extent and lifetime of a disk \citep{Testi2014}, so that this approach might lead to misleading results, in particular when surface chemistry is dominant. This also strongly affects the gas temperature (e.g.~when dust cooling dominates). In order to enable this functionality in the future, \codename{} would need to be coupled with a dedicated code for dust evolution (e.g.~\citealt{Birnstiel2012}).

A similar discussion applies to cosmic rays, which are treated with a single parameter, i.e.~the ionization rate. In principle, their propagation is related to the geometry of the disk \citep{Bai2009, Cleeves2013} and to the topology of the magnetic fields \citep{Padovani2014}, suggesting that, to find the proper ionization rate, we need a non-local treatment, and for this reason this might result in over/underestimating their heating, and their effect on the chemistry, especially where the radiation does not play the main role. Additionally, we do not include any cosmic rays-induced fluorescence, that might have relevant effects on the charge of grains \citep{Ivlev2015} and on the gas chemistry \citep{Visser2018}.

\section{Summary}\label{sect:conclusions}
We present \codename{}, a code for evolving thermochemistry in protoplanetary disks, capable of being coupled with hydrodynamical and multifrequency radiative transfer codes. We discussed the main features of the code, including gas and surface chemistry, photochemistry, and the main cooling and heating processes. We also reported how we tackled different computational bottlenecks, and what are the main algorithms employed. To prove the validity of the results obtained by \codename{}, we presented a set of benchmarks that include photon-dominated regions, slabs illuminated by radiation that include X-ray, and well-established cooling functions evaluated at different temperatures. These tests shows an agreement with the benchmarks both in terms of chemical and thermal structures.

This work represents the first step to provide a tool to explore the thermochemical properties of the wind launching region of protoplanetary disks and to determine the observational features from molecular tracers entrapped in the photoevaporative wind. This specific topic will be discussed in a forthcoming paper where we will show the results obtained by coupling \codename{} with a Monte Carlo radiative transfer code (\textsc{Mocassin}), and then with a hydrodynamical code (\textsc{Pluto}, \citealt{Mignone2007}).

\section*{Acknowledgements}
We acknowledge J.~P.~Ramsey, C.~Rab, M.~{\'A}d{\'a}mkovics, and the referee for useful comments and discussions.
TG acknowledges C.~Clarke for the hospitality during his visit at the IoA in Cambridge in the Framework of the Cambridge-LMU Strategic partnership Grant.
GP acknowledges support from the DFG Research Unit FOR 2634/1, ER 685/8-1.
LSz acknowledges support from the DFG Research Unit FOR 2634/1, CA 1624/1-1.
BE acknowledges support from the DFG cluster of excellence ``Origin and Structure of the Universe''
(\url{http://www.universe-cluster.de/}).
This work was funded by the DFG Research Unit FOR 2634/1 ER685/11-1.

\bibliographystyle{mn2e}
\bibliography{mybib}

\appendix
\section{Reaction rates format}\label{appx:chemical_network}
Here we report examples of the possible expressions that can be employed to represent chemical reactions in \codename{}.

To indicate a custom expression for the rate in the temperature range from zero to 6700~K we use for example
\begin{verbatim}
  H + H+ -> H2+  [,6700]  1.85d-23*Tgas**1.8
\end{verbatim}
where the reaction rate is written using standard \fortran{} syntax, but it can also be written as
\begin{verbatim}
  H + H+ -> H2+  [ ] KIDA
\end{verbatim}
that employs the corresponding rate from the \kida{} database internally stored (including temperature limits), i.e.~the \verb+kida.uva.2014+ version.

Analogously, special reactions not present in \kida{} will be defined by special expression as
\begin{verbatim}
   C++ + E -> C+  [ ] RECOMBINATION
\end{verbatim}
or
\begin{verbatim}
   C  [,4]  ALL_RECOMBINATION
\end{verbatim}
if the user wants to include all the recombination rates of carbon ions up to the fourth ionization level. In this case reactions are taken from the internal database of recombination rates. A detailed description of the databases employed can be found in \sect{sect:database}.

Photochemical rates can be defined with
\begin{verbatim}
  C -> C+ + E    [ ]  PHOTO
\end{verbatim}
so that \codename{} will select the correct cross-section from the database. For molecular chemical reactions without branching ratio specification, or with only a branching ratio, it is possible to indicate this by using a \textsc{JSON} structure as follows
\begin{verbatim}
  H3+ -> H2+ + H  [ ]PHOTO {"branching_ratio": 0.5}
  H3+ -> H2 + H+  [ ]PHOTO {"branching_ratio": 0.5}
\end{verbatim}
where the cross-section for H$_3^+$ is taken from the \textsc{Leiden} database, but the branching ratios are defined by the user.

Chemistry on grains can be defined by using
\begin{verbatim}
  CO_dust -> CO []
  CO -> CO_dust []
  C_dust + O_dust -> CO_dust []
\end{verbatim}
which will result in the pre-processor using the conventions defined in \sect{sect:surface_chemistry}. Binding energies and activation barrier values can be overridden with a \textsc{JSON} data structure similar to defining the branching ratios above.

The code recognises certain decorators for special actions, e.g.~to select a subset of reactions by using \verb+@include_only_species:H, CO, CH, C2+, that selects all the reactions that includes these specific species.
Analogously, blocks can be defined using a \textsc{C}-like format to indicate special reactions to the \python{} preprocessor, for example
\begin{verbatim}
  block_evaluate_once{
   block_cosmic_rays{
    H2 -> H+ + H + E [ ]0.02e+00 * variable_crflux
    H2 -> H + H      [ ]0.10e+00 * variable_crflux
   }
  }
\end{verbatim}
to use these reactions for the cosmic-rays heating, and to force the code to evaluate these rates only once (see \sect{sect:rate_optimization}).

\section{Discretization of integrals to take advantage of vectorization}\label{appx:integrals}
The integrals that are based on variable quantities are written in order to have a vector of pre-computed constants multiplied by the vector with the variables. An example is a photochemical rate, where the radiation flux $J(E)$ is variable, while the rest of the integral can be pre-determined as in
\beq
  k = \int_{E_{\rm th}}^{E_{\rm max}} \frac{\sigma(E) J(E)}{E} \dd E\,,
\eeq
that can be written as
\beq
  k = \int_{E_{\rm th}}^{E_{\rm max}} A(E) \dd E\,,
\eeq
and discretized using the trapezoidal rule in $N$ unequally-spaced energy grid points as
\beq
  k = \frac{1}{2}\sum_{i=2}^{N} \left(A_{i-1} + A_{i}\right) (E_{i}- E_{i-1})\,.
\eeq
The corresponding expanded expression using some algebra and grouping the terms by $A_i$ becomes
\beq
  2 k = A_1 (E_2 - E_1) + \sum_{i=2}^{N-1} A_i (E_{i+1} - E_{i-1}) + A_N(E_N-E_{N-1})
\eeq
that using the definition of $A_i$ becomes
\beq\label{eqn:jsxe}
  2 k = \sum_{i=1}^{N} J_i \frac{\sigma_i x_i}{E_i}\,,
\eeq
where
\beqa
  x_1 & = & E_2 - E_1\\
  x_i & = & E_{i+1} - E_{i-1}\\
  x_N & = & E_N - E_{N-1}\,.
\eeqa
\eqn{eqn:jsxe} is then
\beq
  2 k = \sum_{i=1}^{N} J_i B_i\,,
\eeq
where $J_i$ is a vector that changes at runtime, while $B_i$ can be pre-calculated. \codename{} takes advantage from this pre-calculation and from the fact that the expression can be easily vectorized as \verb+k=0.5*sum(J(:)*B(:))+ using the \fortran{} syntax.

\section{Analytical solution of a linear system with a superdiagonal coefficients matrix and conservation}\label{appx:superdiagonal}
The code frequently tries to find the solution of a linear system $A\times x =b$ with $N$ unknowns, and a matrix $A$ of the form $A_{ij}=0$, except $A_{ii}\neq0$, $A_{i,i+1}\neq0$, and $A_{Ni}=1$, and $b_i=0$, except $b_N\neq0$. The last row of $A$ represents conservation (i.e. $\sum_i x_i=b_N$), and apart from this row, the rest of the matrix is superdiagonal. This can be solved by defining $w_2 = A_{11} / A_{12}$ and $w_i = w_{i-1}A_{i-1,i-1} /A_{i-1,i}$ for $3\leq i\leq N$, the unknowns are then $x_1=b_N/(1+\sum_{i=2}^N w_i)$ and $x_i=w_i/x_1$ for $i>1$. Solving the matrix analytically instead of using e.g.~\texttt{dgesv} from \textsc{LAPACK} reduces considerably the computational time.

\section{Molecular hydrogen cooling tables (low density limit)}\label{appendix:cool_H2}\label{appendix:cool_LDL}
We report the functions employed for the low density limit of the molecular hydrogen cooling \citep{Glover2008,Grassi2014,Glover2015}, where $\Lambda_{H_2}^{\rm low}= \Lambda_{\rm H_2, H}^{\rm low} +  \Lambda_{\rm H_2, H^+}^{\rm low}+ \Lambda_{\rm H_2, H_2}^{\rm low} +  \Lambda_{\rm H_2, e^-}^{\rm low} +  \Lambda_{\rm H_2, He}^{\rm low}$.

\subsection*{H$_2$-H}
If $\tgas\leq10^2$~K
\beqa
    \Lambda_{\rm H_2, H}^{\rm low} &=& \dex\left(-16.818342+ 37.383713 t_3\right.\nonumber\\
                            &+& 58.145166 t_3^2 + 48.656103 t_3^3\nonumber\\
                            &+& 20.159831 t_3^4 + \left.3.8479610 t_3^5\right) n_{\rm H}\,,
\eeqa
if $10^2<\tgas\leq 10^3$~K
\beqa
    \Lambda_{\rm H_2, H}^{\rm low} &=& \dex\left(-24.311209 + 3.5692468 t_3\right.\nonumber\\
                            &-& 11.332860 t_3^2 - 27.850082 t_3^3\nonumber\\
                            &-& 21.328264 t_3^4 + \left.4.2519023 t_3^5\right) n_{\rm H}\,,
\eeqa
if $10^3<\tgas\leq 6\times 10^3$~K
\beqa
    \Lambda_{\rm H_2, H}^{\rm low} &=& \dex\left(-24.311209\right. + 4.6450521 t_3\nonumber\\
                            &-& 3.7209846 t_3^2 + 5.9369081 t_3^3\nonumber\\
                            &-& 5.5108049 t_3^4 + \left.1.5538288 t_3^5\right) n_{\rm H}\,,
\eeqa
and if $\tgas>6\times10^3$~K
\beq
 \Lambda_{\rm H_2, H}^{\rm low} = 1.8623\times10^{-22} w(\tgas) n_{\rm H}\,,
\eeq
where $\dex(x)=10^x$, and  $t_3=\log(\tgas / 10^{3}\,{\rm K})$.

\subsection*{H$_2$-H$^+$}
If $10<\tgas\leq10^4$~K
\beqa
    \Lambda_{\rm H_2, H^+}^{\rm low} &=& \dex\left(-22.089523 + 1.5714711 t_3\right.\nonumber\\
                            &+& 0.015391166 t_3^2 - 0.23619985 t_3^3\nonumber\\
                            &-& 0.51002221 t_3^4 + \left.0.32168730 t_3^5\right) n_{\rm H^+}\,,
\eeqa
and if $\tgas>10^4$~K
\beq
   \Lambda_{\rm H_2, H^+}^{\rm low} = 1.18250913\times10^{-21} w(\tgas) n_{\rm H^+}\,.
\eeq

\subsection*{H$_2$-H$_2$}
If $10^2<\tgas\leq10^4$~K
\beqa
    \Lambda_{\rm H_2, H_2}^{\rm low} &=& \dex\left(-23.962112 + 2.09433740 t_3\right.\nonumber\\
                            &+& 0.77151436 t_3^2 + 0.43693353 t_3^3\nonumber\\
                            &-& 0.14913216 t_3^4 - \left.0.033638326 t_3^5\right)\nonumber\\
                            &\cdot& w(x) n_{\rm H_2}\,.
\eeqa
for which we assume ortho-to-para ratio of 3:1.

\subsection*{H$_2$-e$^-$}
If $\tgas\leq5\times10^2$~K
\beqa
    \Lambda_{\rm H_2, e^-}^{\rm low} &=& \dex\left(-2.1928796 + 16.815730 t_3\right.\nonumber\\
                            &+& 96.743155 t_3^2 + 343.19180 t_3^3\nonumber\\
                            &+& 734.71651 t_3^4 + 983.67576 t_3^5\nonumber\\
                            &+& 801.81247 t_3^6 + 364.14446 t_3^7\nonumber\\
                            &+& 70.609154 t_3^8 \left.\right) n_{\rm e^-}\,,
\eeqa
if $\tgas>5\times10^2$~K
\beqa
    \Lambda_{\rm H_2, e^-}^{\rm low} &=& \dex\left(-22.921189 + 1.6815730 t_3\right.\nonumber\\
                            &+& 0.9331062 t_3^2 + 4.0406627 t_3^3\nonumber\\
                            &-& 4.7274036 t_3^4 - 8.8077017 t_3^5\nonumber\\
                            &+& 8.9167183 t_3^6 + 6.4380698 t_3^7\nonumber\\
                            &-& 6.3701156 t_3^8 \left.\right) w(\tgas) n_{\rm e^-}\,,
\eeqa

\subsection*{H$_2$-He}
If $10<\tgas\leq5\times10^4$~K
\beqa
    \Lambda_{\rm H_2, He}^{\rm low} &=& \dex\left(-23.689237 + 2.1892372 t_3\right.\nonumber\\
                            &-& 0.8152044 t_3^2 + 0.2903628 t_3^3\nonumber\\
                            &-& 0.1659618 t_3^4 + 0.1919138 t_3^5\left.\right) n_{\rm He}\,,
\eeqa
otherwise
\beq
    \Lambda_{\rm H_2, He}^{\rm low} = 1.0025604\times10^{-22} w(\tgas) n_{\rm He}\,.
\eeq

\section{Molecular hydrogen high-density limit}\label{appendix:cool_HDL}
If $\tgas<2\times10^3$~K, the rotational high density cooling function is
\beqa
    \Lambda_{\rm H_2,R}^{\rm high} &=& \frac{9.5\times10^{-22}t_3^{3.76}}{1+0.12 t_3^{2.1}} \exp\left[\left(-\frac{0.13}{T_3}\right)^3\right]\nonumber\\
    &+& 3\times10^{-24} \exp\left(-\frac{0.51}{T_3}\right)\,,
\eeqa
with $T_3=\tgas/10^3$~K, while the vibrational cooling function is
\beqa
    \Lambda_{\rm H_2,V}^{\rm high} &=& 6.7\times10^{-19} \exp\left(\frac{-5.86}{T_3}\right)\nonumber\\
                        &+& 1.6\times10^{-18} \exp\left(\frac{-11.7}{T_3}\right)\,,
\eeqa
so that the total high density is $\Lambda_{\rm H_2}^{\rm high}=\Lambda_{\rm H_2,R}^{\rm high}+\Lambda_{\rm H_2,V}^{\rm high}$, while if $2\times10^3\leq\tgas<10^4$~K
\beqa
    \Lambda_{\rm H_2}^{\rm high} &=& \dex\left(-20.584225 + 5.0194035 t_3\right.\nonumber\\
                            &-& 1.5738805 t_3^2 - 4.7155769 t_3^3\nonumber\\
                            &+& 2.4714161 t_3^4 + 5.4710750 t_3^5\nonumber\\
                            &-& 3.9467356 t_3^6 - 2.2148338 t_3^7\nonumber\\
                            &+& 1.8161874 t_3^8 \left.\right)\,,
\eeqa
otherwise
\beq
    \Lambda_{\rm H_2}^{\rm high} = \frac{5.53133\times10^{-19}}{1+\exp\left[0.0002\left(\tgas-3\times10^4\,{\rm K}\right)\right]}\,.
\eeq

\section{H$_2$ formation on dust grains (benchmark \texttt{V1} and \texttt{V4})}\label{appendix:H2_dust_cfr}
\begin{figure}
 \includegraphics[width=0.48\textwidth]{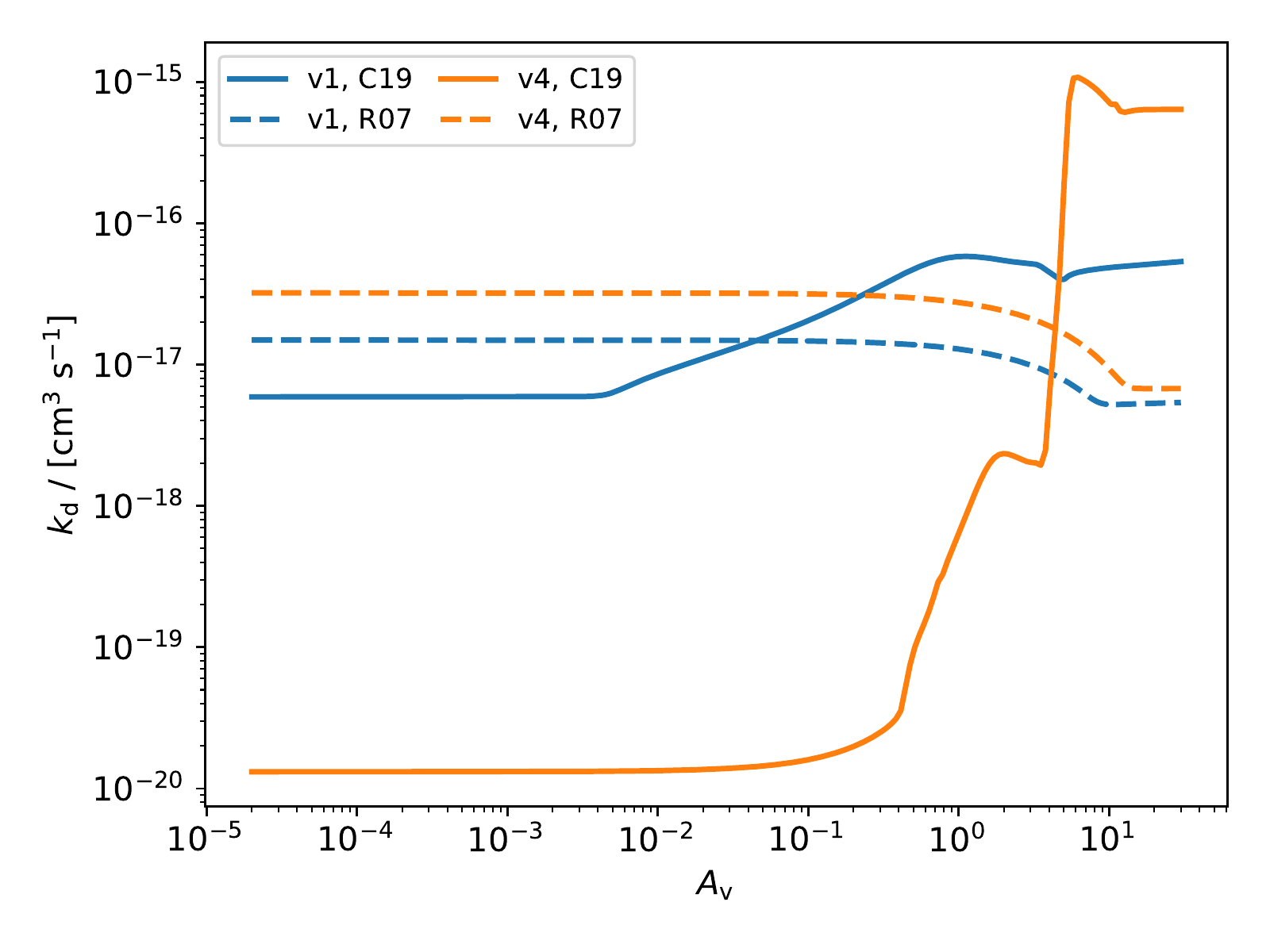}
 \caption{Comparison of the rate coefficients of molecular hydrogen formation on dust grains as a function of visual extinction for the models \texttt{V1} (blue lines) and \texttt{V4} (orange lines). The solid lines labelled C09 employ the model from \citet{Cazaux2009} in \eqn{eqn:H2_dust_k}, while the dashed lines labelled R07 use \citet{Rollig2007}.}\label{fig:H2_dust}
\end{figure}

The impact of molecular hydrogen formation has a relevant impact on the results obtained in benchmark \texttt{V1} and \texttt{V4}. In \fig{fig:H2_dust} for the two benchmarks we report the rate coefficient in \eqn{eqn:H2_dust_k} (labelled C09) compared with the expression from R07 we employed in the tests, i.e.~$k_{\rm d} = 3\times10^{-18}\sqrt{\tdust}$~cm$^3$~s$^{-1}$. While in \texttt{V1} the rate coefficients are similar, the higher gas temperature of \texttt{V4} affects the sticking in \eqn{eqn:stick}, causing a much lower molecular hydrogen formation efficiency. Following these results, since we are more interested in benchmarking the photochemical part of the code, we employ the expression from R07 in the benchmarks in \sect{sect:benchmarks} instead of \eqn{eqn:H2_dust_k}.

\section{Chemical network (tests \texttt{V1} and \texttt{V4})}\label{appendix:network}
We report the chemical network employed for the R07 benchmark tests in \tab{table:network_1}, \tab{table:network_2}, and \tab{table:network_3}.

\begin{table*}
\begin{adjustwidth}{-1cm}{}

\begin{tabular*}{1.1\textwidth}{ll}
\tiny
\centering
\setlength{\tabcolsep}{1pt}
\begin{tabular}{lllll}
\hline

1 & H + CH                                             & C + H$_2$                                          & [300, 2000]                                        & $2.7\times10^{-11} T_{32}^{0.38}$                 \\
2 & H + CH$_2$                                         & CH + H$_2$                                         & [300, 2500]                                        & $6.64\times10^{-11}$                              \\
3 & H + CH$_3$                                         & CH$_2$ + H$_2$                                     & [300, 2500]                                        & $10^{-10} \exp(\frac{-7600 \,{\rm K} }{ T})$\\
4 & H + CH$_4$                                         & CH$_3$ + H$_2$                                     & [300, 2500]                                        & $5.94\times10^{-13} T_{32}^{3.00} \exp(\frac{-4045 \,{\rm K} }{ T})$\\
5 & H + OH                                             & O + H$_2$                                          & [300, 2500]                                        & $6.98\times10^{-14} T_{32}^{2.80} \exp(\frac{-1950 \,{\rm K} }{ T})$\\
6 & H + H$_2$O                                         & OH + H$_2$                                         & [250, 3000]                                        & $1.59\times10^{-11} T_{32}^{1.20} \exp(\frac{-9610 \,{\rm K} }{ T})$\\
7 & H + CO                                             & OH + C                                             & [2590, 41000]                                      & $1.09\times10^{-10} T_{32}^{0.50} \exp(\frac{-77700 \,{\rm K} }{ T})$\\
8 & H + O$_2$                                          & OH + O                                             & [250, 4000]                                        & $2.61\times10^{-10} \exp(\frac{-8156 \,{\rm K} }{ T})$\\
9 & H$_2$ + C                                          & CH + H                                             & [300, 2500]                                        & $6.64\times10^{-10} \exp(\frac{-11700 \,{\rm K} }{ T})$\\
10 & H$_2$ + CH                                         & CH$_2$ + H                                         & [300, 2500]                                        & $5.46\times10^{-10} \exp(\frac{-1943 \,{\rm K} }{ T})$\\
11 & H$_2$ + CH$_2$                                     & CH$_3$ + H                                         & [300, 2500]                                        & $5.18\times10^{-11} T_{32}^{0.17} \exp(\frac{-6400 \,{\rm K} }{ T})$\\
12 & H$_2$ + CH$_3$                                     & CH$_4$ + H                                         & [300, 2500]                                        & $6.86\times10^{-14} T_{32}^{2.74} \exp(\frac{-4740 \,{\rm K} }{ T})$\\
13 & H$_2$ + O                                          & OH + H                                             & [297, 3532]                                        & $3.14\times10^{-13} T_{32}^{2.70} \exp(\frac{-3150 \,{\rm K} }{ T})$\\
14 & H$_2$ + OH                                         & H$_2$O + H                                         & [250, 2581]                                        & $2.04\times10^{-12} T_{32}^{1.52} \exp(\frac{-1736 \,{\rm K} }{ T})$\\
15 & H$_2$ + O$_2$                                      & OH + OH                                            & [300, 2500]                                        & $3.15\times10^{-10} \exp(\frac{-21890 \,{\rm K} }{ T})$\\
16 & C + CH$_2$                                         & CH + CH                                            & [1000, 4000]                                       & $2.69\times10^{-12} \exp(\frac{-23550 \,{\rm K} }{ T})$\\
17 & C + OH                                             & O + CH                                             & [493, 41000]                                       & $2.25\times10^{-11} T_{32}^{0.50} \exp(\frac{-14800 \,{\rm K} }{ T})$\\
18 & C + OH                                             & CO + H                                             & [ , 300]                                           & $1.09\times10^{-10} T_{32}^{0.50}$                \\
19 & C + O$_2$                                          & CO + O                                             & [ , 10000]                                         & $3.3\times10^{-11}$                               \\
20 & CH + O                                             & OH + C                                             & [ , 6000]                                          & $2.52\times10^{-11} \exp(\frac{-2381 \,{\rm K} }{ T})$\\
21 & CH + O                                             & CO + H                                             & [ , 2000]                                          & $6.6\times10^{-11}$                               \\
22 & CH + CH$_4$                                        & CH$_3$ + CH$_2$                                    & [100, 300]                                         & $2.28\times10^{-11} T_{32}^{0.70} \exp(\frac{-3000 \,{\rm K} }{ T})$\\
23 & CH + O$_2$                                         & CO + OH                                            & [298, 298]                                         & $2.6\times10^{-11}$                               \\
24 & CH$_2$ + CH$_2$                                    & CH$_3$ + CH                                        & [300, 2500]                                        & $4\times10^{-10} \exp(\frac{-5000 \,{\rm K} }{ T})$ \\
25 & CH$_2$ + O                                         & OH + CH                                            & [1900, 2300]                                       & $4.98\times10^{-10} \exp(\frac{-6000 \,{\rm K} }{ T})$\\
26 & CH$_2$ + O                                         & CO + H + H                                         & [ , 2500]                                          & $1.33\times10^{-10}$                              \\
27 & CH$_2$ + O                                         & CO + H$_2$                                         & [1900, 2600]                                       & $8\times10^{-11}$                                 \\
28 & CH$_2$ + CH$_4$                                    & CH$_3$ + CH$_3$                                    & [296, 2500]                                        & $7.13\times10^{-12} \exp(\frac{-5050 \,{\rm K} }{ T})$\\
29 & CH$_2$ + OH                                        & O + CH$_3$                                         & [100, 300]                                         & $1.44\times10^{-11} T_{32}^{0.50} \exp(\frac{-3000 \,{\rm K} }{ T})$\\
30 & CH$_2$ + OH                                        & H$_2$O + CH                                        & [100, 300]                                         & $1.44\times10^{-11} T_{32}^{0.50} \exp(\frac{-3000 \,{\rm K} }{ T})$\\
31 & CH$_2$ + O$_2$                                     & CO + H$_2$O                                        & [1000, 1828]                                       & $2.48\times10^{-10} T_{32}^{-3.30} \exp(\frac{-1443 \,{\rm K} }{ T})$\\
32 & CH$_3$ + CH$_3$                                    & CH$_4$ + CH$_2$                                    & [1950, 2300]                                       & $7.13\times10^{-12} \exp(\frac{-5052 \,{\rm K} }{ T})$\\
33 & CH$_3$ + OH                                        & CH$_4$ + O                                         & [298, 2500]                                        & $3.27\times10^{-14} T_{32}^{2.20} \exp(\frac{-2240 \,{\rm K} }{ T})$\\
34 & CH$_3$ + OH                                        & H$_2$O + CH$_2$                                    & [300, 1000]                                        & $1.2\times10^{-10} \exp(\frac{-1400 \,{\rm K} }{ T})$\\
35 & CH$_3$ + H$_2$O                                    & OH + CH$_4$                                        & [300, 3000]                                        & $2.29\times10^{-15} T_{32}^{3.47} \exp(\frac{-6681 \,{\rm K} }{ T})$\\
36 & O + CH$_4$                                         & OH + CH$_3$                                        & [298, 2500]                                        & $2.29\times10^{-12} T_{32}^{2.20} \exp(\frac{-3820 \,{\rm K} }{ T})$\\
37 & O + OH                                             & O$_2$ + H                                          & [ , 10000]                                         & $4.34\times10^{-11} T_{32}^{-0.50} \exp(\frac{-30 \,{\rm K} }{ T})$\\
38 & O + H$_2$O                                         & OH + OH                                            & [300, 2500]                                        & $1.85\times10^{-11} T_{32}^{0.95} \exp(\frac{-8571 \,{\rm K} }{ T})$\\
39 & CH$_4$ + OH                                        & H$_2$O + CH$_3$                                    & [178, 3000]                                        & $3.77\times10^{-13} T_{32}^{2.42} \exp(\frac{-1162 \,{\rm K} }{ T})$\\
40 & OH + OH                                            & H$_2$O + O                                         & [ , 10000]                                         & $1.65\times10^{-12} T_{32}^{1.14} \exp(\frac{-50 \,{\rm K} }{ T})$\\
41 & H + CH$^+$                                         & C$^+$ + H$_2$                                      & [ , 300]                                           & $7.5\times10^{-10}$                               \\
42 & H + CH$_2$$^+$                                     & CH$^+$ + H$_2$                                     & [236, 300]                                         & $10\times10^{-10} \exp(\frac{-7080 \,{\rm K} }{ T})$\\
43 & H + CH$_3$$^+$                                     & CH$_2$$^+$ + H$_2$                                 & [352, 41000]                                       & $6.99\times10^{-10} \exp(\frac{-10560 \,{\rm K} }{ T})$\\
44 & H$_2$ + He$^+$                                     & He + H$^+$ + H                                     & [ , 300]                                           & $3.7\times10^{-14} \exp(\frac{-35 \,{\rm K} }{ T})$ \\
45 & H$_2$ + C$^+$                                      & CH$^+$ + H                                         & [154, 300]                                         & $10^{-10} \exp(\frac{-4640 \,{\rm K} }{ T})$\\
46 & H$_2$ + CH$^+$                                     & CH$_2$$^+$ + H                                     & [ , 300]                                           & $1.2\times10^{-9}$                                \\
47 & He$^+$ + CO                                        & O$^+$ + C + He                                     & [2000, 10000]                                      & $1.4\times10^{-16} T_{32}^{-0.50}$                \\
48 & C$^+$ + OH                                         & CO + H$^+$                                         & [ , 10000]                                         & $7.7\times10^{-10}$                               \\
49 & H$^+$ + H$_2$                                      & H$_2$$^+$ + H                                      & [706, 41000]                                       & $10^{-10} \exp(\frac{-21200 \,{\rm K} }{ T})$\\
50 & H + He$^+$                                         & He + H$^+$                                         & [ , 1000]                                          & $4.84\times10^{-15} T_{32}^{0.18}$                \\

\hline
\end{tabular}

\begin{tabular}{lllll}
\hline

51 & H + O$^+$                                          & O + H$^+$                                          & [ ]                                                & $5.66\times10^{-10} T_{32}^{0.36} \exp(\frac{8.6{\rm K} }{ T})$\\
52 & H$^+$ + O                                          & O$^+$ + H                                          & [ , 10000]                                         & $7.31\times10^{-10} T_{32}^{0.23} \exp(\frac{-225.9{\rm K} }{ T})$\\
53 & H$^+$ + H$_2$O                                     & H$_2$O$^+$ + H                                     & [ , 300]                                           & $6.9\times10^{-9}$                                \\
54 & H$^+$ + O$_2$                                      & O$_2$$^+$ + H                                      & [ , 300]                                           & $2\times10^{-9}$                                  \\
55 & H$_2$ + He$^+$                                     & He + H$_2$$^+$                                     & [ , 300]                                           & $7.19\times10^{-15}$                              \\
56 & He$^+$ + C                                         & C$^+$ + He                                         & [ , 300]                                           & $6.29\times10^{-15} T_{32}^{0.75}$                \\
57 & O$^+$ + CO                                         & CO$^+$ + O                                         & [2000, 10000]                                      & $4.89\times10^{-12} T_{32}^{0.50} \exp(\frac{-4580 \,{\rm K} }{ T})$\\
58 & H$_2$$^+$ + e$^-$                                  & H + H                                              & [ , 300]                                           & $1.6\times10^{-8} T_{32}^{-0.43}$                 \\
59 & H$_3$$^+$ + e$^-$                                  & H + H + H                                          & [ , 1000]                                          & $7.49\times10^{-8} T_{32}^{-0.30}$                \\
60 & H$_3$$^+$ + e$^-$                                  & H$_2$ + H                                          & [ , 1000]                                          & $2.5\times10^{-8} T_{32}^{-0.30}$                 \\
61 & CH$^+$ + e$^-$                                     & C + H                                              & [ , 300]                                           & $1.5\times10^{-7} T_{32}^{-0.42}$                 \\
62 & CH$_2$$^+$ + e$^-$                                 & CH + H                                             & [ , 10000]                                         & $1.6\times10^{-7} T_{32}^{-0.60}$                 \\
63 & CH$_2$$^+$ + e$^-$                                 & C + H + H                                          & [ , 10000]                                         & $4.03\times10^{-7} T_{32}^{-0.60}$                \\
64 & CH$_2$$^+$ + e$^-$                                 & C + H$_2$                                          & [ , 10000]                                         & $7.68\times10^{-8} T_{32}^{-0.60}$                \\
65 & CH$_3$$^+$ + e$^-$                                 & CH$_2$ + H                                         & [ , 10000]                                         & $1.4\times10^{-7} T_{32}^{-0.50}$                 \\
66 & CH$_3$$^+$ + e$^-$                                 & CH + H$_2$                                         & [ , 10000]                                         & $4.9\times10^{-8} T_{32}^{-0.50}$                 \\
67 & CH$_3$$^+$ + e$^-$                                 & CH + H + H                                         & [ , 10000]                                         & $5.6\times10^{-8} T_{32}^{-0.50}$                 \\
68 & CH$_3$$^+$ + e$^-$                                 & C + H$_2$ + H                                      & [ , 10000]                                         & $1.05\times10^{-7} T_{32}^{-0.5}$                 \\
69 & CH$_4$$^+$ + e$^-$                                 & CH$_3$ + H                                         & [ , 300]                                           & $1.75\times10^{-7} T_{32}^{-0.50}$                \\
70 & CH$_4$$^+$ + e$^-$                                 & CH$_2$ + H + H                                     & [ , 300]                                           & $1.75\times10^{-7} T_{32}^{-0.50}$                \\
71 & OH$^+$ + e$^-$                                     & O + H                                              & [ , 300]                                           & $3.74\times10^{-8} T_{32}^{-0.50}$                \\
72 & CH$_5$$^+$ + e$^-$                                 & CH$_3$ + H$_2$                                     & [ , 300]                                           & $5.5\times10^{-7} T_{32}^{-0.30}$                 \\
73 & CH$_5$$^+$ + e$^-$                                 & CH$_4$ + H                                         & [ , 300]                                           & $5.5\times10^{-7} T_{32}^{-0.30}$                 \\
74 & H$_2$O$^+$ + e$^-$                                 & O + H + H                                          & [ , 10000]                                         & $2.44\times10^{-7} T_{32}^{-0.50}$                \\
75 & H$_2$O$^+$ + e$^-$                                 & O + H$_2$                                          & [ , 10000]                                         & $3.59\times10^{-8} T_{32}^{-0.50}$                \\
76 & H$_2$O$^+$ + e$^-$                                 & OH + H                                             & [ , 10000]                                         & $7.91\times10^{-8} T_{32}^{-0.50}$                \\
77 & H$_3$O$^+$ + e$^-$                                 & O + H$_2$ + H                                      & [ , 10000]                                         & $5.59\times10^{-9} T_{32}^{-0.50}$                \\
78 & H$_3$O$^+$ + e$^-$                                 & OH + H + H                                         & [ , 10000]                                         & $2.58\times10^{-7} T_{32}^{-0.50}$                \\
79 & H$_3$O$^+$ + e$^-$                                 & OH + H$_2$                                         & [ , 10000]                                         & $6.45\times10^{-8} T_{32}^{-0.50}$                \\
80 & H$_3$O$^+$ + e$^-$                                 & H$_2$O + H                                         & [ , 10000]                                         & $1.08\times10^{-7} T_{32}^{-0.50}$                \\
81 & CO$^+$ + e$^-$                                     & O + C                                              & [ , 10000]                                         & $2\times10^{-7} T_{32}^{-0.48}$                   \\
82 & HCO$^+$ + e$^-$                                    & CO + H                                             & [ , 300]                                           & $1.1\times10^{-7} T_{32}^{-1.00}$                 \\
83 & O$_2$$^+$ + e$^-$                                  & O + O                                              & [ , 300]                                           & $1.95\times10^{-7} T_{32}^{-0.70}$                \\
84 & H$^+$ + e$^-$                                      & H                                                  & [ , 20000]                                         & $3.5\times10^{-12} T_{32}^{-0.75}$                \\
85 & He$^+$ + e$^-$                                     & He                                                 & [ , 300]                                           & $2.36\times10^{-12} T_{32}^{-0.64}$               \\
86 & C$^+$ + e$^-$                                      & C                                                  & [ , 7950]                                          & $4.67\times10^{-12} T_{32}^{-0.60}$               \\
87 & CH$_3$$^+$ + e$^-$                                 & CH$_3$                                             & [ , 300]                                           & $1.09\times10^{-10} T_{32}^{-0.50}$               \\
88 & O$^+$ + e$^-$                                      & O                                                  & [ ]                                                & $3.24\times10^{-12} T_{32}^{-0.66}$               \\
89 & H$^+$ + H                                          & H$_2$$^+$                                          & [200, 4000]                                        & $5.13\times10^{-19} T_{32}^{1.85}$                \\
90 & H + C                                              & CH                                                 & [ , 300]                                           & $10^{-17}$                                \\
91 & H + C$^+$                                          & CH$^+$                                             & [ , 300]                                           & $1.69\times10^{-17}$                              \\
92 & H + O                                              & OH                                                 & [ , 300]                                           & $9.9\times10^{-19} T_{32}^{-0.38}$                \\
93 & H + OH                                             & H$_2$O                                             & [20, 300]                                          & $5.26\times10^{-18} T_{32}^{-5.22} \exp(\frac{-90 \,{\rm K} }{ T})$\\
94 & H$_2$ + C                                          & CH$_2$                                             & [ , 300]                                           & $10^{-17}$                                \\
95 & H$_2$ + C$^+$                                      & CH$_2$$^+$                                         & [ , 300]                                           & $4\times10^{-16} T_{32}^{-0.20}$                  \\
96 & H$_2$ + CH                                         & CH$_3$                                             & [ , 300]                                           & $5.09\times10^{-18} T_{32}^{-0.71} \exp(\frac{-11.6{\rm K} }{ T})$\\
97 & H$_2$ + CH$_3$$^+$                                 & CH$_5$$^+$                                         & [ , 300]                                           & $1.3\times10^{-14} T_{32}^{-1.00}$                \\
98 & C + O                                              & CO                                                 & [ , 300]                                           & $2.1\times10^{-19}$                               \\
99 & C$^+$ + O                                          & CO$^+$                                             & [ , 300]                                           & $2.5\times10^{-18}$                               \\
100 & O + O                                              & O$_2$                                              & [ , 300]                                           & $4.9\times10^{-20} T_{32}^{1.58}$                 \\

\hline
\end{tabular}

\end{tabular*}
\end{adjustwidth}
\caption{Chemical network employed in \texttt{V1} and \texttt{V4} benchmark tests. Temperature limits are in K, $[\,]$ indicates no temperature limits, while e.g.~$[,\,300]$ and $[50,\,]$ indicate that there are no lower and upper limits, respectively. Note that in the benchmark we arbitrarily extended the temperature limits reported here for completeness. Two-body reaction rate coefficients are in units of cm$^3$~s$^{-1}$, photochemical and cosmic-rays reactions in s$^{-1}$. $T_{32}=T/300$~K. A detailed analysis of the network will be presented in Paper~II. Table \underline{continues} in \tab{table:network_2}.}\label{table:network_1}
\end{table*}

\clearpage
\begin{table*}

\begin{tabular*}{1.1\textwidth}{ll}
\tiny
\centering
\setlength{\tabcolsep}{1pt}
\begin{tabular}{lllll}
\hline

101 & H + H$_2$                                          & H + H + H                                          & [1833, 41000]                                      & $4.67\times10^{-7} T_{32}^{-1.00} \exp(\frac{-55000 \,{\rm K} }{ T})$\\
102 & H + CH                                             & C + H + H                                          & [1340, 41000]                                      & $6\times10^{-9} \exp(\frac{-40200 \,{\rm K} }{ T})$ \\
103 & H + OH                                             & O + H + H                                          & [1696, 41000]                                      & $6\times10^{-9} \exp(\frac{-50900 \,{\rm K} }{ T})$ \\
104 & H + H$_2$O                                         & OH + H + H                                         & [1763, 41000]                                      & $5.8\times10^{-9} \exp(\frac{-52900 \,{\rm K} }{ T})$\\
105 & H + O$_2$                                          & O + O + H                                          & [1743, 41000]                                      & $6\times10^{-9} \exp(\frac{-52300 \,{\rm K} }{ T})$ \\
106 & H$_2$ + e$^-$                                      & H + H + e$^-$                                      & [3400, 41000]                                      & $3.22\times10^{-9} T_{32}^{0.35} \exp(\frac{-102000 \,{\rm K} }{ T})$\\
107 & H$_2$ + H$_2$                                      & H$_2$ + H + H                                      & [2803, 41000]                                      & $10^{-8} \exp(\frac{-84100 \,{\rm K} }{ T})$\\
108 & H$_2$ + CH                                         & C + H$_2$ + H                                      & [1340, 41000]                                      & $6\times10^{-9} \exp(\frac{-40200 \,{\rm K} }{ T})$ \\
109 & H$_2$ + OH                                         & O + H$_2$ + H                                      & [1696, 41000]                                      & $6\times10^{-9} \exp(\frac{-50900 \,{\rm K} }{ T})$ \\
110 & H$_2$ + H$_2$O                                     & OH + H$_2$ + H                                     & [1763, 41000]                                      & $5.8\times10^{-9} \exp(\frac{-52900 \,{\rm K} }{ T})$\\
111 & H$_2$ + O$_2$                                      & O + O + H$_2$                                      & [1743, 41000]                                      & $6\times10^{-9} \exp(\frac{-52300 \,{\rm K} }{ T})$ \\
112 & CH + O                                             & HCO$^+$ + e$^-$                                    & [ , 1750]                                          & $2\times10^{-11} T_{32}^{0.44}$                   \\
113 & H + H                                              & H$_2$                                              & [ ]                                                & see text                 \\
114 & H$^+$ + CH$_2$                                     & CH$^+$ + H$_2$                                     & [ ]                                                & $1.4\times10^{-9}$                                \\
115 & H$^+$ + CH$_4$                                     & CH$_3$$^+$ + H$_2$                                 & [ ]                                                & $2.29\times10^{-9}$                               \\
116 & H + CH$_4$$^+$                                     & CH$_3$$^+$ + H$_2$                                 & [ ]                                                & $10^{-11}$                                \\
117 & H + CH$_5$$^+$                                     & CH$_4$$^+$ + H$_2$                                 & [ ]                                                & $2\times10^{-11}$                                 \\
118 & H$_2$$^+$ + H$_2$                                  & H$_3$$^+$ + H                                      & [ ]                                                & $2.07\times10^{-9}$                               \\
119 & H$_2$$^+$ + C                                      & CH$^+$ + H                                         & [ ]                                                & $2.4\times10^{-9}$                                \\
120 & H$_2$$^+$ + CH                                     & CH$_2$$^+$ + H                                     & [ ]                                                & $7.1\times10^{-10}$                               \\
121 & H$_2$$^+$ + CH$_2$                                 & CH$_3$$^+$ + H                                     & [ ]                                                & $10^{-9}$                                \\
122 & H$_2$ + CH$_2$$^+$                                 & CH$_3$$^+$ + H                                     & [ ]                                                & $1.6\times10^{-9}$                                \\
123 & H$_2$$^+$ + O                                      & OH$^+$ + H                                         & [ ]                                                & $1.5\times10^{-9}$                                \\
124 & H$_2$ + O$^+$                                      & OH$^+$ + H                                         & [ ]                                                & $1.69\times10^{-9}$                               \\
125 & H$_2$$^+$ + CH$_4$                                 & CH$_5$$^+$ + H                                     & [ ]                                                & $1.13\times10^{-10}$                              \\
126 & H$_2$ + CH$_4$$^+$                                 & CH$_5$$^+$ + H                                     & [ ]                                                & $3.3\times10^{-11}$                               \\
127 & H$_2$$^+$ + CH$_4$                                 & CH$_3$$^+$ + H$_2$ + H                             & [ ]                                                & $2.29\times10^{-9}$                               \\
128 & H$_2$$^+$ + OH                                     & H$_2$O$^+$ + H                                     & [ ]                                                & $7.6\times10^{-10}$                               \\
129 & H$_2$ + OH$^+$                                     & H$_2$O$^+$ + H                                     & [ ]                                                & $1.01\times10^{-9}$                               \\
130 & H$_2$$^+$ + H$_2$O                                 & H$_3$O$^+$ + H                                     & [ ]                                                & $3.39\times10^{-9}$                               \\
131 & H$_2$ + H$_2$O$^+$                                 & H$_3$O$^+$ + H                                     & [ ]                                                & $6.4\times10^{-10}$                               \\
132 & H$_2$$^+$ + CO                                     & HCO$^+$ + H                                        & [ ]                                                & $2.15\times10^{-9}$                               \\
133 & H$_2$ + CO$^+$                                     & HCO$^+$ + H                                        & [ ]                                                & $1.79\times10^{-9}$                               \\
134 & H$_3$$^+$ + C                                      & CH$^+$ + H$_2$                                     & [ ]                                                & $2\times10^{-9}$                                  \\
135 & H$_3$$^+$ + CH                                     & CH$_2$$^+$ + H$_2$                                 & [ ]                                                & $1.2\times10^{-9}$                                \\
136 & H$_3$$^+$ + CH$_2$                                 & CH$_3$$^+$ + H$_2$                                 & [ ]                                                & $1.69\times10^{-9}$                               \\
137 & H$_3$$^+$ + CH$_3$                                 & CH$_4$$^+$ + H$_2$                                 & [ ]                                                & $2.1\times10^{-9}$                                \\
138 & H$_3$$^+$ + O                                      & OH$^+$ + H$_2$                                     & [ ]                                                & $8\times10^{-10}$                                 \\
139 & H$_3$$^+$ + CH$_4$                                 & CH$_5$$^+$ + H$_2$                                 & [ ]                                                & $2.4\times10^{-9}$                                \\
140 & H$_3$$^+$ + OH                                     & H$_2$O$^+$ + H$_2$                                 & [ ]                                                & $1.3\times10^{-9}$                                \\
141 & H$_3$$^+$ + H$_2$O                                 & H$_3$O$^+$ + H$_2$                                 & [ ]                                                & $5.9\times10^{-9}$                                \\
142 & H$_3$$^+$ + CO                                     & HCO$^+$ + H$_2$                                    & [ ]                                                & $1.69\times10^{-9}$                               \\
143 & He$^+$ + CH                                        & C$^+$ + He + H                                     & [ ]                                                & $1.09\times10^{-9}$                               \\
144 & He$^+$ + CH$_2$                                    & C$^+$ + He + H$_2$                                 & [ ]                                                & $7.5\times10^{-10}$                               \\
145 & He$^+$ + CH$_2$                                    & CH$^+$ + He + H                                    & [ ]                                                & $7.5\times10^{-10}$                               \\
146 & He$^+$ + CH$_3$                                    & CH$^+$ + He + H$_2$                                & [ ]                                                & $1.79\times10^{-9}$                               \\
147 & He$^+$ + CH$_4$                                    & CH$^+$ + He + H$_2$ + H                            & [ ]                                                & $2.4\times10^{-10}$                               \\
148 & He$^+$ + CH$_4$                                    & CH$_2$$^+$ + He + H$_2$                            & [ ]                                                & $9.5\times10^{-10}$                               \\
149 & He$^+$ + CH$_4$                                    & CH$_3$ + He + H$^+$                                & [ ]                                                & $4.8\times10^{-10}$                               \\
150 & He$^+$ + CH$_4$                                    & CH$_3$$^+$ + He + H                                & [ ]                                                & $8.5\times10^{-11}$                               \\

\hline
\end{tabular}

\begin{tabular}{lllll}
\hline

151 & He$^+$ + OH                                        & O$^+$ + He + H                                     & [ ]                                                & $1.09\times10^{-9}$                               \\
152 & He$^+$ + H$_2$O                                    & OH + He + H$^+$                                    & [ ]                                                & $2.04\times10^{-10}$                              \\
153 & He$^+$ + H$_2$O                                    & OH$^+$ + He + H                                    & [ ]                                                & $2.86\times10^{-10}$                              \\
154 & He$^+$ + CO                                        & O + C$^+$ + He                                     & [ ]                                                & $1.6\times10^{-9}$                                \\
155 & He$^+$ + O$_2$                                     & O$^+$ + O + He                                     & [ ]                                                & $10^{-9}$                                \\
156 & C + OH$^+$                                         & O + CH$^+$                                         & [ ]                                                & $1.2\times10^{-9}$                                \\
157 & C$^+$ + OH                                         & CO$^+$ + H                                         & [ ]                                                & $7.7\times10^{-10}$                               \\
158 & C + CH$_5$$^+$                                     & CH$_4$ + CH$^+$                                    & [ ]                                                & $1.2\times10^{-9}$                                \\
159 & C + H$_2$O$^+$                                     & OH + CH$^+$                                        & [ ]                                                & $1.09\times10^{-9}$                               \\
160 & C$^+$ + H$_2$O                                     & HCO$^+$ + H                                        & [ ]                                                & $9\times10^{-10}$                                 \\
161 & C + H$_3$O$^+$                                     & HCO$^+$ + H$_2$                                    & [ ]                                                & $10^{-11}$                                \\
162 & C + HCO$^+$                                        & CO + CH$^+$                                        & [ ]                                                & $1.09\times10^{-9}$                               \\
163 & C$^+$ + O$_2$                                      & CO$^+$ + O                                         & [ ]                                                & $3.8\times10^{-10}$                               \\
164 & C$^+$ + O$_2$                                      & CO + O$^+$                                         & [ ]                                                & $6.2\times10^{-10}$                               \\
165 & C + O$_2$$^+$                                      & CO$^+$ + O                                         & [ ]                                                & $5.2\times10^{-11}$                               \\
166 & CH$^+$ + O                                         & CO$^+$ + H                                         & [ ]                                                & $3.49\times10^{-10}$                              \\
167 & CH + O$^+$                                         & CO$^+$ + H                                         & [ ]                                                & $3.49\times10^{-10}$                              \\
168 & CH + OH$^+$                                        & O + CH$_2$$^+$                                     & [ ]                                                & $3.49\times10^{-10}$                              \\
169 & CH$^+$ + OH                                        & CO$^+$ + H$_2$                                     & [ ]                                                & $7.5\times10^{-10}$                               \\
170 & CH + CH$_5$$^+$                                    & CH$_4$ + CH$_2$$^+$                                & [ ]                                                & $6.9\times10^{-10}$                               \\
171 & CH$^+$ + H$_2$O                                    & H$_3$O$^+$ + C                                     & [ ]                                                & $5.8\times10^{-10}$                               \\
172 & CH + H$_2$O$^+$                                    & OH + CH$_2$$^+$                                    & [ ]                                                & $3.4\times10^{-10}$                               \\
173 & CH$^+$ + H$_2$O                                    & HCO$^+$ + H$_2$                                    & [ ]                                                & $2.9\times10^{-9}$                                \\
174 & CH + H$_3$O$^+$                                    & H$_2$O + CH$_2$$^+$                                & [ ]                                                & $6.8\times10^{-10}$                               \\
175 & CH + CO$^+$                                        & HCO$^+$ + C                                        & [ ]                                                & $3.2\times10^{-10}$                               \\
176 & CH + HCO$^+$                                       & CO + CH$_2$$^+$                                    & [ ]                                                & $6.3\times10^{-10}$                               \\
177 & CH$^+$ + O$_2$                                     & CO$^+$ + OH                                        & [ ]                                                & $10^{-11}$                                \\
178 & CH$^+$ + O$_2$                                     & HCO$^+$ + O                                        & [ ]                                                & $9.69\times10^{-10}$                              \\
179 & CH + O$_2$$^+$                                     & HCO$^+$ + O                                        & [ ]                                                & $3.1\times10^{-10}$                               \\
180 & CH$_2$$^+$ + O                                     & HCO$^+$ + H                                        & [ ]                                                & $7.5\times10^{-10}$                               \\
181 & CH$_2$ + OH$^+$                                    & O + CH$_3$$^+$                                     & [ ]                                                & $4.8\times10^{-10}$                               \\
182 & CH$_2$ + CH$_5$$^+$                                & CH$_4$ + CH$_3$$^+$                                & [ ]                                                & $9.6\times10^{-10}$                               \\
183 & CH$_2$ + H$_2$O$^+$                                & OH + CH$_3$$^+$                                    & [ ]                                                & $4.7\times10^{-10}$                               \\
184 & CH$_2$ + H$_3$O$^+$                                & H$_2$O + CH$_3$$^+$                                & [ ]                                                & $9.4\times10^{-10}$                               \\
185 & CH$_2$ + CO$^+$                                    & HCO$^+$ + CH                                       & [ ]                                                & $4.3\times10^{-10}$                               \\
186 & CH$_2$ + HCO$^+$                                   & CO + CH$_3$$^+$                                    & [ ]                                                & $8.6\times10^{-10}$                               \\
187 & CH$_2$$^+$ + O$_2$                                 & HCO$^+$ + OH                                       & [ ]                                                & $9.1\times10^{-10}$                               \\
188 & CH$_3$$^+$ + O                                     & HCO$^+$ + H$_2$                                    & [ ]                                                & $4\times10^{-10}$                                 \\
189 & O$^+$ + CH$_4$                                     & OH + CH$_3$$^+$                                    & [ ]                                                & $1.09\times10^{-10}$                              \\
190 & O + CH$_4$$^+$                                     & OH + CH$_3$$^+$                                    & [ ]                                                & $10^{-9}$                                \\
191 & O$^+$ + OH                                         & O$_2$$^+$ + H                                      & [ ]                                                & $3.59\times10^{-10}$                              \\
192 & O + OH$^+$                                         & O$_2$$^+$ + H                                      & [ ]                                                & $7.1\times10^{-10}$                               \\
193 & O + CH$_5$$^+$                                     & H$_3$O$^+$ + CH$_2$                                & [ ]                                                & $2.19\times10^{-10}$                              \\
194 & O + H$_2$O$^+$                                     & O$_2$$^+$ + H$_2$                                  & [ ]                                                & $4\times10^{-11}$                                 \\
195 & CH$_4$$^+$ + CH$_4$                                & CH$_5$$^+$ + CH$_3$                                & [ ]                                                & $1.5\times10^{-9}$                                \\
196 & CH$_4$ + OH$^+$                                    & CH$_5$$^+$ + O                                     & [ ]                                                & $1.95\times10^{-10}$                              \\
197 & CH$_4$ + OH$^+$                                    & H$_3$O$^+$ + CH$_2$                                & [ ]                                                & $1.31\times10^{-9}$                               \\
198 & CH$_4$$^+$ + H$_2$O                                & H$_3$O$^+$ + CH$_3$                                & [ ]                                                & $2.6\times10^{-9}$                                \\
199 & CH$_4$ + H$_2$O$^+$                                & H$_3$O$^+$ + CH$_3$                                & [ ]                                                & $1.4\times10^{-9}$                                \\
200 & CH$_4$$^+$ + CO                                    & HCO$^+$ + CH$_3$                                   & [ ]                                                & $1.4\times10^{-9}$                                \\

\hline
\end{tabular}
\end{tabular*}
\caption{Chemical network employed in \texttt{V1} and \texttt{V4} benchmark tests. Continued from \tab{table:network_1}, \underline{continues} in \tab{table:network_3}. More details in \tab{table:network_1}.}\label{table:network_2}
\end{table*}

\clearpage
\begin{table*}
\begin{adjustwidth}{-0.7cm}{}

\begin{tabular*}{1.1\textwidth}{ll}
\tiny
\centering
\setlength{\tabcolsep}{8pt}
\begin{tabular}[t]{lllll}
\hline

201 & CH$_4$ + CO$^+$                                    & HCO$^+$ + CH$_3$                                   & [ ]                                                & $4.55\times10^{-10}$                              \\
202 & OH$^+$ + OH                                        & H$_2$O$^+$ + O                                     & [ ]                                                & $6.99\times10^{-10}$                              \\
203 & OH + CH$_5$$^+$                                    & H$_2$O$^+$ + CH$_4$                                & [ ]                                                & $6.99\times10^{-10}$                              \\
204 & OH$^+$ + H$_2$O                                    & H$_3$O$^+$ + O                                     & [ ]                                                & $1.3\times10^{-9}$                                \\
205 & OH + H$_2$O$^+$                                    & H$_3$O$^+$ + O                                     & [ ]                                                & $6.9\times10^{-10}$                               \\
206 & OH$^+$ + CO                                        & HCO$^+$ + O                                        & [ ]                                                & $1.05\times10^{-9}$                               \\
207 & OH + CO$^+$                                        & HCO$^+$ + O                                        & [ ]                                                & $3.1\times10^{-10}$                               \\
208 & OH + HCO$^+$                                       & CO + H$_2$O$^+$                                    & [ ]                                                & $6.2\times10^{-10}$                               \\
209 & CH$_5$$^+$ + H$_2$O                                & H$_3$O$^+$ + CH$_4$                                & [ ]                                                & $3.7\times10^{-9}$                                \\
210 & CH$_5$$^+$ + CO                                    & HCO$^+$ + CH$_4$                                   & [ ]                                                & $10^{-9}$                                \\
211 & H$_2$O$^+$ + H$_2$O                                & H$_3$O$^+$ + OH                                    & [ ]                                                & $2.1\times10^{-9}$                                \\
212 & H$_2$O$^+$ + CO                                    & HCO$^+$ + OH                                       & [ ]                                                & $5\times10^{-10}$                                 \\
213 & H$_2$O + CO$^+$                                    & HCO$^+$ + OH                                       & [ ]                                                & $8.84\times10^{-10}$                              \\
214 & H$_2$O + HCO$^+$                                   & CO + H$_3$O$^+$                                    & [ ]                                                & $2.5\times10^{-9}$                                \\
215 & H + H$_2$$^+$                                      & H$_2$ + H$^+$                                      & [ ]                                                & $6.4\times10^{-10}$                               \\
216 & H$^+$ + CH                                         & CH$^+$ + H                                         & [ ]                                                & $1.9\times10^{-9}$                                \\
217 & H$^+$ + CH$_2$                                     & CH$_2$$^+$ + H                                     & [ ]                                                & $1.4\times10^{-9}$                                \\
218 & H$^+$ + CH$_3$                                     & CH$_3$$^+$ + H                                     & [ ]                                                & $3.39\times10^{-9}$                               \\
219 & H$^+$ + CH$_4$                                     & CH$_4$$^+$ + H                                     & [ ]                                                & $1.5\times10^{-9}$                                \\
220 & H$^+$ + OH                                         & OH$^+$ + H                                         & [ ]                                                & $2.1\times10^{-9}$                                \\
221 & H + CO$^+$                                         & CO + H$^+$                                         & [ ]                                                & $7.5\times10^{-10}$                               \\
222 & H$_2$$^+$ + CH                                     & CH$^+$ + H$_2$                                     & [ ]                                                & $7.1\times10^{-10}$                               \\
223 & H$_2$$^+$ + CH$_2$                                 & CH$_2$$^+$ + H$_2$                                 & [ ]                                                & $10^{-9}$                                \\
224 & H$_2$$^+$ + CH$_4$                                 & CH$_4$$^+$ + H$_2$                                 & [ ]                                                & $1.4\times10^{-9}$                                \\
225 & H$_2$$^+$ + OH                                     & OH$^+$ + H$_2$                                     & [ ]                                                & $7.6\times10^{-10}$                               \\
226 & H$_2$$^+$ + H$_2$O                                 & H$_2$O$^+$ + H$_2$                                 & [ ]                                                & $3.9\times10^{-9}$                                \\
227 & H$_2$$^+$ + CO                                     & CO$^+$ + H$_2$                                     & [ ]                                                & $6.4\times10^{-10}$                               \\
228 & H$_2$$^+$ + O$_2$                                  & O$_2$$^+$ + H$_2$                                  & [ ]                                                & $8\times10^{-10}$                                 \\
229 & He$^+$ + CH                                        & CH$^+$ + He                                        & [ ]                                                & $5\times10^{-10}$                                 \\
230 & He$^+$ + CH$_4$                                    & CH$_4$$^+$ + He                                    & [ ]                                                & $5.1\times10^{-11}$                               \\
231 & He$^+$ + H$_2$O                                    & H$_2$O$^+$ + He                                    & [ ]                                                & $6.05\times10^{-11}$                              \\
232 & He$^+$ + O$_2$                                     & O$_2$$^+$ + He                                     & [ ]                                                & $3.3\times10^{-11}$                               \\
233 & C$^+$ + CH                                         & CH$^+$ + C                                         & [ ]                                                & $3.8\times10^{-10}$                               \\
234 & C$^+$ + CH$_2$                                     & CH$_2$$^+$ + C                                     & [ ]                                                & $5.19\times10^{-10}$                              \\
235 & C + CO$^+$                                         & CO + C$^+$                                         & [ ]                                                & $1.09\times10^{-10}$                              \\
236 & C + O$_2$$^+$                                      & O$_2$ + C$^+$                                      & [ ]                                                & $5.2\times10^{-11}$                               \\
237 & CH + O$^+$                                         & O + CH$^+$                                         & [ ]                                                & $3.49\times10^{-10}$                              \\
238 & CH + OH$^+$                                        & OH + CH$^+$                                        & [ ]                                                & $3.49\times10^{-10}$                              \\
239 & CH + H$_2$O$^+$                                    & H$_2$O + CH$^+$                                    & [ ]                                                & $3.4\times10^{-10}$                               \\
240 & CH + CO$^+$                                        & CO + CH$^+$                                        & [ ]                                                & $3.2\times10^{-10}$                               \\
241 & CH + O$_2$$^+$                                     & O$_2$ + CH$^+$                                     & [ ]                                                & $3.1\times10^{-10}$                               \\
242 & CH$_2$ + O$^+$                                     & O + CH$_2$$^+$                                     & [ ]                                                & $9.69\times10^{-10}$                              \\
243 & CH$_2$ + OH$^+$                                    & OH + CH$_2$$^+$                                    & [ ]                                                & $4.8\times10^{-10}$                               \\
244 & CH$_2$ + H$_2$O$^+$                                & H$_2$O + CH$_2$$^+$                                & [ ]                                                & $4.7\times10^{-10}$                               \\
245 & CH$_2$ + CO$^+$                                    & CO + CH$_2$$^+$                                    & [ ]                                                & $4.3\times10^{-10}$                               \\
246 & CH$_2$ + O$_2$$^+$                                 & O$_2$ + CH$_2$$^+$                                 & [ ]                                                & $4.3\times10^{-10}$                               \\
247 & O$^+$ + CH$_4$                                     & CH$_4$$^+$ + O                                     & [ ]                                                & $8.9\times10^{-10}$                               \\
248 & O$^+$ + OH                                         & OH$^+$ + O                                         & [ ]                                                & $3.59\times10^{-10}$                              \\
249 & O$^+$ + H$_2$O                                     & H$_2$O$^+$ + O                                     & [ ]                                                & $3.2\times10^{-9}$                                \\
250 & O + CO$^+$                                         & CO + O$^+$                                         & [ ]                                                & $1.4\times10^{-10}$                               \\

\hline
\end{tabular}

\begin{tabular}[t]{lllll}
\hline

251 & O$^+$ + O$_2$                                      & O$_2$$^+$ + O                                      & [ ]                                                & $1.9\times10^{-11}$                               \\
252 & CH$_4$ + CO$^+$                                    & CO + CH$_4$$^+$                                    & [ ]                                                & $7.92\times10^{-10}$                              \\
253 & CH$_4$$^+$ + O$_2$                                 & O$_2$$^+$ + CH$_4$                                 & [ ]                                                & $4\times10^{-10}$                                 \\
254 & OH$^+$ + H$_2$O                                    & H$_2$O$^+$ + OH                                    & [ ]                                                & $1.59\times10^{-9}$                               \\
255 & OH + CO$^+$                                        & CO + OH$^+$                                        & [ ]                                                & $3.1\times10^{-10}$                               \\
256 & OH$^+$ + O$_2$                                     & O$_2$$^+$ + OH                                     & [ ]                                                & $5.9\times10^{-10}$                               \\
257 & H$_2$O + CO$^+$                                    & CO + H$_2$O$^+$                                    & [ ]                                                & $1.72\times10^{-9}$                               \\
258 & H$_2$O$^+$ + O$_2$                                 & O$_2$$^+$ + H$_2$O                                 & [ ]                                                & $4.59\times10^{-10}$                              \\
259 & CO$^+$ + O$_2$                                     & O$_2$$^+$ + CO                                     & [ ]                                                & $1.2\times10^{-10}$                               \\
260 & H$_2$  + $\gamma$                                  & H + H                                              & [ ]                                                & see text                                          \\
261 & CO  + $\gamma$                                     & C + O                                              & [ ]                                                & see text                                          \\
262 & H$_2$$^+$  + $\gamma$                              & H$^+$ + H                                          & [ ]                                                & see text                                          \\
263 & H$_3$$^+$  + $\gamma$                              & H$_2$$^+$ + H                                      & [ ]                                                & see text (branch ratio 0.5, \textsc{KIDA}) \\
264 & H$_3$$^+$  + $\gamma$                              & H$_2$ + H$^+$                                      & [ ]                                                & see text (branch ratio 0.5, \textsc{KIDA}) \\
265 & C  + $\gamma$                                      & C$^+$ + e$^-$                                      & [ ]                                                & see text                                          \\
266 & CH  + $\gamma$                                     & CH$^+$ + e$^-$                                     & [ ]                                                & see text                                          \\
267 & CH  + $\gamma$                                     & C + H                                              & [ ]                                                & see text                                          \\
268 & CH$^+$  + $\gamma$                                 & C$^+$ + H                                          & [ ]                                                & see text                                          \\
269 & CH$_2$  + $\gamma$                                 & CH$_2$$^+$ + e$^-$                                 & [ ]                                                & see text                                          \\
270 & CH$_2$  + $\gamma$                                 & CH + H                                             & [ ]                                                & see text                                          \\
271 & CH$_2$$^+$  + $\gamma$                             & CH$^+$ + H                                         & [ ]                                                & see text                                          \\
272 & CH$_3$  + $\gamma$                                 & CH$_3$$^+$ + e$^-$                                 & [ ]                                                & see text                                          \\
273 & CH$_3$  + $\gamma$                                 & CH$_2$ + H                                         & [ ]                                                & see text                                          \\
274 & CH$_3$  + $\gamma$                                 & CH + H$_2$                                         & [ ]                                                & see text                                          \\
275 & CH$_4$  + $\gamma$                                 & CH$_3$ + H                                         & [ ]                                                & see text                                          \\
276 & CH$_4$  + $\gamma$                                 & CH$_2$ + H$_2$                                     & [ ]                                                & see text                                          \\
277 & CH$_4$  + $\gamma$                                 & CH + H$_2$ + H                                     & [ ]                                                & see text                                          \\
278 & OH  + $\gamma$                                     & OH$^+$ + e$^-$                                     & [ ]                                                & see text                                          \\
279 & OH  + $\gamma$                                     & O + H                                              & [ ]                                                & see text                                          \\
280 & OH$^+$  + $\gamma$                                 & O + H$^+$                                          & [ ]                                                & see text                                          \\
281 & H$_2$O  + $\gamma$                                 & OH + H                                             & [ ]                                                & see text                                          \\
282 & H$_2$O  + $\gamma$                                 & H$_2$O$^+$ + e$^-$                                 & [ ]                                                & see text                                          \\
283 & O$_2$  + $\gamma$                                  & O + O                                              & [ ]                                                & see text                                          \\
284 & O$_2$  + $\gamma$                                  & O$_2$$^+$ + e$^-$                                  & [ ]                                                & see text                                          \\
285 & H$_2$  + CR                                        & H$^+$ + H + e$^-$                                  & [ ]                                                & $2\times10^{-2} \zeta$                            \\
286 & H$_2$  + CR                                        & H + H                                              & [ ]                                                & $10^{-1} \zeta$                           \\
287 & H$_2$  + CR                                        & H$_2$$^+$ + e$^-$                                  & [ ]                                                & $8.79\times10^{-1} \zeta$                         \\
288 & H  + CR                                            & H$^+$ + e$^-$                                      & [ ]                                                & $4.59\times10^{-1} \zeta$                         \\
289 & He  + CR                                           & He$^+$ + e$^-$                                     & [ ]                                                & $5\times10^{-1} \zeta$                            \\

\hline
\end{tabular}
\end{tabular*}
\end{adjustwidth}

\caption{Chemical network employed in \texttt{V1} and \texttt{V4} benchmark tests. Continued from \tab{table:network_2}, more details in \tab{table:network_1}.}\label{table:network_3}
\end{table*}

\section{\texttt{V1} model with enhanced density}\label{appendix:vhigh}

\begin{figure}
   \includegraphics[width=.48\textwidth]{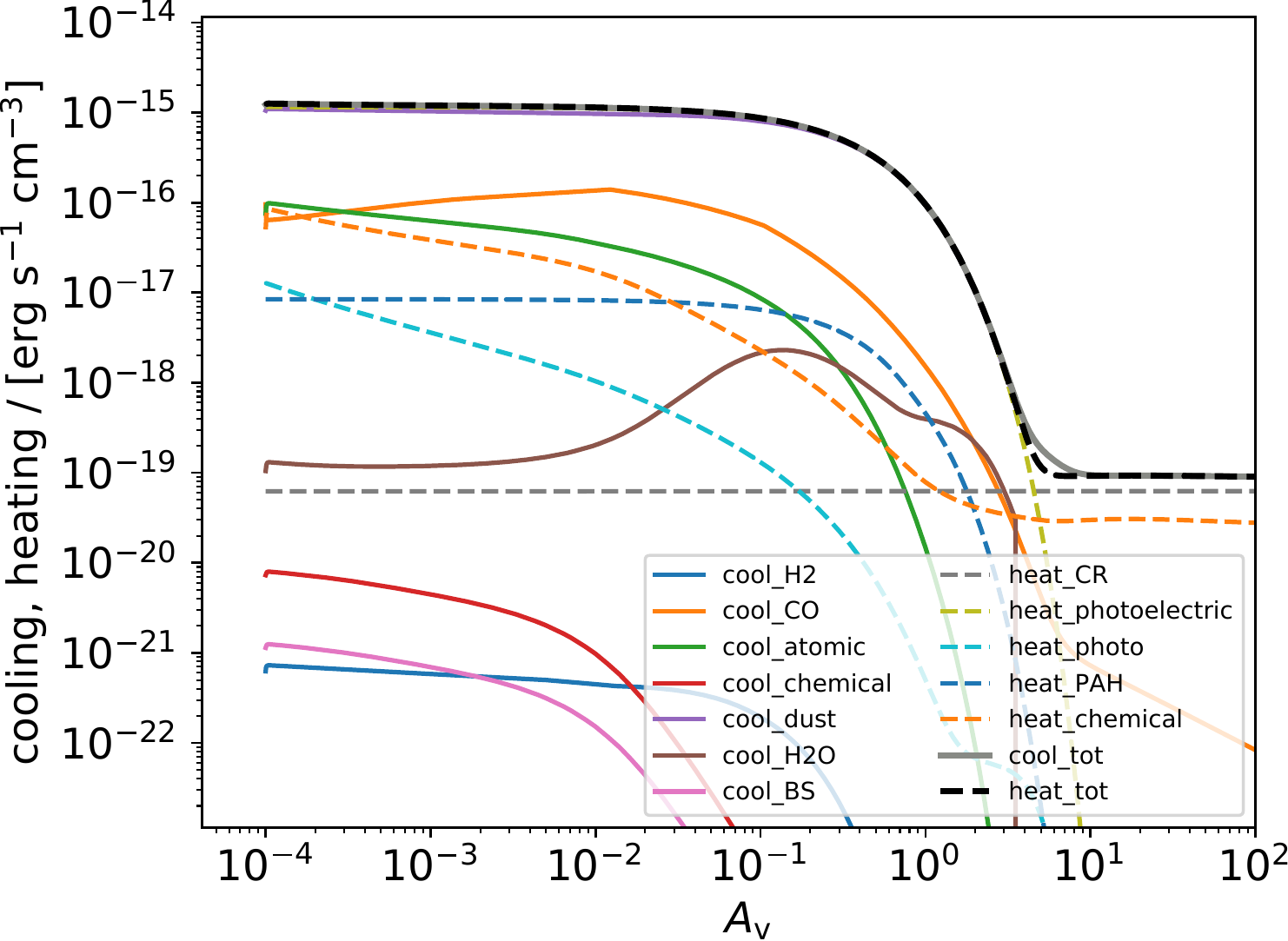}
 \caption{Model \texttt{V1} with enhanced density. Cooling (solid) and heating (dashed) functions at different visual extinction values. Grey solid and black dashed lines indicate the sum of the cooling (\texttt{cool\_tot}) and heating terms (\texttt{heat\_tot}), respectively. In this model $\ngas=10^8$~cm$^{-3}$ and $A_{\rm v}=100$.}
 \label{fig:bench_vhigh_cooling_array_at}
\end{figure}

\begin{figure}
       \includegraphics[width=.48\textwidth]{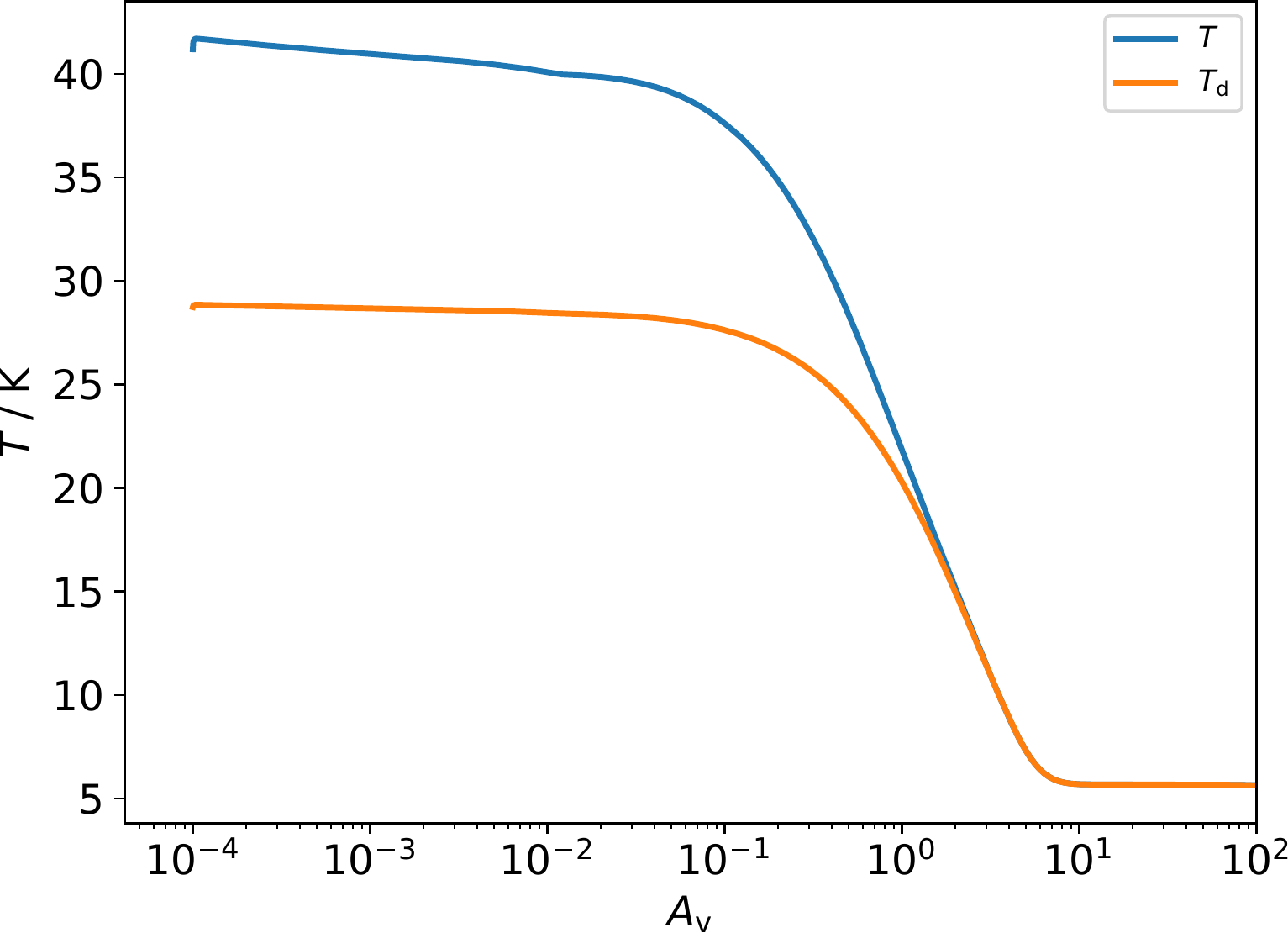}
 \caption{Model \texttt{V1} with enhanced density. Gas (blue) and dust (orange) temperature profiles as a function of the visual extinction $A_{\rm v}$. In this model $\ngas=10^8$~cm$^{-3}$ and $A_{\rm v}=100$.}
 \label{fig:bench_vhigh_Tgas__Tdust}
\end{figure}

To determine the behaviour of the code in an environment similar to what can be found for example in the deeper region of a molecular cloud, we increased the density of the model \texttt{V1} to $\ngas=10^8$~cm$^{-3}$, and extended the maximum visual extinction to $A_{\rm v}=100$, with $1000$~spatial grid points. As shown in \fig{fig:bench_vhigh_cooling_array_at}, at higher $A_{\rm v}$ the radiation becomes almost ineffective, and the only sources of heating are cosmic rays (since $\zeta$ is assumed constant), and the chemical heating due to the formation of molecular hydrogen on dust grains. Analogously, dust cooling dominates the thermal budget, and from \eqn{eqn:lambda_dg} and \eqn{eqn:lambda_dg_gamma} the solution of the heating-cooling balance is then $\tgas\simeq\tdust$, as shown in \fig{fig:bench_vhigh_Tgas__Tdust}.

\clearpage
\section{X-ray molecular reactions}\label{appendix:xray_mols}
The photochemical reactions of  molecular species when X-ray photons are present, are computed summing the cross-sections of individual atomic species, following the criteria of \citet{Adamkovics2011}. The cross sections are reported in \tab{tab:xray_molecules}, where the atomic cross sections $\sigma_{i,j}$ are for the \ith species and the \jth electronic shell.

\begin{table}
    \centering
    \begin{tabular}{ll}
        \hline
            Molecular cross-section & Mixing criterion\\
        \hline
        ${\rm CO \to C^+ + O^+ + 2e^-}$ & ${0.5\, ( \sigma_{\rm C,K} +\sigma_{\rm C,L} + \sigma_{\rm O,K} + \sigma_{\rm O,L})}$\\
        ${\rm H_2O \to O^{++} + 2H + 2e^-}$ & ${\sigma_{\rm H,K} + \sigma_{\rm H,K} + \sigma_{\rm O,K}}$\\
        ${\rm H_2O \to O^+ + 2H + e^-}$ & ${\sigma_{\rm O,L}}$\\
        ${\rm OH \to O^{++} + H + 2e^-}$ & ${\sigma_{\rm O,K} + \sigma_{\rm H,K}}$\\
        ${\rm OH \to O^+ + H + e^-}$ & ${\sigma_{\rm O,L}}$\\
        ${\rm N_2 \to N^+ + N + e^-}$ & ${\sigma_{\rm N,L} + \sigma_{\rm N,L}}$\\
        ${\rm N_2 \to N^{++} + N + 2e^-}$ & ${\sigma_{\rm N,K} + \sigma_{\rm N,K}}$\\
        ${\rm O_2 \to O^+ + O + e^-}$ & ${\sigma_{\rm O,L} + \sigma_{\rm O,L}}$\\
        ${\rm O_2 \to O^{++} + O + 2e^-}$ & ${\sigma_{\rm O,K} + \sigma_{\rm O,K}}$\\
        ${\rm CH_2 \to C^{++} + 2H + 2e^-}$ & ${\sigma_{\rm C,K}}$\\
        ${\rm NH_2 \to N^{++} + 2H + 2e^-}$ & ${\sigma_{\rm N,K}}$\\
        ${\rm CH3 \to C^{++} + 3H + 2e^-}$ & ${\sigma_{\rm C,K}}$\\
        ${\rm NH3 \to N^{++} + 3H + 2e^-}$ & ${\sigma_{\rm N,K}}$\\
        ${\rm CH4 \to C^{++} + 4H + 2e^-}$ & ${\sigma_{\rm C,K}}$\\
        ${\rm NH4 \to N^{++} + 4H + 2e^-}$ & ${\sigma_{\rm N,K}}$\\
        ${\rm CN \to C^{++} + N + 2e^-}$ & ${\sigma_{\rm C,K}}$\\
        ${\rm CN \to C + N^{++} + 2e^-}$ & ${\sigma_{\rm N,K}}$\\
        ${\rm CN \to C^+ + N^+ + 2e^-}$ & ${\sigma_{\rm C,L} + \sigma_{\rm N,L}}$\\
        ${\rm NO \to N^{++} + O + 2e^-}$ & ${\sigma_{\rm N,K}}$\\
        ${\rm NO \to N + O^{++} + 2e^-}$ & ${\sigma_{\rm O,K}}$\\
        ${\rm NO \to N^+ + O^+ + 2e^-}$ & ${\sigma_{\rm N,L} + \sigma_{\rm O,L}}$\\
        ${\rm SiH \to Si^+ + H + e^-}$ & ${\sigma_{\rm Si,K}}$\\
        ${\rm SiO \to Si^+ + O^+ + 2e^-}$ & ${\rm Si_L + \sigma_{\rm O,L}}$\\
        ${\rm SiO \to Si + O^{++} + 2e^-}$ & ${\sigma_{\rm O,K}}$\\
        ${\rm HCN \to C^{++} + NH + 2e^-}$ & ${\sigma_{\rm C,K}}$\\
        ${\rm HCN \to N^{++} + CH + 2e^-}$ & ${\sigma_{\rm N,K}}$\\
        ${\rm CO_2 \to C^{++} + O_2 + 2e^-}$ & ${\sigma_{\rm C,K}}$\\
        ${\rm CO_2 \to O^{++} + CO + 2e^-}$ & ${\sigma_{\rm O,K} + \sigma_{\rm O,K}}$\\
        \hline
    \end{tabular}\caption{Mixing criteria for the cross-sections employed for molecular photochemistry in the presence of X-rays, where $\sigma_{i,j}$ is the cross-section of the \jth electronic shell of the \ith atomic species.}\label{tab:xray_molecules}
\end{table}

\bsp

\label{lastpage}

\end{document}